\documentclass[11pt, letterpaper]{article}
\usepackage[utf8]{inputenc}
\usepackage[T1]{fontenc}
\usepackage{hyperref}
\usepackage{url}
\usepackage{pgfplots}
\usepackage{booktabs}
\usepackage{amsmath}
\usepackage{amsfonts}
\usepackage{nicefrac}
\usepackage{graphicx}
\usepackage{microtype}
\usepackage{amssymb}

\usepackage{float}
\usepackage{caption}
\usepackage{lmodern}
\usepackage{mathtools, nccmath}
\usepackage{authblk}
\usepackage[letterpaper, total={6.5in, 9in}]{geometry}
\usepackage{titling}

\usepackage[backend=biber, style=ieee, sorting=nty]{biblatex}
\usepackage{wrapfig}
\usepackage{hhline}

\usepackage{chngcntr}
\counterwithin{figure}{section}

\usepackage[utf8]{inputenc} 
\usepackage[T1]{fontenc}    
\usepackage{hyperref}       
\usepackage{url}            
\usepackage{booktabs}       
\usepackage{amsfonts}       
\usepackage{nicefrac}       
\usepackage{microtype}      
\usepackage[ruled,vlined]{algorithm2e}

\title{Scalable, Decentralized Multi-Agent Reinforcement Learning Methods Inspired by Stigmergy and Ant Colonies}

\author{%
  Austin A. Nguyen\thanks{Primary Author} \\
  Department of Computer Science\\
  University of California, Berkeley\\
  Berkeley, CA 94709 \\
  \texttt{austinnguyen517@berkeley.edu} \\
}

\date{May 2021}
\begin{document}

\maketitle
\setlength{\parindent}{0pt}
\setlength{\parskip}{0.5em}

\begin{abstract}
  Bolstering multi-agent learning algorithms to tackle complex coordination and control tasks has been a long-standing challenge of on-going research. Numerous methods have been proposed to help reduce the effects of non-stationarity and unscalability. In this work, we investigate a novel approach to decentralized multi-agent learning and planning that attempts to address these two challenges. In particular, this method is inspired by the cohesion, coordination, and behavior of ant colonies. As a result, these algorithms are designed to be naturally scalable to systems with numerous agents. While no optimality is guaranteed, the method is intended to work well in practice and scale better in efficacy with the number of agents present than others. The approach combines single-agent RL and an ant-colony-inspired decentralized, stigmergic algorithm for multi-agent path planning and environment modification. Specifically, we apply this algorithm in a setting where agents must navigate to a goal location, learning to push rectangular boxes into holes to yield new traversable pathways. It is shown that while the approach yields promising success in this particular environment, it may not be as easily generalized to others. The algorithm designed is notably scalable to numerous agents but is limited in its performance due to its relatively simplistic, rule-based approach. Furthermore, the composability of RL-trained policies is called into question, where, while policies are successful in their training environments, applying trained policies to a larger-scale, multi-agent framework results in unpredictable behavior.
\end{abstract}

\section{Introduction}
Devising scalable yet robust learning algorithms for multi-agent reinforcement learning (MARL) has proven to be a challenging endeavor due to complications such as scalability and non-stationarity. The former refers to how state and action spaces grow exponentially with respect to the number of agents. The latter refers to how, from the perspective of local observations by an agent, observation-action pairs do not correspond to fixed transition function distributions nor global reward functions, as these functions also depend on other agents' states and actions as well. While single agent RL has shown promise in narrow control and planning domains, directly applying these approaches to multi-agent environments, particularly those involving more complex coordination tasks, has shown limited success. Because state and action spaces grow exponentially with the number of agents present, such single-agent RL algorithms are bottle-necked by such complexity, hindering sample efficiency. On the other hand, more scalable, decentralized MARL adaptations that address these issues encounter non-stationarity between agent observation spaces, causing high variance learning signals and, as a result, unstable learning. 

Numerous approaches have been proposed to tackle non-stationarity and scalability. In particular, Multi-Agent Deep Deterministic Policy Gradients (MADDPG) \cite{MADDPG}, and Mini-max Multi-Agent Deep Deterministic Policy Gradients (M3DDPG) \cite{M3DDPG} invoke the idea of centralized training ad decentralized execution. A value function is trained in the global, joint space while policies are administered for each agent, using actor critic updates to train agent policies. In this way, global objective learning signals are used to update decentralized policies. Similarly, Counterfactual Policy Gradients (COMA) \cite{Counterfactual} uses the same principle but also uses a counterfactual baseline to more clearly attribute an agent's contribution to the task at hand. While these approaches have had extensive success, they are limited in their scalability and sample efficiency, because value functions are iteratively approximated and learned in joint space during training. 

Other reinforcement learning approaches have attempted to learn both value functions and policies directly in decentralized, local spaces as opposed to the joint space, such as Independent Q-Learning (IQL) \cite{IQL}. While it has worked well in practice, IQL struggles when learning more difficult multi-agent coordination and control tasks due to non-stationarity instability. Strategies such as Nash-Q learning \cite{NashQ, DeepNashQ, Survey1, Survey2, Survey3}, Minimax \cite{Minimax, Survey1, Survey2, Survey3}, and Friend or Foe Q-Learning \cite{FriendOrFoe, Survey1, Survey2, Survey3} have been proposed to solve stochastic games by finding a Nash Equilibrium policy. While these methods work in stochastic game settings, the complexity of the task at hand becomes a significant bottleneck as non-stationarity causes unstable learning. Convergence towards a Nash Equilibrium may take an exorbitant amount of iterations due to high variance updates, assuming that an equilibrium exists.

Other decentralized RL approaches have attempted to reduce the effects of non-stationarity by emphasizing communication between agents. Relevant information regarding the task at hand is encoded and used by other agents. Approaches utilizing opponent modeling such as DRON-Concat and DRON-MoE \cite{Opponent} reduce the effects of non-stationarity by encoding relevant information of other agents and accounting for such information in their own policies. Similarly, other approaches such as Self-Play \cite{Self-play} encourage agents to predict other agents' subsequent actions to achieve more optimal policies themselves. 

Alternatively, hierarchical approaches such as Feudal MADRL \cite{FeudalMARL} attempt to boost scalability of multi-agent systems by specifying a manager policy and numerous worker policies, each representing an agent. The manager learns in joint space to delegate relevant sub-goals to each of the workers, while workers learn in local space to complete assigned sub-goals appropriately. Other hierarchical methods such as Multi-Agent MAXQ \cite{HierarchicalMARL} draw inspiration from mini-max algorithms by decomposing tasks into subtasks. Then, agents share information at each sub-task level as opposed to the joint level, improving the algorithm's scalability. Hierarchical approaches are intuitive, powerful and relatively robust but sometimes suffer due to their complexity, causing instability as well.

Lastly, there are exclusively decentralized algorithms inspired by ant colonies, where agents have access solely to local information and relatively simple policies. Agents communicate indirectly with each other through simple rules and pheromones, real-numbered values that other agents account for in their policies hence stimulating subsequent actions. While these algorithms are highly scalable, simple and interpretable, they typically require domain information and are tailored to solve particular tasks. Malley (et al.) demonstrated the design of a soft robot, coupled with a rule-based policy, to mimic ant bridging behavior \cite{EcitonBridge}. Furthermore, stigmergic, decentralized learning or planning algorithms have been shown to be effective in distance mapping problems \cite{RFID} and iteratively solving intractable problems such as the Traveling Salesman Problem \cite{AntTSP}. Approaches such as Stigmergic Independence Reinforcement Learning \cite{SIRL} have combined the use of pheromones and RL to accomplish narrow coordination tasks.

In this work, we explore an approach that allows a group of automobiles to learn to navigate towards a goal location using environment modification. In particular, environment modification in this case refers to pushing rectangular boxes into holes, effectively creating new traversable paths for the automobiles to travel across. The approach can be viewed as a multi-level hierarchical approach. The automobile agents use hierarchical single-agent RL to learn how to maneuver, reorient, and push a rectangular box across various terrains. Then, using this trained, hierarchical control policy as a primitive, a stigmergic learning algorithm is used to coordinate multiple agents, inducing a navigation plan towards a predetermined location. By separating low-level control from multi-agent coordination, scalability is improved.

In Section 2, a single-agent control policy is trained using hierarchical single-agent RL. The automobile learns how to push a rectangular box across terrain to particular locations and holes. Sections 2.1 through 2.2 provide background on single-agent RL while Section 2.3 and 2.4 introduce the environments used to train the policy. Sections 2.5 and 2.6 outline the single-agent RL algorithm used, and the remainder of Section 2 assesses the approach through ablation tests and performance analysis. 

In Section 3, the trained policy from the preceding section is integrated into a multi-agent planning setting where multiple agents learn how to best navigate to a goal location, whether it be by pushing boxes into holes to create new paths or determining the shortest path to a location. This stigmergic learning algorithm facilitates coordination between agents by developing a pheromone-induced, rule-based policy, using the trained hierarchical control policy from Section 2 as an incorporated action primitive. In this sense, the overall approach is a multi-level hierarchy, with a hierarchical, single-agent RL policy in the middle and a stigmergic coordination algorithm at the top. Sections 3.1 and 3.2 construct the background for the algorithm. Next, Sections 3.3 and 3.4 enumerate the various pheromones the agents use to communicate indirectly amongst themselves, describe the local policy used by all agents, and show how the policy from Section 2 is integrated. Finally, Section 3.5 reports the performance of both the stigmergic algorithm independently and its policy-integrated counterpart. This extended experiment demonstrates the advantages and downfalls of decentralized multi-agent learning. While division of learning controls from learning multi-agent coordination seems promising, the experiments show that success in these domains are relatively limited and multi-agent coordination, without generalizable algorithms, can be explicit in nature.

\section{Hierarchical Single-Agent Reinforcement Learning}
In this section, we use single-agent RL to iteratively train wheeled automobiles to push rectangular boxes to varying locations across flat and sloped terrains. We employ a hierarchical learning approach similar to that introduced in T. Li et al.'s work \cite{Bohg}, where a policy is learned using a finite number of preset low-level controls as its action space. 

The trained policy in this section will be used as a primitive in the multi-agent coordination algorithm introduced in Section 3. As a result, robotic control and multi-agent interaction are trained independently from each other to improve scalability. In this section, we train a two-level control hierarchy for RL training, with fixed, low-level controllers at the bottom and a trainable, mid-level RL policy above it. As a result, the results of this section will be part of a three-level hierarchy in Section 3, where the stigmergic algorithm is placed at the top of the hierarchy.

The automobile's movement is dictated by differential drive inputs. In other words, we specify two angular frequencies, one for the wheels on the left side of the car and another for the wheels on the right. Using a similar method as that introduced in \cite{Bohg}, instead of applying reinforcement learning in continuous action space, we reduce the action space to a discrete space of hand-designed controls. As a result, we can effectively reduce sample complexity and yield faster training. 

First, we outline the background and formulations of single-agent reinforcement learning. Then, we list the variegated environments used in the reinforcement learning process to train the agent. Next, we delineate each of the hand-crafted controls that constitute the automobile's action space to interact with said environments and utilize in the RL algorithm. Lastly, we describe our adaptation to the hierarchical reinforcement learning algorithm to appropriately learn policies, synthesizing all of the previous steps.

\subsection{Related Work}
Single-agent deep reinforcement learning has witnessed significant strides over the past decade as variegated approaches have been proposed to balance or prioritize metrics such as training stability, state space exploration, and the ability to handle sparse rewards. Popular approaches such as DQN \cite{DQN} and DDQN \cite{DDQN} iteratively learn value functions and directly use Q-values in their policies. Deep Deterministic Policy Gradients (DDPG) \cite{DDPG}, Twin Delayed Deep Deterministic Policy Gradients (TD3) \cite{TD3}, and Asynchronous Advantage Actor Critic (A3C) \cite{A3C} have had extensive success in a variety of control problems in continuous state and action spaces. Model-based reinforcement learning (MBRL) for control using MPC \cite{Image-Conditioned} have been extensively used for their sample efficiency and stability. Furthermore, previous works have combined MBRL methods with model-free algorithms for a balance between sample efficiency and robustness \cite{ModelFreeTuning}.

Minorize-maximization algorithms such as Trust Region Policy Optimization (TRPO) \cite{TRPO} and Proximal Policy Optimization (PPO) \cite{PPO} are widely popular for their training stability and computational performance. Furthermore, entropy maximization algorithms such as Soft Actor Critic (SAC) \cite{SAC} has demonstrated promising results particularly for its state space coverage and exploration properties. Hierarchical methods have been proposed to directly tackle sparse reward contexts. Stochastic neural networks \cite{SNN} and unsupervised learning have been used to generate a low-level action space and subsequently learn a high-level policy. Other hierarchical approaches such as HIRO \cite{HIRO} and Feudal Networks \cite{FeudalNets} learn how to decompose tasks by training a manager to delegate sub-goals to a worker who, in turn, learns how to complete given sub-goals.  

\subsection{Background}

In reinforcement learning, we model the environment as a Markov Decision Process (MDP) characterized by the tuple ($\mathcal{S}$, $\mathcal{A}$, $r$, P, $\gamma$, $p_0$). $\mathcal{S} = \{\text{}s_1 , \dots\ ,S\}\ $denotes the state space, $\mathcal{A}$ = $\{\text{}1 , \dots\ ,A\}\ $denotes the action space, $r: \mathcal{S} \times \mathcal{A} \rightarrow \mathbb{R}$ corresponds to the reward function, P: $\mathcal{S} \times \mathcal{A} \times \mathcal{S} \rightarrow \mathbb{R}$ describes the transition dynamics, $\gamma$ $\text{} \in [0, 1)$ is the discount factor, and $p_0$ is the starting state distribution.

The variable $\pi_\theta : \mathcal{S} \times \mathcal{A} \rightarrow [0, 1]$ denotes a policy parameterized by $\theta$, mapping a state action pair to a probability. The policy characterizes the agent's movement and interaction with the environment. The goal of single-agent reinforcement learning is to iteratively derive a policy that maximizes the expected accumulated reward across all trajectories in the environment, as shown in Equation \ref{AccumulatedReward}.

\begin{equation}
    \begin{split}
        J(\theta) & = E_{\tau \sim \pi_{\theta}(\tau)}[R(\tau)] \\
        where \; \pi_{\theta}(\tau) &= p(s_1, a_1, s_2, a_2, ..., ) = p(s_1)\pi_{\theta}(a_1|s_1)\prod_{t=2}^{\infty} p(s_t|s_{t-1}, a_{t-1})\pi_{\theta}(a_t|s_t) \\
        R(\tau) &= R(s_1, a_1, s_2, a_2, ..., ) = \sum_{t=1}^{\infty} r(s_t, a_t)
    \end{split}
    \label{AccumulatedReward}
\end{equation}

For each state-action pair, we define the Q value under a particular policy. The Q value denotes the expected, discounted future accumulated reward under a policy if the agent decides to execute a specific action in a given state. Similarly, a state can be described by a value function, denoting the expected accumulated reward under a policy given the agent is in a particular state. These variables are defined in Equation \ref{QandVDefinition}, where $\gamma$ is the discount factor.

\begin{equation}
    \begin{split}
        Q^{\pi}(s_t, a_t) & = E_{\tau \sim \pi_{\theta}(\tau)}[r_{t}, \gamma r_{t+1} + \gamma^2 r_{t+2}... | s_t, a_t] = \sum_{s'}^{} p(s'|s,a)(r(s, a, s') + \gamma V^{\pi}(s')) \\
        V^{\pi}(s_t) & =  E_{\tau \sim \pi_{\theta}(\tau)}[r_{t}, \gamma r_{t+1} + \gamma^2 r_{t+2}... | s_t] = \sum_{a}\pi(a|s)Q^{\pi}(s_t, a_t)\\
    \end{split}
    \label{QandVDefinition}
\end{equation}

After training, convergence of the Q value and value functions are indicated when the definitions in Equation \ref{ConvergeQV} hold. 

\begin{equation}
    \begin{split}
        Q^{*}(s_t, a_t) & = \sum_{s'}^{} p(s'|s,a)(r(s, a, s') + \gamma V^{*}(s')) \\
        V^{*}(s_t) & = \max_{a_t}Q^{*}(s_t, a_t)\\
    \end{split}
    \label{ConvergeQV}
\end{equation}

\subsection{Training Environments}
To allow varied training environments and quick real-time simulation, we use V-REP/Coppelia Sim software. In particular, we use the Robotnik Summit XL automobile as our agent and a primitive cuboid to simulate the box. We train the agent to push this rectangular box to locations across different terrain. 

\begin{figure}[!ht]
    \setlength{\tabcolsep}{1pt} 
    \begin{tabular}{ccc}
         \includegraphics[width=.3\textwidth, height=4cm]{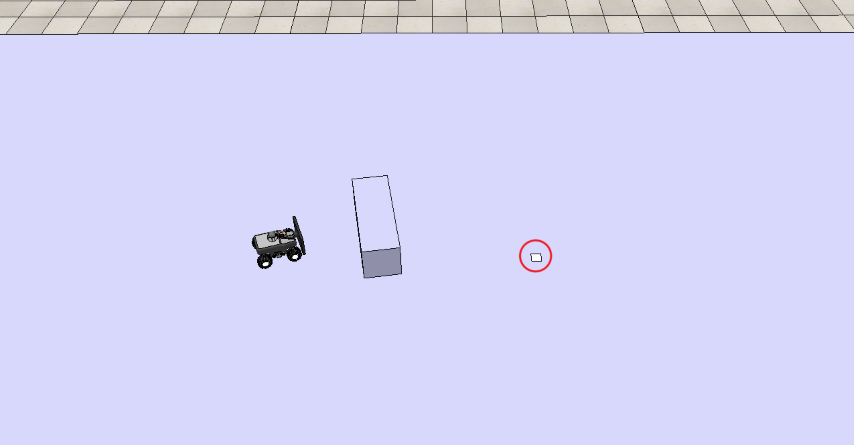} &  
         \includegraphics[width=.3\textwidth, height=4cm]{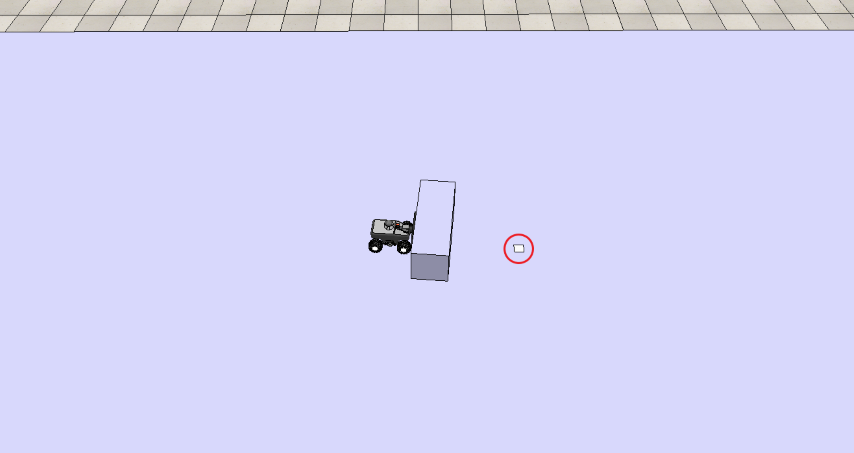} & 
         \includegraphics[width=.3\textwidth, height=4cm]{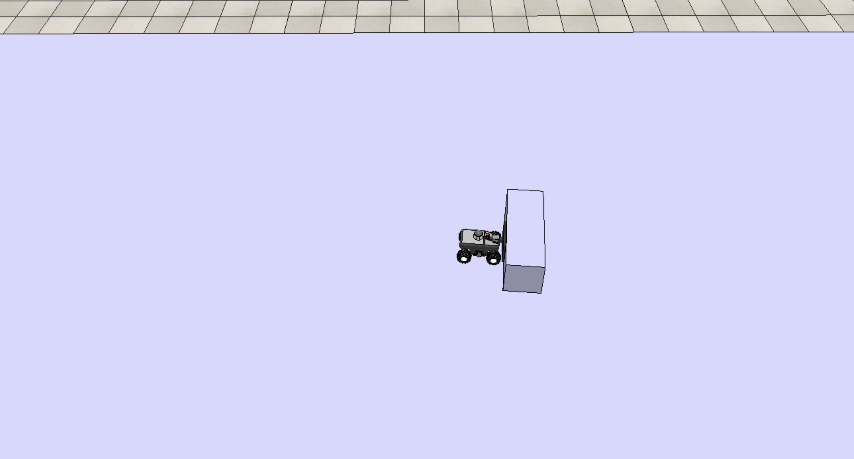}\\
          \includegraphics[width=.3\textwidth, height=4cm]{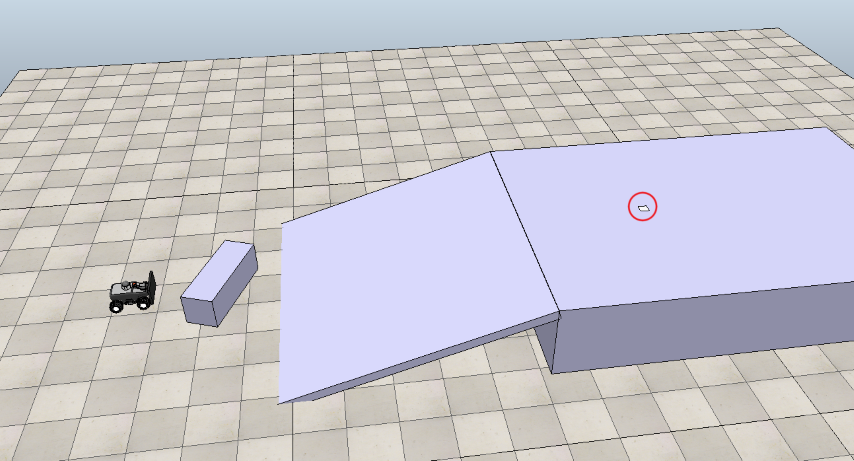} &  
         \includegraphics[width=.3\textwidth, height=4cm]{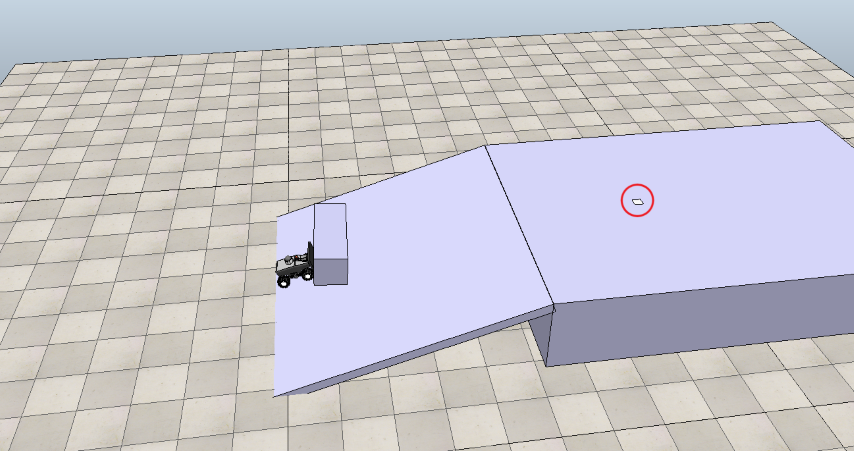} & 
         \includegraphics[width=.3\textwidth, height=4cm]{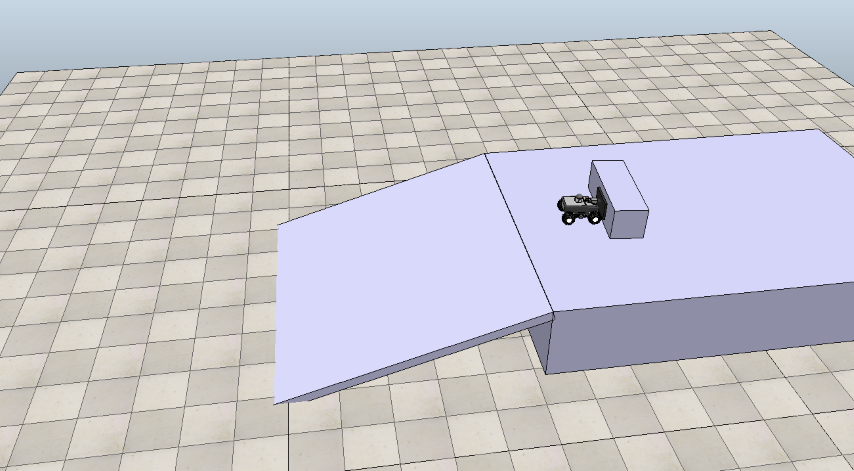} \\
          \includegraphics[width=.3\textwidth, height=4cm]{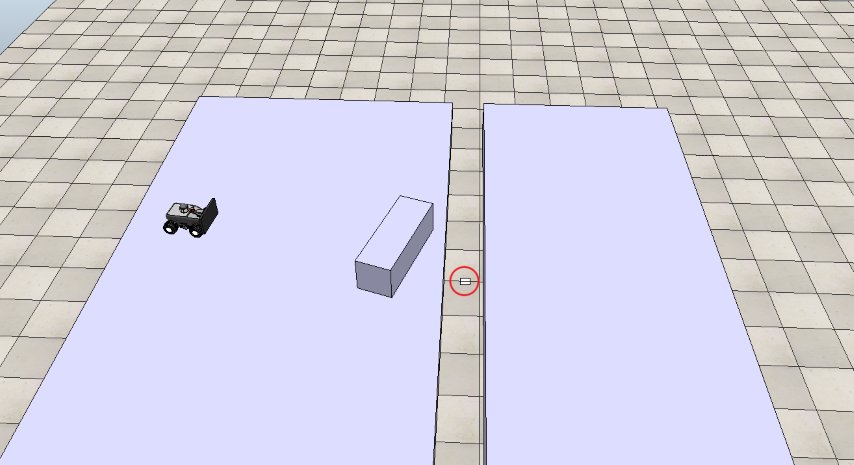} &  
         \includegraphics[width=.3\textwidth, height=4cm]{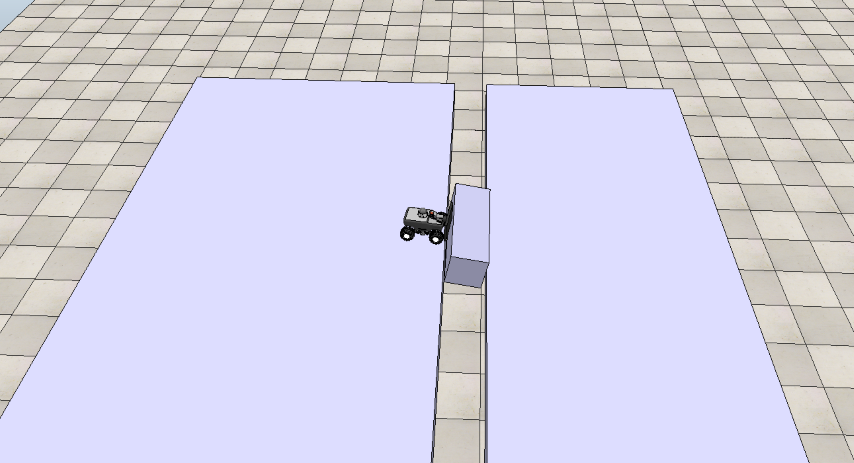} & 
         \includegraphics[width=.3\textwidth, height=4cm]{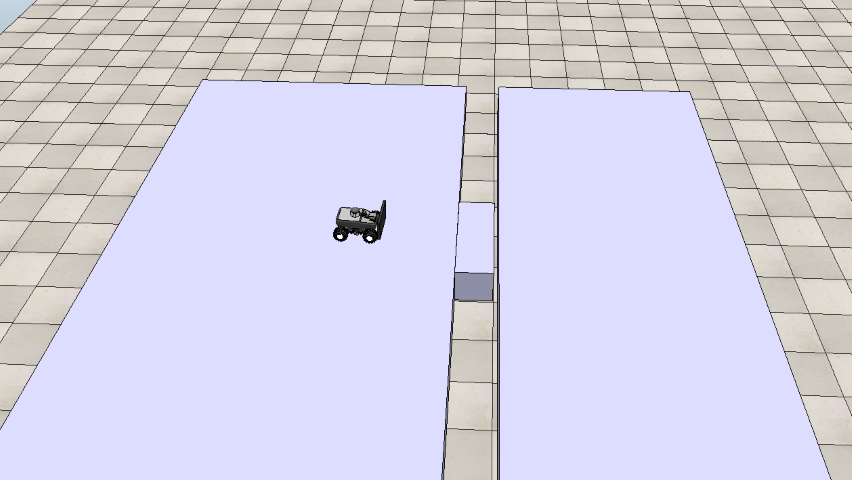}
    \end{tabular}
    \caption{Flat, Slope and Hole environment examples. The goal node is circled in red.}
    \label{SingleTrainEnvs}
\end{figure}

In the first environment, the agent's goal is to autonomously push a box to a given location, denoted by a white marker, on \textbf{flat} terrain as shown in the top row of Figure \ref{SingleTrainEnvs}. In the second, the agent's seeks to push a box in a similar fashion up a \textbf{sloped} terrain as shown in the middle row of Figure \ref{SingleTrainEnvs}. Lastly, the agent's goal is to push a box in a similar fashion into a \textbf{hole} as shown in the bottom row of Figure \ref{SingleTrainEnvs}. Initial conditions for each of the environments are randomized for sufficient state space coverage. In other words, different training episodes have different initial configurations for the agent, box and goal location.

\subsection{Hierarchical Controls}

\begin{figure}
    \centering
    \includegraphics[width=1\linewidth]{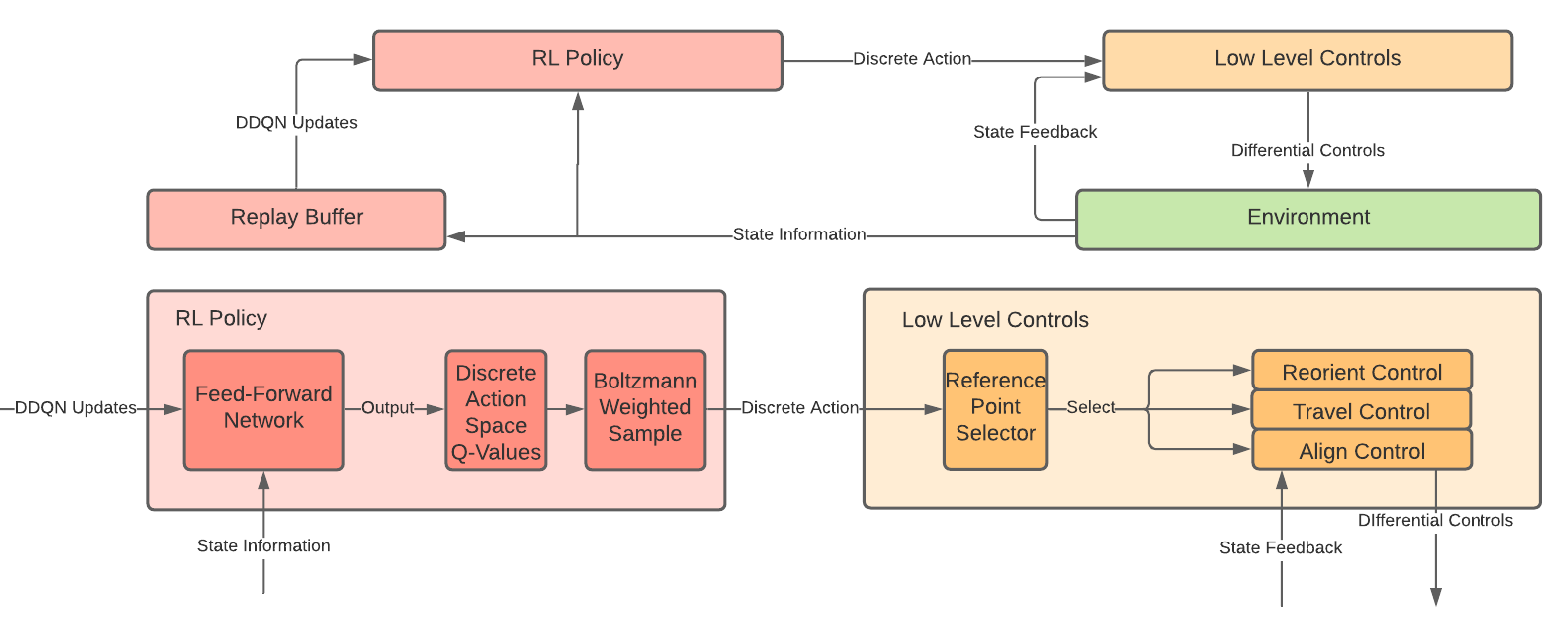}
    \caption{Training Architecture}
    \label{TrainingArchitecture}
\end{figure}
As shown in Figure \ref{TrainingArchitecture}, we use lower level controls as our action space for a reinforcement learning policy. More specifically, when the RL policy, shown in red, returns an action, the lower level controller executes the associated feedback control for $T$ time steps, shown in the yellow State Feedback loop. After $T$ steps of the associated action, the policy receives updated state information, updates its policy, and returns a new discrete action. Here, we outline the information the agent has access to and each of the automobile's differential controls used in the action space for single-agent box-pushing. 

Regarding state space information, the agent has access to position and orientation information as provided by the V-REP simulation. Specifically, all state information variables are measured from the reference frame of the automobile. The state space is characterized by a vector of length 9. We denote $p_{box}^T = (x_{box}, y_{box}, z_{box})$ as the vector location of the box relative to the agent. We denote $a_{box}^T = (\alpha_{box}, \psi_{box})$ as the pitch and yaw of the box. We denote $p_{node}^T = (x_{node}, y_{node}, z_{node})$ as the position of the goal node. Lastly, we denote $\psi_{agent}$ as the automobile's yaw with respect to a line connecting the starting and goal nodes. The state vector enumeration is provided in Equation \ref{StateSpace}. All further variables utilized in subsequent control derivations can be calculated using this state information.

\begin{equation}
    s = 
    \begin{pmatrix} p_{box} \\ a_{box} \\ p_{node} \\ \psi_{agent} \end{pmatrix}
    = \begin{pmatrix} x_{box} \\ y_{box} \\ z_{box} \\ \alpha_{box} \\ \psi_{box} \\ x_{node} \\ y_{node} \\ z_{node} \\ \psi_{agent} \end{pmatrix}
    \label{StateSpace}
\end{equation}

Each control specifies individual frequencies for the wheels on the left and right sides of the car. In other words, each control specifies two real numbers. Furthermore, the controllers are characterized by one of three control hyper-parameters (specified by the user) corresponding to gains of each control. We will call these three gains $gain_{rotate}$, $gain_{travel}$, and $gain_{align}$, denoting their respective purposes.

The discrete action space consists of 8 total controls, each deriving its differential frequencies from one of three controls below. A summary outlining all 8 controls is shown in Table \ref{table:AllControlsTable}. After an agent selects an action corresponding to one of the controls, all $L$ subsequent differential calculations use that same control. Only after these $L$ differentials are finished is another action selected.

\subsubsection{Orientation}

\begin{figure}[!ht]
    \begin{minipage}{0.48\textwidth}
        \centering
        \includegraphics{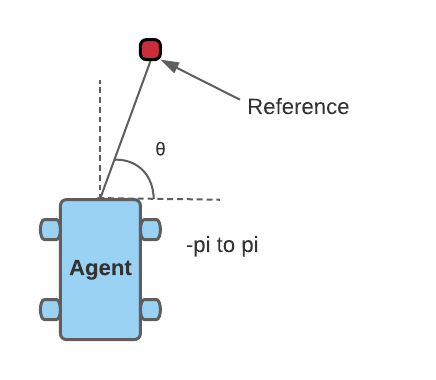} 
        \label{fig:subim1}
    \end{minipage}
    \begin{minipage}{0.48\textwidth}
        \centering
        \begin{equation}
        \begin{pmatrix} \omega_{L} \\ \omega_{R} \end{pmatrix}
        = gain_{rotate} * \begin{pmatrix} cos \,\theta \\ -cos \, \theta \end{pmatrix}
        \end{equation}
    \end{minipage}
    
    \caption{Reorient Control}
    \label{ReorientDiagram}
\end{figure}

The first kind of control corresponds to reorienting the automobile to face a specified point in the environment. The necessary variable corresponds to a single angle we denote as $\theta$, measured between $-\pi$ and $\pi$, with corresponding control denoted by Figure \ref{ReorientDiagram}.

\subsubsection{Traveling}

\begin{figure}[!ht]
    \begin{minipage}{0.48\textwidth}
        \centering
        \includegraphics{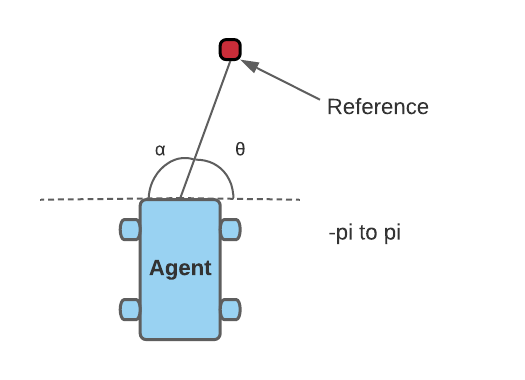} 
        \label{fig:subim1}
    \end{minipage}
    \begin{minipage}{0.48\textwidth}
        \centering
        \begin{equation}
        \begin{pmatrix} \omega_{L} \\ \omega_{R} \end{pmatrix}
        = gain_{travel} * \begin{pmatrix} \alpha \\ \theta \end{pmatrix}
        \end{equation}
    \end{minipage}
    
    \caption{Travel Control}
    \label{TravelDiagram}
\end{figure}

The next kind of control corresponds to traveling towards a specified point in the environment. From the perspective of the automobile's coordinate frame, we derive two angles for the two control frequencies, respectively. We call these angles $\alpha$ and $\theta$, measured between $-\pi$ and $\pi$. The control is shown in Figure \ref{TravelDiagram}.

\subsubsection{Alignment}

\begin{figure}[!ht]
    \begin{minipage}{0.48\textwidth}
        \centering
        \includegraphics{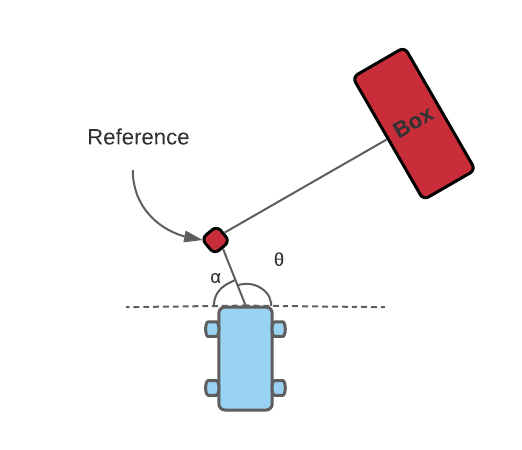} 
        \label{fig:subim1}
    \end{minipage}
    \begin{minipage}{0.48\textwidth}
        \centering
        \begin{equation}
        \begin{pmatrix} \omega_{L} \\ \omega_{R} \end{pmatrix}
        = gain_{align} * \begin{pmatrix} \alpha * dist_{ref} \\ \theta * dist_{ref} \end{pmatrix}
        \end{equation}
    \end{minipage}
    
    \caption{Alignment Control}
    \label{AlignmentDiagram}
\end{figure}

The last control corresponds to aligning the automobile to the closest point that is on the line connecting the box's location and the goal location. This control finds the this corresponding point, uses equation B and inserts a P-controller to restrict movement as the automobile is more closely aligned. This P-controller adds a dampener that is inversely proportional to the distance to the specified, alignment point. The control is shown in Figure \ref{AlignmentDiagram}.

\begin{figure}[!ht]
    \centering
    \includegraphics{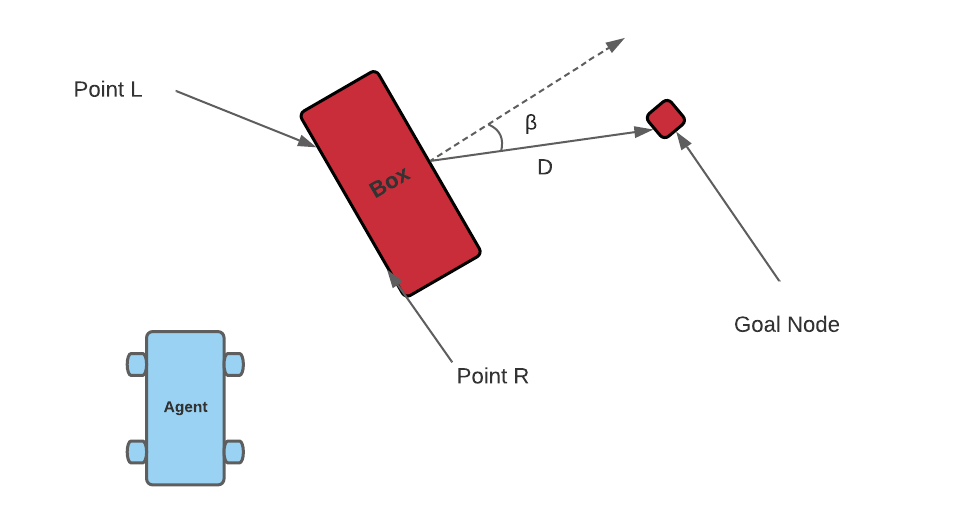} 
    \caption{Reference Points and Box Location}    \label{ReferencePoints}
\end{figure}

\begin{table}[!ht]
    \centering
    \begin{tabular}{|c||c|c|}
     \hline
     \multicolumn{3}{|c|}{\textbf{Controls}} \\
     \hline
     \textbf{Action Name} & \textbf{Control} & \textbf{Reference Point} \\
     \hline
     Move Backwards & \multicolumn{2}{|c|}{$\omega_L = \omega_R = -\,gain_{travel}$}\\
     \hline
     Approach & Equation 6 & Point Box \\
     \hline
     Angle Towards the Box & Equation 5 & Point Box \\
     \hline
     Align & Equation 7 & Point shown Figure 5 \\
     \hline
     Push In & Equation 6 & Point Node \\
     \hline
     Push Left & Equation 6 & Point L \\
     \hline
     Push Right & Equation 6 & Point R \\
     \hline
     Angle Towards Goal & Equation 6 & Point Node \\
     \hline
    \end{tabular}
    \caption{Controls and Associated Calculations}
    \label{table:AllControlsTable}
\end{table}

\subsection{Reward Function}

To reiterate, our automobile agent uses reinforcement learning to discover how to push a box to various locations. One of the requisites of RL is defining an appropriate reward function. 

We define a reward function that encourages two things: pushing the box to the goal location and orienting the box towards the location. As a result, we define two variables for a given state: $D$ and $\beta$. The former denotes the distance of the box to the goal location. The latter denotes the yaw heading from the box to the goal location. An illustration is provided in Figure \ref{ReferencePoints}.

To prevent numerical instability in Q-value estimation, rewards are calculated using the differences between the previous and current values of these variables. A scalar is added to encourage the agent to push the box to the location more quickly. This reward function is shown in Equation \ref{RewardFunction}.

\begin{equation}
r(s) = (D_{prev} - D_{curr}) * 5 + (abs(\beta_{prev}) - abs(\beta_{curr})) * 2 - .1
\label{RewardFunction}
\end{equation}

The episode is considered successful and restarts accordingly when any of the Success Conditions outlined in Table \ref{SuccessFailureConditions} are fulfilled. Similarly, the episode is considered a failure and restarts accordingly when any of the Failure Conditions outlined are fulfilled. 

\begin{table}[!ht]
    \centering
    \begin{tabular}{ |p{.8cm}||p{7.1cm}| p{7.2cm}|}
     \hline
     \multicolumn{2}{|c|}{Success Conditions} & Failure Conditions \\
     \hline
     Hole & $D \leq .2 $ \& $ height_{box} < .2 $ \& $height_{agent} > .2$ & ($D>.2 $ \& $ height_{box} < .2 $) | $height_{agent} < .2$ \\
     Flat & $D \leq .2 $ \& $ yaw_{box} < .2$ & $D>5 $ | $ yaw_{box} > .3$\\
     Slope & $D \leq .2 $ \& $ yaw_{box} < .2$ & $D>5 $ | $ yaw_{box} > .3$\\
     \hline
    \end{tabular}
    \caption{Conditions necessary for an episode to be considered a success or failure in the three training environments.}
    \label{SuccessFailureConditions}
\end{table}

If neither the success nor failure conditions are met in any transition, the we limit the number of steps for one episode in a given environment. Because the sloped environment naturally takes more time to complete due to its increased length between starting and ending positions, we allow relatively more steps in this environment. These maximum steps are shown in Table \ref{MaxSteps}.

\begin{table}[!ht]
    \begin{center}
    \begin{tabular}{ |p{2cm}||p{1cm}|}
     \hline
     \multicolumn{2}{|c|}{Max Steps} \\
     \hline
     Hole & 50 \\
     Flat & 50 \\
     Slope & 100 \\
     \hline
    \end{tabular}
    \caption{Maximum number of steps per episode in each environment.}
    \label{MaxSteps}
    \end{center}
\end{table}

\subsection{Double DQN}

Ubiquitous in reinforcement learning formulations, the Q-value is defined as the expected accumulated rewards of all subsequent observations given a policy and a current state-action pair as given in Equation \ref{QandVDefinition}.

As a result, a policy revolving around this value would ideally select actions with higher Q-values given the current state. The Deep Q-Network algorithm (DQN) \cite{DQN} uses a feed-forward neural network, parameterized by $\theta$ to estimate Q-values given a state. The algorithm uses temporal difference (TD) learning to, over time, approximate these Q-values. Furthermore, DQN also uses a target network for more stability. 

Denoting $Q$ as Q-values estimated by our training network and $Q'$ as Q-values estimated by our target network, TD learning updates and gradient descent are used to update network parameters $\theta$. The algorithm explores the environment and slowly fills an experience replay buffer consisting of state action transitions. Then, every action step, we sample a finite batch of size $N$ from the experience replay and update the policy according to Equation \ref{SingleDQNUpdate}. 

\begin{equation}
    \begin{split}
        TD(s_t, a_t, s_{t+1}) & = (r(s_t, a_t, s_{t+1}) + \max_{a_{t+1}}Q(s_{t+1}, a_{t+1})) - Q'(s_t, a_t) \\
        J(\theta) & = (1/N)\sum_{s, a, s'}TD(s, a, s') \\
        \theta & = \theta - \nabla J(\theta)
    \end{split}
    \label{SingleDQNUpdate}
\end{equation}

However, DQN is susceptible to overestimation of Q-values, resulting in unstable, slower learning. As a result, the Double Deep Q Network algorithm \cite{DDQN} proposes a different update as shown in Equation \ref{SingleDDQNUpdate}.

\begin{equation}
    \begin{split}
        TD(s_t, a_t, s_{t+1}) & = (r(s_t, a_t, s_{t+1}) + Q(s_{t+1}, argmax_{a_{t+1}}Q'(s_{t+1}, a_{t+1}))) - Q(s_t, a_t) \\
        J(\theta) & = (1/N)\sum_{s, a, s'}TD(s, a, s') \\
        \theta & = \theta - \nabla J(\theta)
    \end{split}
    \label{SingleDDQNUpdate}
\end{equation}

In discrete action spaces, Q-learning typically dictates that the agent greedily chooses the action corresponding to the highest Q-value. However, in practice, this causes issues regarding exploration and stagnation of the agent, particularly when chosen actions cause the agent to remain in place. As a result, we decide to choose subsequent actions using a Boltzmann variable, $\beta$, shown in Equation \ref{BoltzmannQPolicy}.  

\begin{equation}
    \pi_{\theta}^{\beta}(s, a) = \frac{\exp(\beta * Q(s, a))}{\sum_{a'} \exp(\beta * Q(s, a'))}
    \label{BoltzmannQPolicy}
\end{equation}

The Boltzmann constant corresponds to the certainty or fidelity we place upon the Q-values. We further encourage exploration by adding a finite probability $\epsilon$ in which the agent chooses an action at uniform random. This probability is decayed exponentially throughout training by factor $d_{\epsilon}$ every time the agent invokes exploration.

We denote the distribution of varied training environments as $\mathcal{E}$. In Section 2.7.2, we delineate the two ways in which we define this distribution. The full Hierarchical DDQN algorithm employed is outlined in Algorithm \ref{HierarchicalDDQNAlg}. 

\begin{algorithm}[H]
\SetAlgoLined
\textbf{Initialize} primary network $Q_{\theta}$, target network $Q'_{\theta'}$, learning rate $\alpha$, discount factor $\gamma$, Boltzmann $\beta$, exploration probability $\epsilon$, exploration decay $d_{\epsilon}$, feedback loop length $T$, replay buffer $\mathcal{D}$, environment distribution $\mathcal{E}$, uniform distribution $U$, target network update frequency $M$, batch size $b$, initial exploration steps $S$\\
\For{each episode}{
    Sample environment $e_i \sim \mathcal{E}$ with initial state distribution $p_e$\\
    Sample initial state $s_t \sim p_e$\\
    Set done to False\\
    
    \While{not done}{
        \eIf{with probability $1- \epsilon$ and $|\mathcal{D}| \geq S$}{
            Select $a_t \sim \pi_{\theta}^{\beta}(s_t, a_t)$\\}
            {
            Select $a_t \sim U(a_t)$\\
            $\epsilon = \epsilon * d_{\epsilon}$ if $|\mathcal{D}| \geq S$\\
            }
        Compute and execute ($\omega_L$, $\omega_R$) with $a_t$ for $T$ steps\\
        Check if done, observe $s_{t+1}$ \\
        Determine reward $r_t$ = R($s_t$, $a_t$, $s_{t+1}$)\\
        Store ($s_t$, $a_t$, $r_t$, $s_{t+1}, done$) in $\mathcal{D}$\\
        
        Sample $b$ batches $e_t = (s_t$, $a_t$, $r_t$, $s_{t+1}, done) \sim \mathcal{D}$\\ 
        Compute $Q^* = r_t + \gamma Q(s_{t+1}, argmax_{a_{t+1}}Q'(s_{t+1}, a_{t+1}))$\\
        
        $J(\theta) = (1/N)\sum_(Q^* - Q_{\theta}(s, a))^2$ \\
        $\theta = \theta - \alpha \nabla J(\theta)$ \\
        
        \If{$M$ steps since last update}{
            $\theta' = \theta$
        }
    }
}
 \caption{Hierarchical Double DQN Algorithm}
 \label{HierarchicalDDQNAlg}
\end{algorithm}

\subsection{Experiments and Results}
\subsubsection{Ablation Tests}

\begin{table}[!ht]
\centering
\begin{tabular}{ |p{5cm}||p{1.8cm}|}
 \hline
 \multicolumn{2}{|c|}{Training Parameters} \\
 \hline
 Input Nodes & 10  \\
 Hidden Width & 200\\
 Hidden Depth & 3 \\
 Output Nodes & 8 \\
 Optimizer & Adam  \\
 Learning Rate ($\alpha$) & 3e-4 \\
 Discount ($\gamma$) & .975 \\
 Buffer Size ($|\mathcal{D}|$) & 10000 \\
 Boltzmann ($\beta$) & 8 \\
 Max Number of Steps & 50 \\
 Feedback Loop Length ($T$) & 50 \\
 Batch Size ($b$) & $-$ \\
 Initial Exploration ($S$) & $-$ \\
 Target Update Frequency ($M$) & $-$ \\
 \hline
\end{tabular}
\caption{DDQN training parameters. The last three parameters are unspecified as they are varied throughout ablation tests.}
\label{DDQNTrainingParameters}
\end{table}

To determine a strong set of hyperparameters for the aforementioned environments, we perform ablation tests with respect to three parameters: batch size $b$, initial exploration steps $S$, and target network update frequency $M$. Afterwards, we choose final hyperparameters prioritizing both end performance and sample efficiency. After $M$ steps under a uniform policy (i.e. $a_t \sim U(a_t)$), the agent alternates between 20 episodes of training and 20 episodes of testing. In training, the process outlined in Algorithm \ref{HierarchicalDDQNAlg} is adhered to verbatim. During testing, $\epsilon = 0$, $\mathcal{D}$ is stagnated, gradient updates to $Q_{\theta}$ are skipped, and observation steps are not accounted for when checking steps since target network updates, $M$. All parameters and processes are returned to normal once training continues again.

\begin{figure}[!ht]
    \begin{tabular}{cc}
     \includegraphics[width=.45\textwidth]{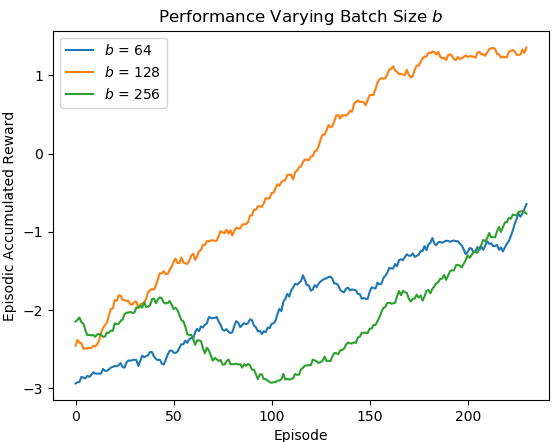} &
      \includegraphics[width=.45\textwidth]{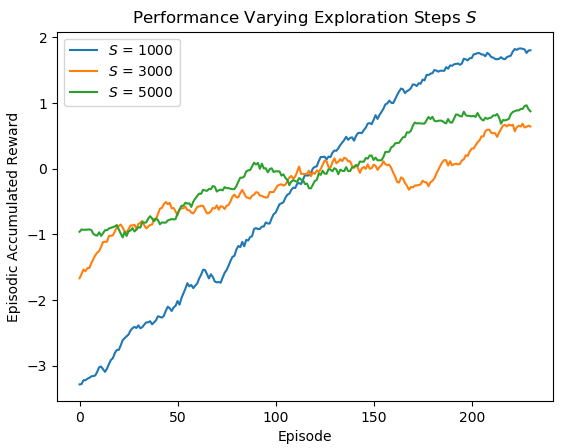} \\
      \multicolumn{2}{c}{
       \includegraphics[width=.45\textwidth]{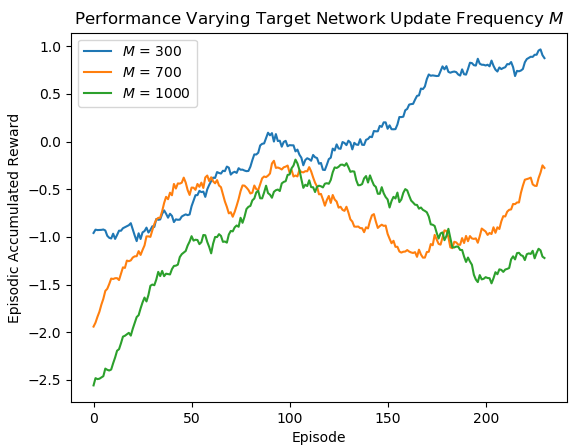}}\\
    \end{tabular}
    \caption{Ablation tests over a) batch size, b) exploration steps and c) target network frequency. For each set of hyperparameters, unless varied by the respective ablation, $b = 128$, $S = 1000$, $M = 300$. Episodes that executed actions as part of exploration $S$ are not included in the graphs.}
    \label{SingleDDQNAblation}
\end{figure}

We restrict ablation tests to the hole environment as it is, objectively, the most difficult environment for the agent to navigate. After 800 total episodes, implying 400 training episodes and 400 testing episodes, we analyze the accumulated episodic reward in the testing phases, where plots are averaged over 150 episodes. We gauge performance by including accumulated reward data only from the testing episodes. In other words, graphs shown in Figure \ref{SingleDDQNAblation} only represent the moving average performance of the 400 testing episodes. In this way, the plots represent the policy's performance in the environment decoupled from the exploration aspect of the algorithm. All episodes that executed actions as part of the initial exploration steps $S$ are not included in the graphs in Figure \ref{SingleDDQNAblation}. Table \ref{DDQNTrainingParameters} show parameters used throughout training.

Firstly, DDQN performance varied widely with respect to batch size, $b$. A relatively low batch size $b = 64$ generated weaker performance than the optimal batch size $b = 128$. This was likely because smaller batch sizes typically yield high variance gradient updates which, unless learning rate is tuned accordingly, sometimes yields unstable training. In contrast, a high batch size $b = 256$ provides possibly more stable updates, but slower learning and worse end performance. As a result, a batch size of $b = 128$ yields the best performance of the three batches sizes tested. These findings are shown in Figure \ref{SingleDDQNAblation}. 

Analyzing initial exploration steps, $S$, we found that, although larger exploration steps such as $S = 3000$ or $S = 5000$ provide better initial performance, the effects are transitory, as lower exploration iterations such as $S = 1000$ provide faster learning and better end performance. This effect can be credited to the quality or usefulness of transitions stored in the experience replay, $\mathcal{D}$. When $S$ is large, $\mathcal{D}$ will contain a larger proportion of transitions that follow a uniform, exploration policy as opposed to the current policy $\pi_{\theta}$, particularly in earlier stages of training. This exploration data may not be as useful as exploitation data provided by policies similar to $\pi_{\theta}$, because exploitation allows training in more vital states farther down state trajectories. Keeping $S$ small allows $\mathcal{D}$ to have a higher proportion of transitions with these vital states.

Finally, target network update frequency $M$ shows varying performance between different parameters. Although higher values of $M$ are known to have better training stability and are less susceptible to Q-value overestimation, it was found that, among the three values tested, the smallest update frequency $M = 300$ yielded the fastest training. This aligns with intuition, as lower values of $M$ allow DDQN to more quickly account for Q-values in future states with the risk of calculating inaccurate target Q-values. However, setting $M$ to a relatively lower value did not detract from end performance in this experiment, allowing faster training with minimal detriments.

In summary, we finalized the batch size to $b = 128$, initial exploration steps to $S = 1000$, and target network update frequency to $M = 300$ based on empirical data shown in Figure \ref{SingleDDQNAblation}.

\subsubsection{Policy Style Tests}
Lastly, we consider whether it is more beneficial to train disparate, individual policies for each environment (i.e. $|\mathcal{E}| = 1$ has a single element and we run Algorithm \ref{HierarchicalDDQNAlg} three times, each with different environments) or train a single policy to navigate all three environments (i.e. $|\mathcal{E}| = 3$ where each environment is associated with a probability $p_{hole}$, $p_{flat}$, or $p_{slope}$ and Algorithm 1 is run once). In this section, we compare performance between the two policy training styles, denoting the former as Individual and the latter as All-in-One. Notice, the former training method yields a total of three Q network policies while the latter yields only one. 

\begin{figure}[!ht]
    \centering
    \begin{tabular}{cc}
        \includegraphics[width=.42\textwidth]{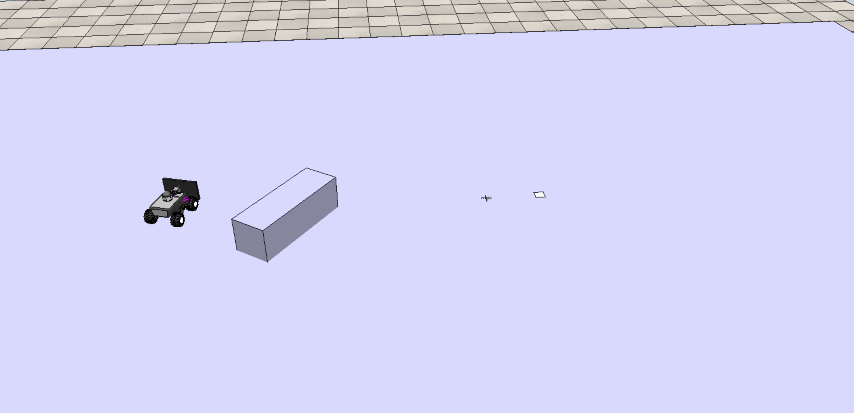} &
        \includegraphics[width=.42\textwidth]{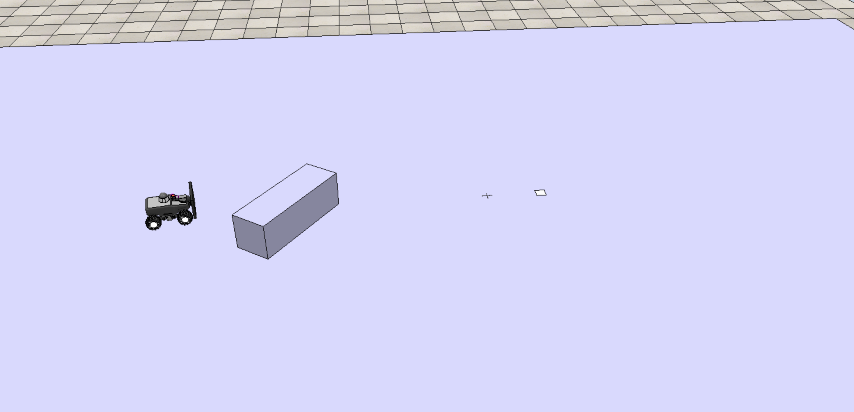} \\
        \includegraphics[width=.42\textwidth]{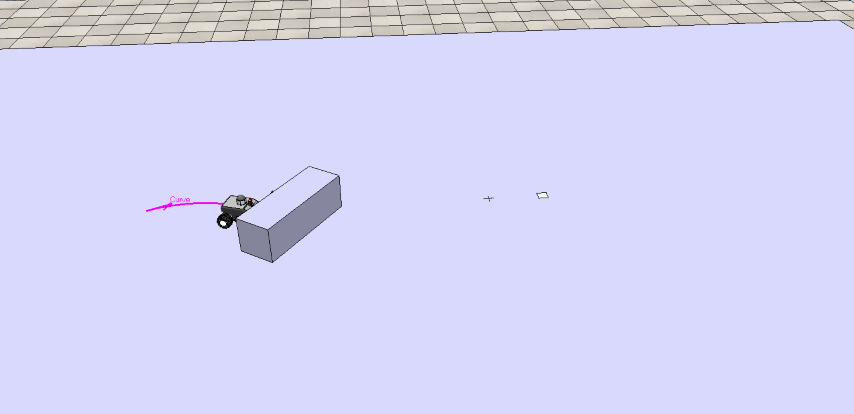} &
        \includegraphics[width=.42\textwidth]{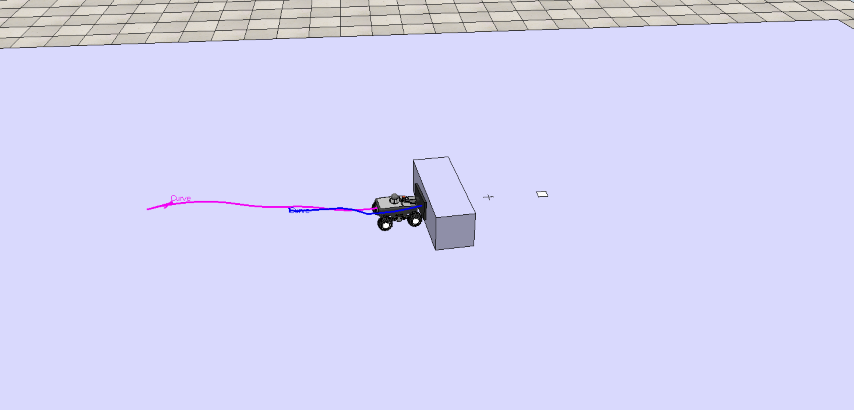} \\
        \includegraphics[width=.42\textwidth]{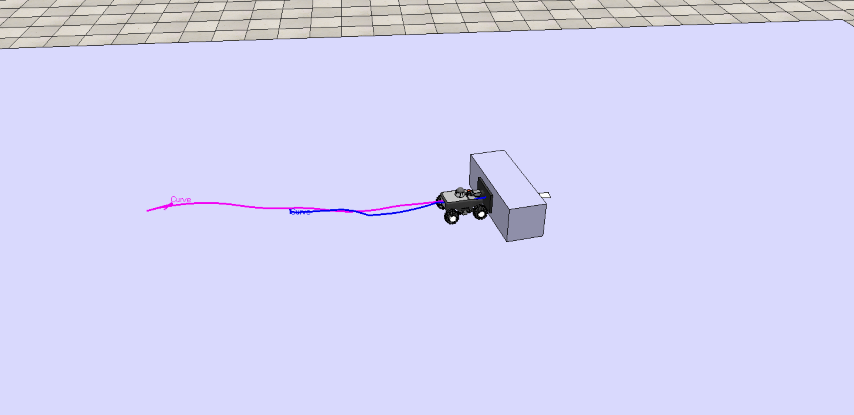} &
        \includegraphics[width=.42\textwidth]{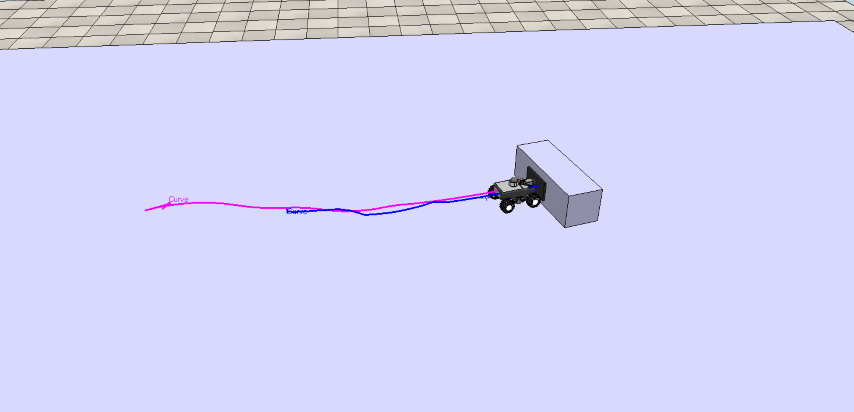} \\
    \end{tabular}
    \caption{A sample trajectory of the trained policy in the flat environment. Pink line represents the agent trajectory and the blue line represents the box trajectory.}
    \label{TrajectoryFlat}
\end{figure}

\begin{figure}[!ht]
    \centering
    \begin{tabular}{cc}
        \includegraphics[width=.42\textwidth]{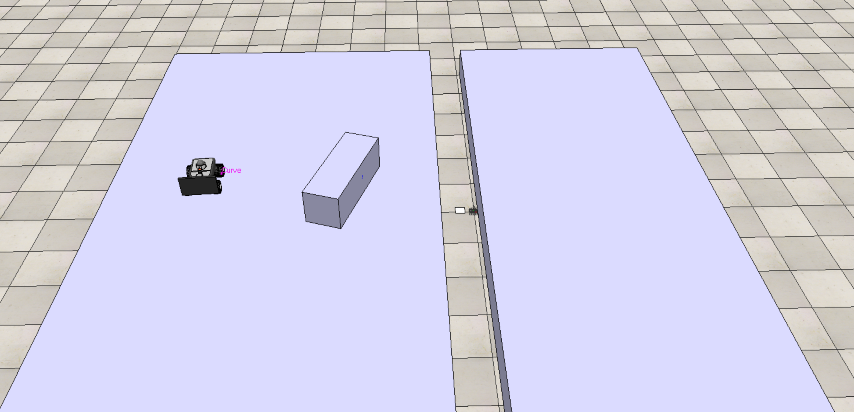} &
        \includegraphics[width=.42\textwidth]{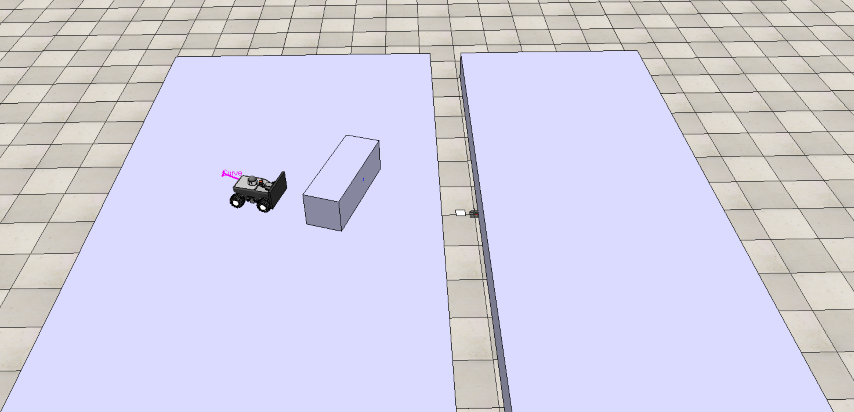} \\
        \includegraphics[width=.42\textwidth]{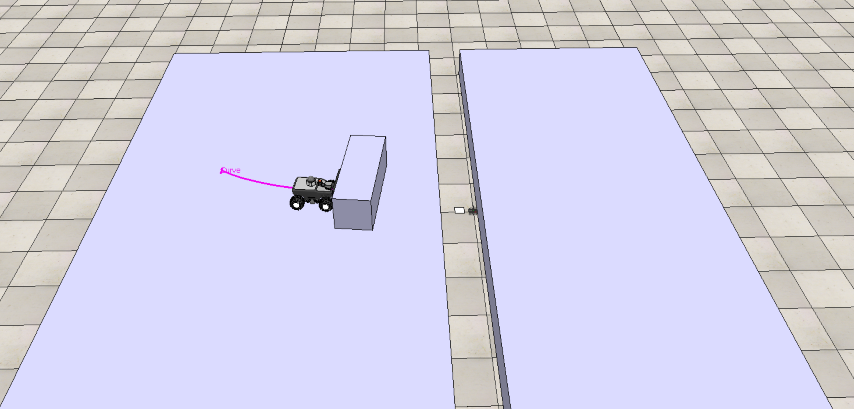} &
        \includegraphics[width=.42\textwidth]{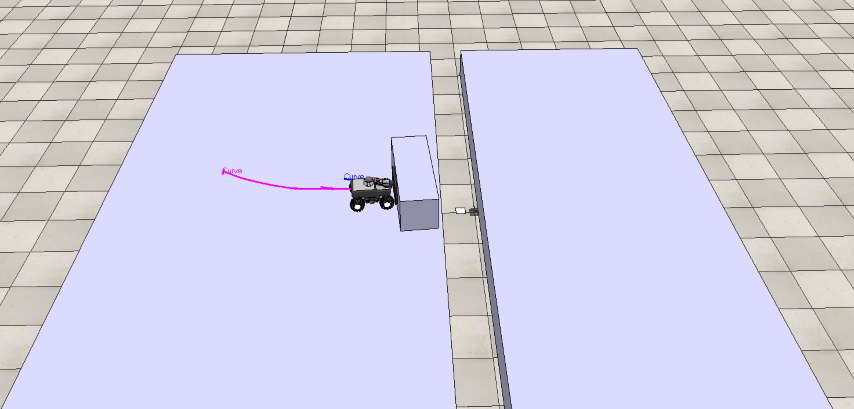} \\
        \includegraphics[width=.42\textwidth]{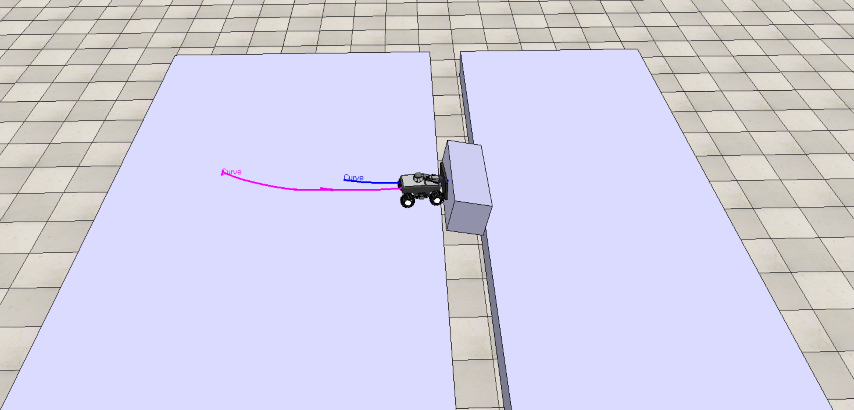} &
        \includegraphics[width=.42\textwidth]{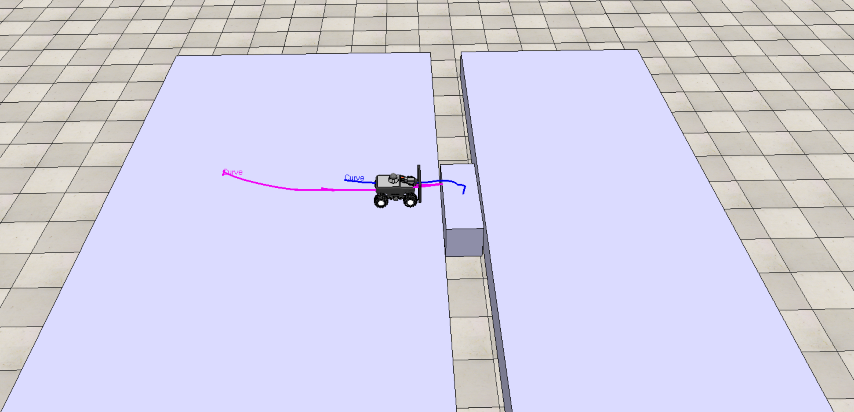} \\
    \end{tabular}
    \caption{A sample trajectory of the trained policy in the hole environment. Pink line represents the agent trajectory and the blue line represents the box trajectory.}
    \label{TrajectoryHole}
\end{figure}

We determine the probabilities $p_{hole}$, $p_{flat}$, and $p_{slope}$ based on environment difficulty and time required to complete the task. In other words, tasks that take more actions to complete are given lower distribution probabilities and vice versa. That way, replay samples are more uniform across all three environments. Environment difficulty and length were determined qualitatively and respective probabilities were adjusted accordingly as shown in Figure \ref{PolicyStyleTest}. The flat and hole environments were given 50 steps to achieve the task, while the sloped environment was delegated 100, as it was an inherently longer task to complete. As a result, to compensate for this imbalance, we set $p_{flat} = .3$, $p_{hole} = .5$, and $p_{slope} = .2$. We set $p_{hole} > p_{flat}$ because the hole environment shares less in its state space with the other two. The sloped and flat environments have overlap in state space, since the sloped environment eventually encounters flat terrain near the goal node.

\begin{figure}[!ht]
    \centering
    \begin{tabular}{cc}
        \includegraphics[width=.42\textwidth]{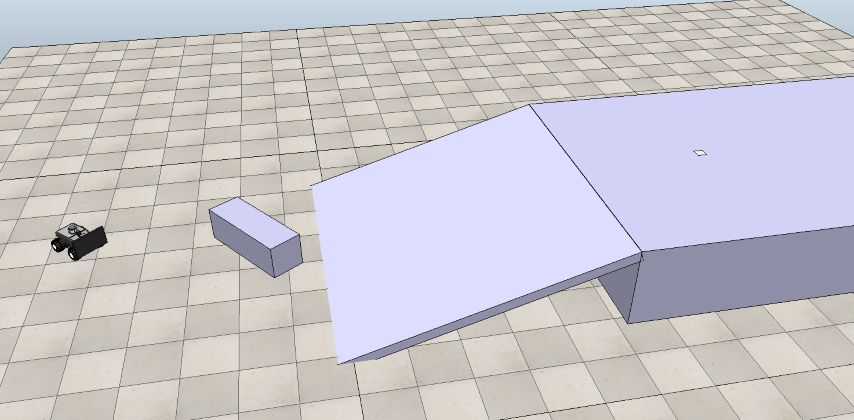} &
        \includegraphics[width=.42\textwidth]{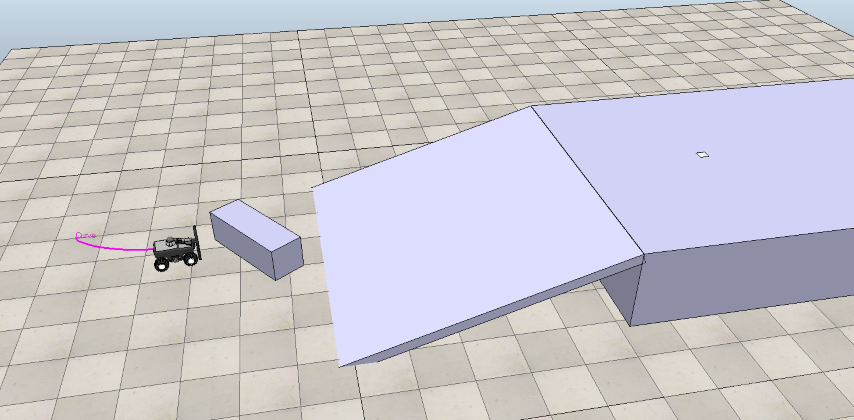} \\
        \includegraphics[width=.42\textwidth]{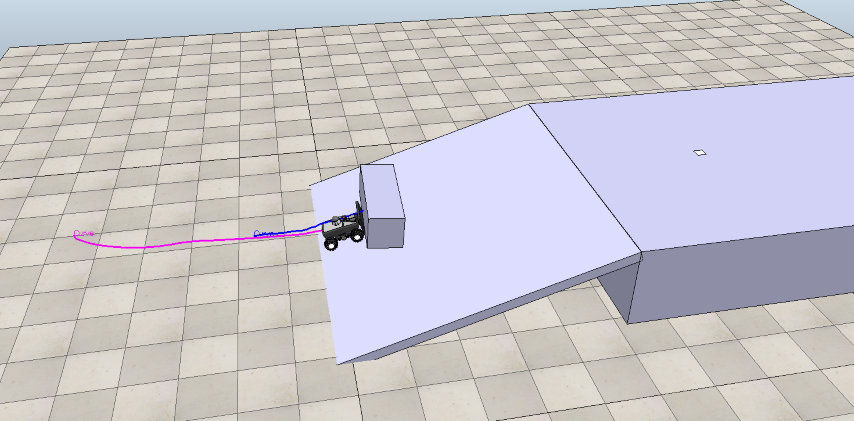} &
        \includegraphics[width=.42\textwidth]{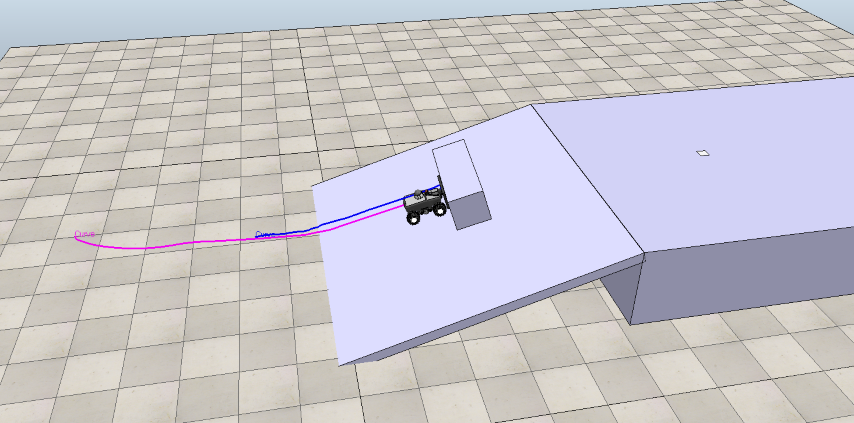} \\
        \includegraphics[width=.42\textwidth]{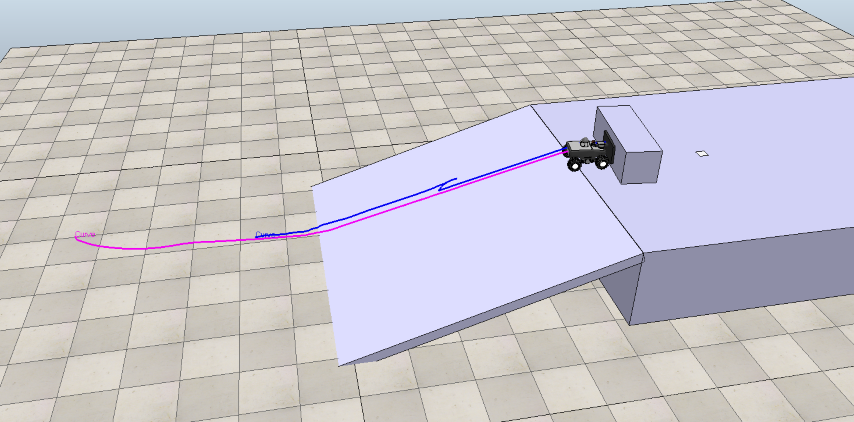} &
        \includegraphics[width=.42\textwidth]{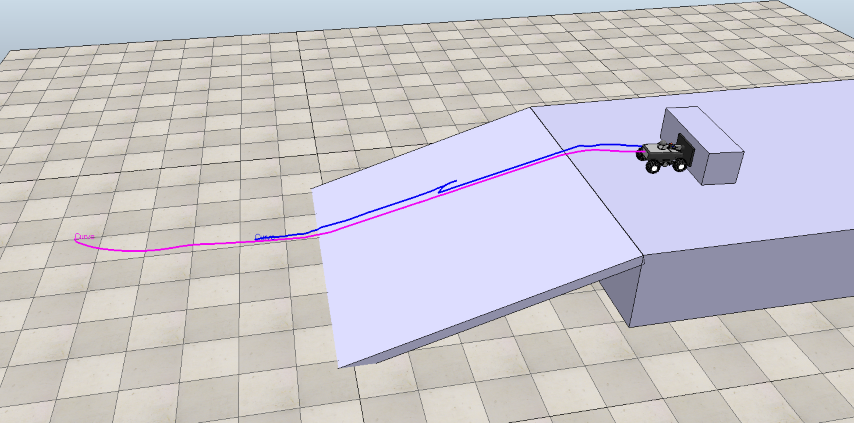} \\
    \end{tabular}
    \caption{A sample trajectory of the trained policy in the sloped environment. Pink line represents the agent trajectory and the blue line represents the box trajectory.}
    \label{TrajectorySlope}
\end{figure}

\begin{table}[!ht]
    \begin{minipage}{.55\textwidth}
        \begin{tabular}{ |c|c|c|c|}
             \hline
             \multicolumn{4}{|c|}{Performance in Rewards} \\
             \hline & Flat & Sloped & Hole  \\
             \hline Individual & 4.77$\pm$\textbf{1.93} & 3.99$\pm$4.18 & 2.69$\pm$3.43  \\
             \hline All-in-One & \textbf{4.84}$\pm$2.62 & \textbf{5.64}$\pm$1.14 & \textbf{2.95}$\pm$3.30\\
             \hhline{|=|=|=|=|}
             \multicolumn{4}{|c|}{Performance in Success Rate} \\
             \hline Individual & \textbf{.97}$\pm$.17 & .93$\pm$.26 & .82$\pm$.38  \\
             \hline All-in-One & .94$\pm$.24 & \textbf{1.0}$\pm$0.0 & \textbf{.9}$\pm$.3\\
             \hhline{|=|=|=|=|}
             \multicolumn{4}{|c|}{Environment Parameters} \\
             \hline Steps & 50 & 100 & 50\\
             \hline Probability & .3 & .2 & .5\\
             \hline
        \end{tabular}
    \end{minipage}
    \begin{minipage}{.35\textwidth}
        \begin{tabular}{|c|c|c|c|c|}
         \hline
         \multicolumn{5}{|c|}{Training Episodes} \\
         \hline & Flat & Sloped & Hole & Total \\
         \hline Individual & 270 & 220 & 400 & 890\\
         \hline All-in-One & \multicolumn{3}{|c|}{400} & \textbf{400}\\
         \hline
        \end{tabular}
    \end{minipage}
    \caption{Policy style comparison results. The all-in-one policy performs more reliably in all three environments, coupled with more sample efficiency. This can be credited to similarities between the environments. Learning in one environment may help learning in another.}
    \label{PolicyStyleTest}
\end{table}

In the same manner as described in the previous, the agent explores for $S$ action steps and then alternates between training and testing, each for 20 episodes. 

Because of training methodology differences, comparing performance graphs is not useful. As a result, after training is complete, we individually tested both policies in each of the three environments. We recorded and compared the accumulated rewards for both trained policies over 200 episodes in each environment. Furthermore, we tracked the number of episodes all policies took to reach optimal performance, (i.e. number of training episodes before performance plummeted due to over-training and instability). In doing so, we can compare sample efficiency and training time between the two styles of training. These results are shown in Figure \ref{PolicyStyleTest}. 

It is evident that the all-in-one policy provides more sample efficiency and stronger ending performance than the individualized policies. This can be reasonably credited to the fact that all three environments are similar in objective and state space. As a result, experience from one environment can be used to help better learn policies in another.

In terms of accumulated episodic rewards, the All-in-One policy achieves dominating performance in the sloped and hole environments. However, the specialized Individual policy provides more reliability in performance in the flat environment than the All-in-One. While these discrepancies may be due to training or testing variance, such differences are not as significant as those in the sloped and hole environments. As a result, ultimately, the All-in-One policy provides a superior policy. These disparity in performance is reflected in the success rates for the two policies While the All-in-One policy provides more reliable success rates in the sloped and hole environments, the Individual policy achieves better reliability in the flat environment.

The preference towards the All-in-One policy is further corroborated by its sample efficiency as compared to the Individual policy. Whereas the Individual policy used 890 episodes of training, the All-in-One policy used less than half of that amount, 400 episodes. As mentioned before, this boost in sample-efficiency can be accredited to the similarities across these three environments. Gaining adeptness in one environment can translate to other, similar tasks. Sample trajectories of the trained, All-in-One policy are shown in Figures \ref{TrajectoryFlat}, \ref{TrajectoryHole}, and \ref{TrajectorySlope}.

\subsection{Reflection}

The performance of single-agent hierarchical RL when learning how to maneuver, reorient and push a rectangular box achieved variable performance with respect to various hyper-parameters. In particular, batch size $b$, target update frequency $M$, and exploration steps $S$ provided variegated results as shown in Figure \ref{SingleDDQNAblation}. After determining a relatively strong subset of hyper-parameters, further tests suggested that training a single policy to tackle all three environments, as opposed to three, disparate policies, yielded better performance in its sample efficiency and reliability, as shown in Table \ref{PolicyStyleTest}. 

Even though this policy shows promising reliability in this particular domain, there is no guarantee that it will be as effective when integrated into a multi-agent coordination algorithm. In particular, the multi-agent environment may require an agent to push a box along a path discretized as multiple goal locations. As a result, the trained primitive must be invoked multiple, consecutive times. These discretizations can be viewed as chains of training environment episode scenarios. Even though the agent may have strong performance when invoking the primitive once, there is no guarantee it will perform well when executing it multiple, consecutive times. This issue of composability does not reveal itself in metrics provided in Section 2, as we primarily concern ourselves with single episode performance as shown in Figure \ref{SingleDDQNAblation} and Table \ref{PolicyStyleTest}. However, these issues become more apparent in Section 3, where the trained policy is used as a mid-level control primitive. By adding one more level to the overall, hierarchical structure, the weaknesses of the mid-level, trained primitive become more evident.

\newpage

\section{Pheromone-Induced Multi-Agent Learning and Planning}
In order to extend our trained, single-agent policy to multi-agent coordination, we combine it with a stigmergic, rule-based algorithm inspired by ant colonies. The goal of this section is to design an algorithm that allows groups of agents to plan a path to a goal location using environment modification, or pushing boxes across terrain and into holes. 

\subsection{Related Work}
Environment modification and path planning is a popular problem that has seen the advent of numerous approaches. In particular, many single-agent approaches have utilized classical planning techniques such as hierarchical task networks \cite{Planner9} to help determine how an agent should modify the environment. Approaches such like those of Magnenat et al. \cite{ScarceBox} uses this approach and arms robots with pre-designed controllers.

As shown in Section 2, this work uses mid-level controllers that are learned as opposed to given. Furthermore, multiple agents, as opposed to one, must navigate to a goal location. As a result, such a goal lends itself to stigmergic coordination. Stigmergic algorithms provide local, simple rules to individual agents such that the collective group successfully achieves some task. Algorithms have been provided for distance mapping towards a goal location \cite{RFID} or navigating towards locations with obstructing obstacles \cite{CuldeSac}. Algorithms have been designed for large groups of agents to iteratively solve originally intractable problems such as the Traveling Salesman Problem \cite{AntTSP}. 

Other stigmergic algorithms have addressed more control and coordination based tasks. To mimic ant behavior when crossing large gaps, Malley et. al. \cite{EcitonBridge} designed soft, adhesive robots armed with a stigmergic rule to cooperatively traverse a ravine. Similar algorithms have been designed to collectively sort and group large objects \cite{Frisbee} or assemble into formations autonomously \cite{SIRL}.

\subsection{Background}

\begin{figure}[!ht]
    \centering
    \includegraphics[width=1\linewidth]{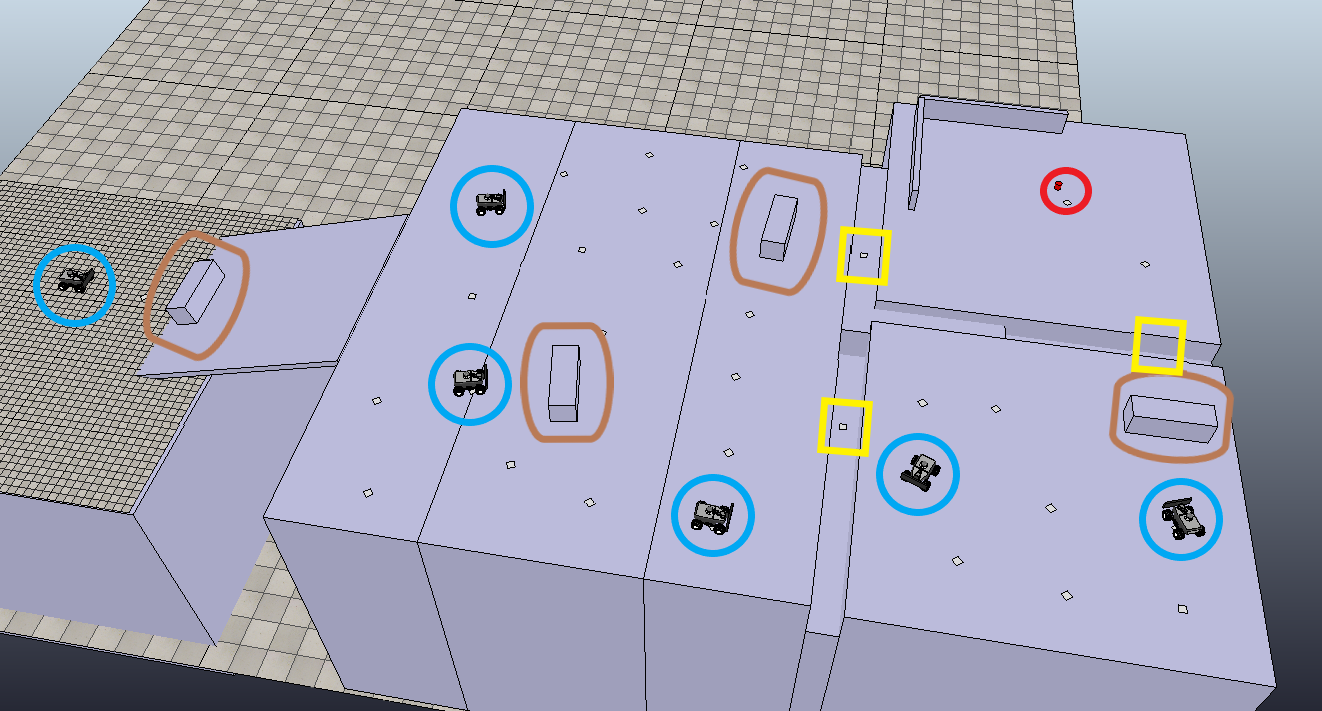}
    \caption{Example environment where locations represented by nodes (white squares). The goal location is denoted by the red circle on the upper right platform. Yellow squares show that could be filled, brown ovals show boxes that could be pushed, and blue circles show several agents whose goal is to reach the red circle.}
    \label{EnvWithCircles}
\end{figure}

We discretize environment locations into nodes as shown in Figure \ref{EnvWithCircles}. At the beginning of each episode, boxes and agents are each initialized at different, specified nodes. Each agent observes only local information. In other words, it can only access environment details and agents within a fixed radius. Similar to other works inspired by ant colonies, each agent is able to place a variety of pheromones onto nodes to indirectly communicate with other agents. In practical terms, these pheromones can be simulated in real life using RFID tags \cite{RFID}.  

Each node and box in the environment has a bank of pheromones, each with possibly different combinations and concentrations, represented as a set of real numbers. As a general note, these pheromones give information to future agents that travel to or use the same node or box, respectively. With appropriate pheromone updates, the collective team of agents can induce an intelligent policy, each only using local information.

\begin{figure}[!ht]
\begin{tabular}{ c c}
 \includegraphics[width=3.1in]{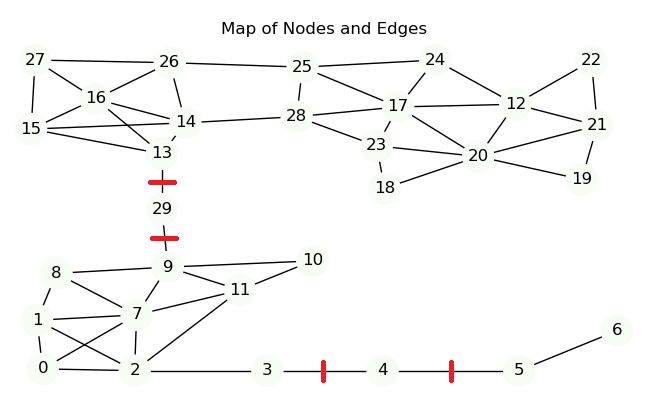} &
  \includegraphics[width=3.2in]{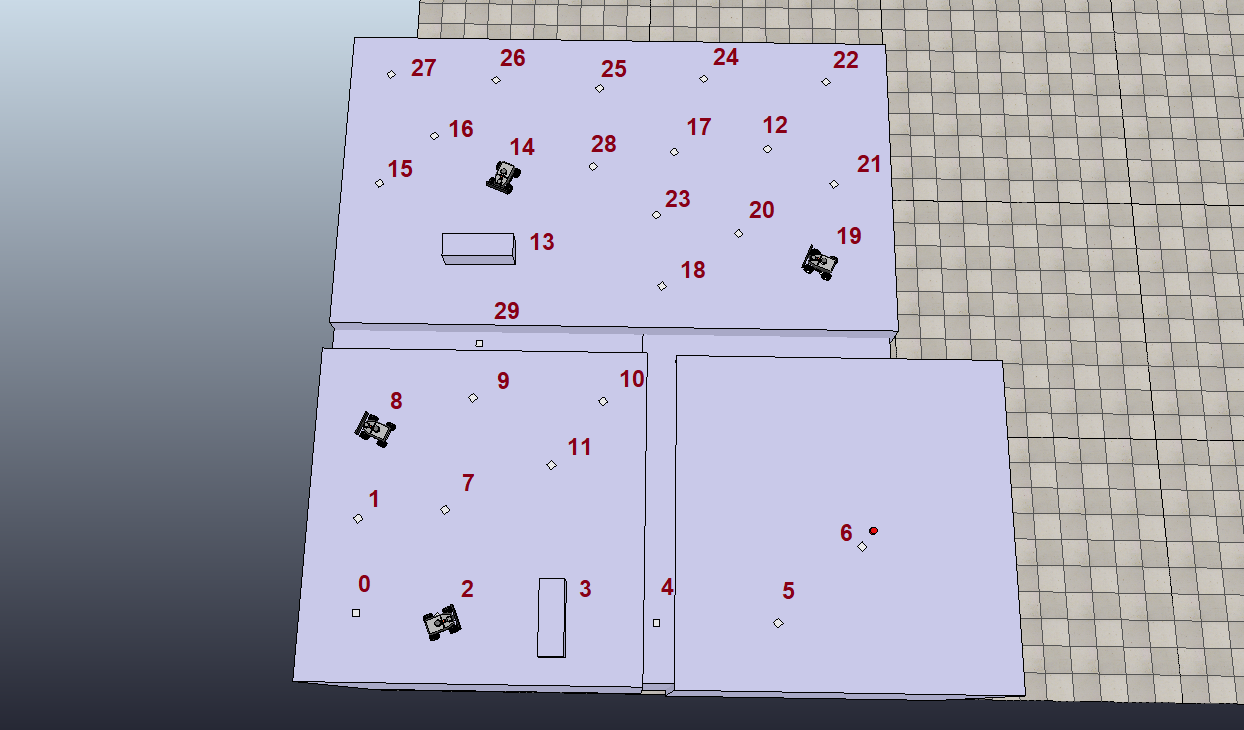}\\
   (a) & (b)\\
\end{tabular}
\begin{tabular}{ c c c }
 \includegraphics[width=2.1in, height=1.5in]{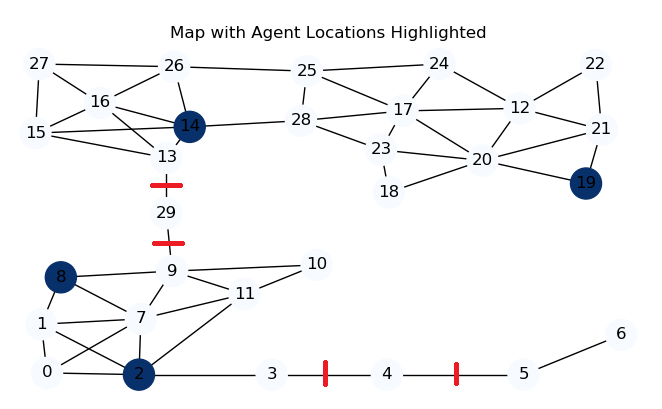} &
  \includegraphics[width=2.1in, height=1.5in]{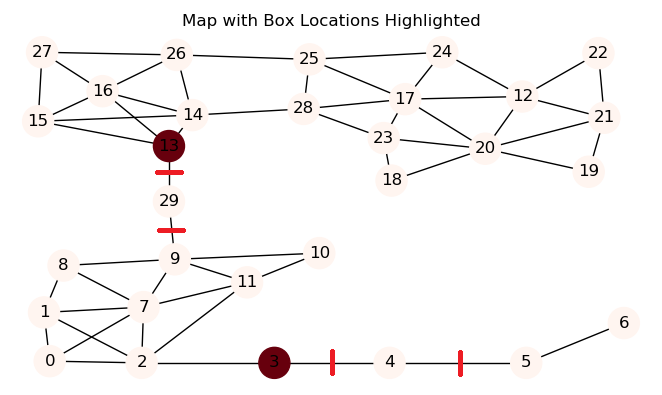} & 
    \includegraphics[width=2.1in, height=1.5in]{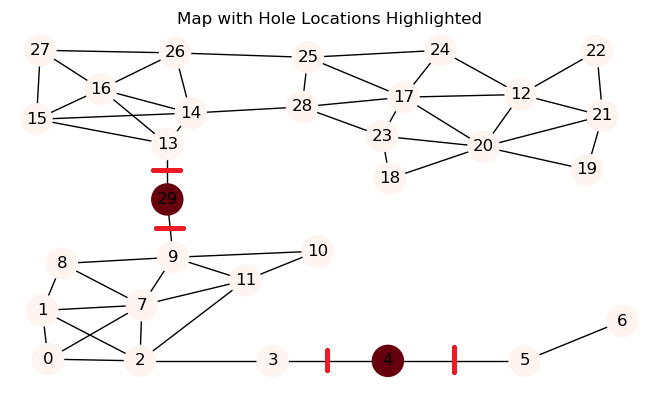}\\
   (c) & (d) & (e)\\
\end{tabular}

\caption{Converting a V-REP environment into a node environment, where red slashes through edges indicate untraversable paths due to holes at certain nodes. Filling up the holes appropriately will make these edges traversable, removing the red slashes.}
\label{EasyEnvNodes}
\end{figure}

In this section, we first demonstrate the effectiveness of the stigmergic algorithm independently by constructing a grid world version of the V-REP simulation in Figure \ref{EasyEnvNodes}. Then, we outline how the algorithm is combined with the aforementioned policy to tackle robotic simulation and analyze its performance.

\subsection{Pheromones}
Each node is designated a label $n_i$, denoted as the $i^{th}$ node. Each $n_i$ is dedicated a bank of pheromones. This bank is represented as a set $\{c_j\}_{j=1}^{J} \in (\mathbb{R}^+)^J$, where $J$ denotes number of kinds of pheromones. Throughout this work, for clarity, we will describe each pheromone $c_j$ as a capital letter associated with an index (Ex: $D_i \in \mathbb{R}^+$). The letter will typically refer to the pheromone type. The index association is variable and will be described in upcoming sections.

Agents can only modify the pheromones concentrations of nodes they currently occupy. Here, we list and describe pheromones the agents use for indirect coordination and planning.

\subsubsection{Distance Pheromone}
The first pheromone instance is named the Distance Pheromone (\textbf{D-pheromone}). We denote the D-pheromone concentration at the node $n_i$ as $D_i$. This pheromone provides a heuristic or measure of how close the $n_i$ is to the goal node, $n_{goal}$. Denoting the neighboring nodes of $n_i$ as set $N(n_i)$, an agent at node $n_i$ can only observe $D_i$ and $\{D_{j}, n_j \in N(n_i)\}$, or the D-pheromones at its current and neighboring nodes.

When an agent travels to and occupies node $n_i$, it is able to modify $D_i$. We design the algorithm to mimic ant behavior, encouraging agents to travel towards nodes with higher pheromone concentration $D_i$. Therefore, higher values of $D_i$ should correspond to shorter path distances from $n_i$ to $n_{goal}$. However, in the beginning of the training process, the agent only has knowledge of the goal location, not the path to it. As a result, the agent initially assigns $D_i$ a value inversely proportional to the Euclidian distance to the goal, as opposed to path distance. This value is temporary, and will be modified later to reflect path distance. 

In addition to $D_i$, the $i^{th}$ node stores its current distance to the goal, $d_i \in \mathbb{R}^+$. Over time, a node $n_i$ can be deemed to have Official Distance, marked by the indicator $f_i \in \{0, 1\}$. An Official node signals that the node's distance and D-pheromone are calculated using some updated path distance to the goal, not Euclidian.

\begin{figure}[!ht]
    \centering
    \includegraphics[width=1\linewidth]{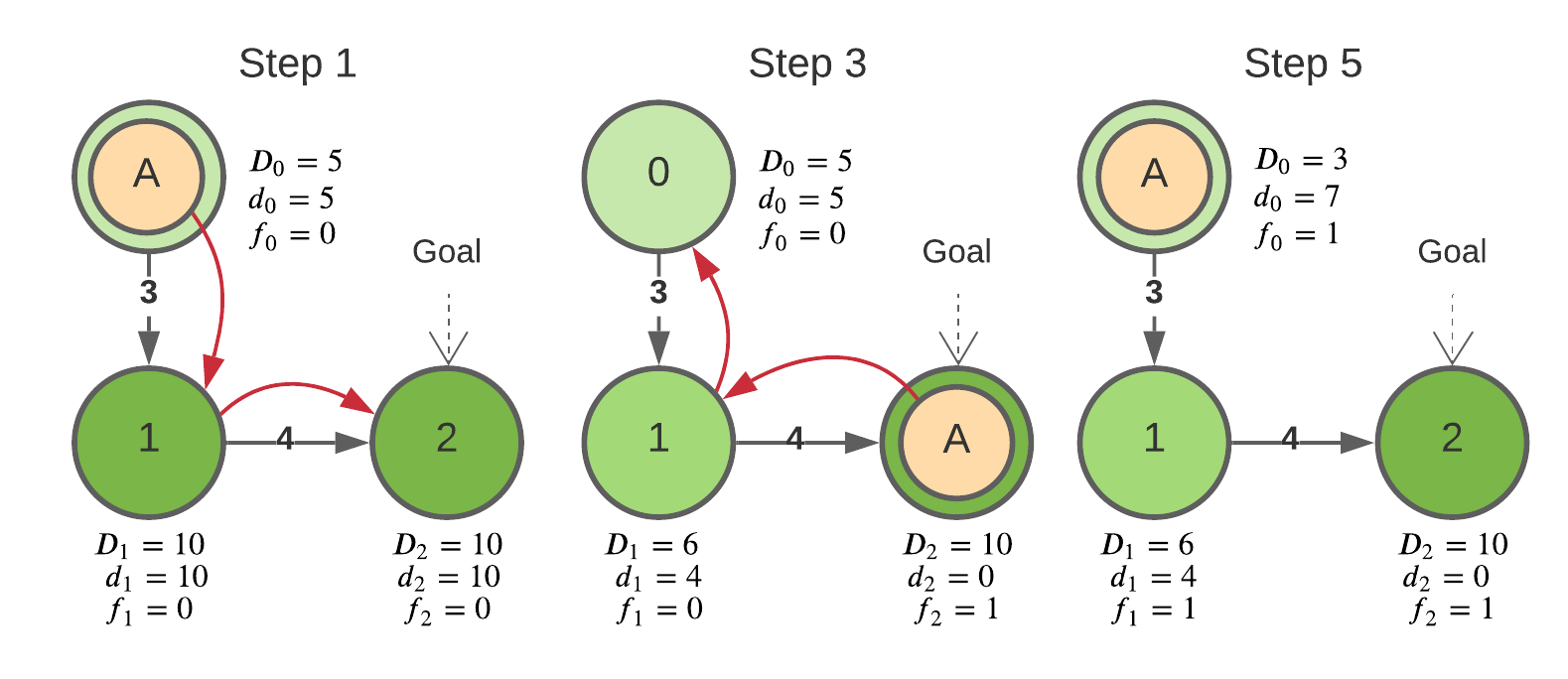}
    \caption{D-Pheromone Update Example. $\textbf{A}$ represents the agent and each node has been labeled with its D-pheromone concentration $D_i$, distance to goal $d_i$, and officiality $f_i$. The goal node is $n_2$. $d_{max}$ has been set, arbitrarily, to 10 for simplicity. In Step 1, agent A starts at $n_0$. Euclidian distance $d_0$ is set to be 5. Accordingly, $D_0 = d_{max} - d_0 = 5$. Then, agent A moves to $n_1$ and $n_2$. In Step 3, $n_1$ is shown with $d_1=4$ using Euclidian distance. $D_1=d_{max} - d_1 = 6$ is calculated and updated. Because $n_2$ is the goal node, agent A deems the node official $f_2 = 1$, sets $d_2 = 0$ and $D_2 = d_{max}$. Afterwards, agent A moves back to $n_1$ and $n_0$. Firstly, $n_1$ gets updated using path distance, which happens to also equal Euclidian distance $d_1=4$. $n_1$ becomes official with $f_1=1$. Lastly, $n_0$ becomes official, $f_0=1$, as it also is calculated using the path distance from $n_1$. $d_0$ gets updated using $d_0=d_1+edge(n_0, n_1)=3+4=7$. Then, $D_0$ gets updated accordingly.}
    \label{DPheromoneExample}
\end{figure}

More explicitly, to update $D_i$ and $d_i$ of node $n_i$, an agent observes the distance stored in immediately neighboring, reachable nodes, denoted as set $\{d_{j}, n_j \in N(n_i)\}$. A node $n_j$ is considered \textit{reachable} from $n_i$ if the single-length path from $n_j$ to $n_i$ is unobstructed by a hole or hill. Let the subset of official nodes be denoted as $F$. If the current node is the goal node, it automatically becomes official $f_i = 1$, $d_{i}=0$ and D-pheromone is updated accordingly (shown in Algorithm \ref{DPheromoneUpdate}).

Otherwise, denote the subset of official, reachable neighbors as $O_i = F\cap N(n_i)$. If $O_i \neq \varnothing$, then we let $d_{i} = \min(\{edge(n_i, n_j) + d_j, n_j \in O_i\})$ where $edge(\cdot)$ is the edge length. Then, we update the D-pheromone accordingly (Algorithm 2). Since $d_i$ was calculated using an official node, it then becomes official itself $f_i = 1$, as it was calculated using a path distance. Otherwise, if $O_i = \varnothing$ we use Euclidian distance to update $d_i$. This approach is reminiscent of an iterative version of Dijkstra's algorithm. 

Over time, all node distances converge to the shortest path distance. In effect, $D_i, \forall i $ converge as well. The update rule for the D-pheromone is shown in Algorithm \ref{DPheromoneUpdate}

\begin{algorithm}[H]
\SetAlgoLined
 Max distance across map $d_{max}$, current node $n_i$, goal node $n_{goal}$, $i^{th}$ node distance to goal $d_i$, D-pheromone on node $D_i$, and set of official nodes $F$\\

 Get reachable neighbors set $N(n_i)$ \\
 Get official neighbors set $O_i = F \cap N(n_i)$ \\
 
 \eIf{$n_i$ is $n_{goal}$}{
        set $d_i$ = 0 and $D_i$ = $d_{max}$ \\
        $F = F \cup n$\\
        return \\
 }
 {
    \eIf{$O_i \neq \varnothing$}{
        set $d_i$ to $\min(edge(n_i, n_j) + d_j, n_j \in O_i)$\\
        $F = F \cup n$ \\
     } {
         $d_i$ = $Euclidian(n_i, n_{goal})$ \;
     }
     $D_i$ = $d_{max}$ - $d_i$
 }
 \caption{Update D-Pheromone of Current Node}
 \label{DPheromoneUpdate}
\end{algorithm}

Intuitively, assuming that there are no holes to be filled, obstacles to cross and all nodes are Official, an agent ought to travel to nodes $n_i$ where $D_i$ is highest to navigate to the goal location. However, in addition to other factors, we must account for nodes that are not reachable by the agent and hence require environment modification to traverse. As a result, we must provide additional functionality for box pushing.

\subsubsection{Box Pheromones}
An environment may contain numerous boxes and numerous holes. Some may be extraneous, meaning they are irrelevant to the path to the goal location, while others are vital to the task. In this section, we describe mechanisms to help agents realize over time which boxes to use and where to push them for path planning. In this work, we assume that all boxes have identical width and length but varied height. Similarly, all holes have varying depth but identical length.

Each box in the environment is initialized to different node locations. Furthermore, we denote each box as $b_k$, describing the $k^{th}$ box in the environment. Each hole is denoted as $h_j$, describing the $j^{th}$ hole in the environment. Additionally, each box contains a modifiable list of holes in which a nearby agent may choose to push it to. The box $b_k$ has a list represented by a map from a hole to a value called Box Pheromones, $B_k(h_j) \in \mathbb{R}^+$, interpreted as the value of placing box $b_k$ in hole $h_j$. The higher $B_k(h_j)$ is, the more likely a nearby agent is to push the box to hole $h_j$. The process of determining candidate holes is discussed later.

\begin{figure}[!ht]
    \centering
    \includegraphics[width=1\linewidth]{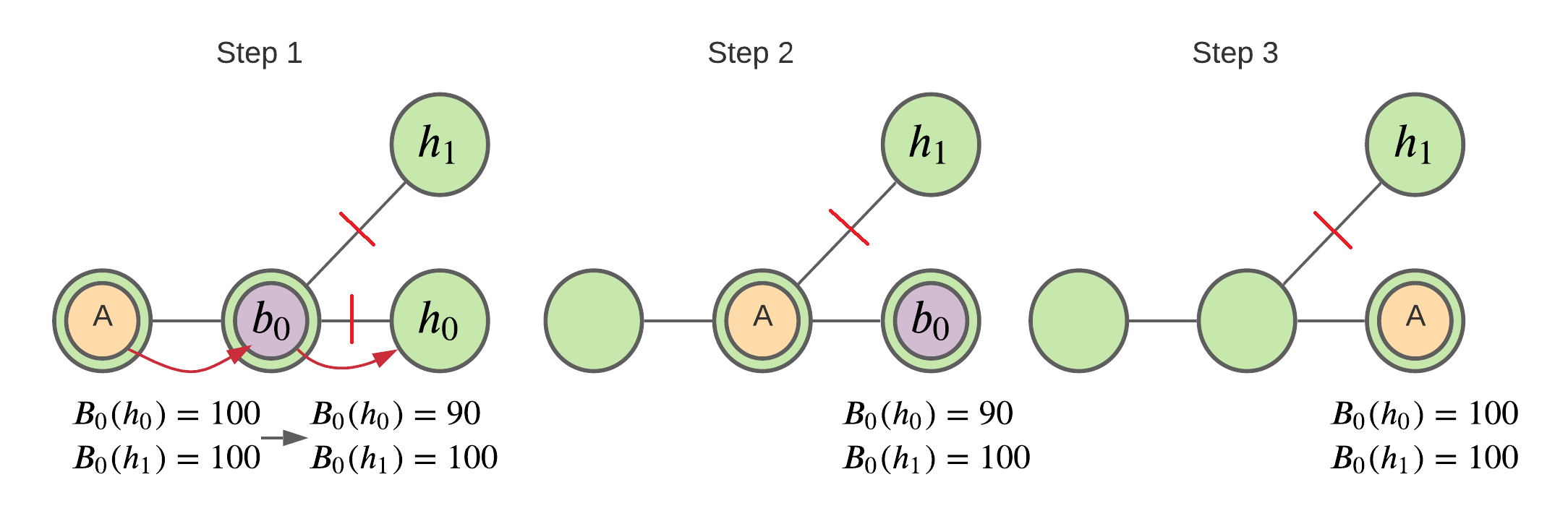}
    \caption{B-Pheromone Update Example. An agent A claims box $b_0$ and targets hole $h_0$. In doing so, it decays the corresponding B-Pheromone $B_0(h_0)$. Then, after placing $b_0$ into $h_0$, it traverses over the node housing $h_0$, multiplying $B_0(h_0)$ by the inverse of the decay.}
\end{figure}

An agent may only decide to push a box if the box occupies a neighboring node. Once a box is claimed by an agent to be pushed to hole $h_j$, we decay $B_k(h_j)$ by a set hyper-parameter factor, $B_{decay} \in (0, 1)$. The agent determines the shortest path from the box to the node hole $h_j$ occupies, if the shortest path distance is available. The determination of path using H-Pheromones is discussed later. Then, the agent pushes the box along that path until it reaches the hole.

After placing box $b_k$ in hole $h_j$, $B_k(h_j)$ may still be modified. When an agent steps over box $b_k$, or any other box stacked above it, to cross hole $h_j$, $B_k(h_j)$ is multiplied by $\frac{1}{B_{decay}}$. $B_k(h_j)$ does not change if an agent does not successfully step over the box.

With these pheromones, we intend to mimic ant pheromone placement behavior on physical objects. If an agent uses the box in a useful way (i.e. stepping over it), the value $B_k(h_j)$ of the box and hole pair is increased. Through these pheromones, the holes of the following character are discouraged from being filled: holes that are unreachable and holes not useful to reaching the goal. Conversely, holes that are useful will be encouraged.

The process of updating box Placement Value is shown in Algorithm \ref{BPheromoneUpdate}.

\begin{algorithm}[H]
\SetAlgoLined
 \For{all boxes $b_k$ in environment}{
    \If{$b_k$ was just claimed}{
        Denote $h_j$ as target hole index \\
        $B_k(h_j) = B_k(h_j) * B_{decay}$\\
    }
    \If{$b_k$ was stepped over}{
        Denote $h_j$ as current hole index\;
        $B_k(h_j) = \frac{B_k(h_j)}{B_{decay}}$\\
    }
 }
 \caption{B-Pheromone Update Example}
 \label{BPheromoneUpdate}
\end{algorithm}

\subsubsection{Hole Pheromones}
In this section, we establish local rules for two things. Firstly, we discuss how an agent determines which boxes it is able to push. Then, we discuss where an agent is able to push a particular box. 

We equip each agent with a finite detection radius, $R \in \mathbb{R}^+$. Agents are only able to detect holes and boxes within this radius. To determine which boxes an agent is able to push, the algorithm enforces a significant restriction to simplify this decision-making process: an agent may only travel towards or push the closest (through path distance) box within radius $R$ at that given time step. All other boxes are ignored. This process is outlined formally in Algorithm \ref{GetBoxCandidateAlg}.

Next, we outline how the algorithm decides candidate nodes [with holes] to push the box to. Firstly, all holes within radius are candidates. The agent is able to determine that shortest path to holes within detection radius. However, the algorithm must also account for holes outside of the agent's detection radius $R$.

\begin{figure}[h]
    \centering
    \tiny
    \includegraphics[width=1\linewidth]{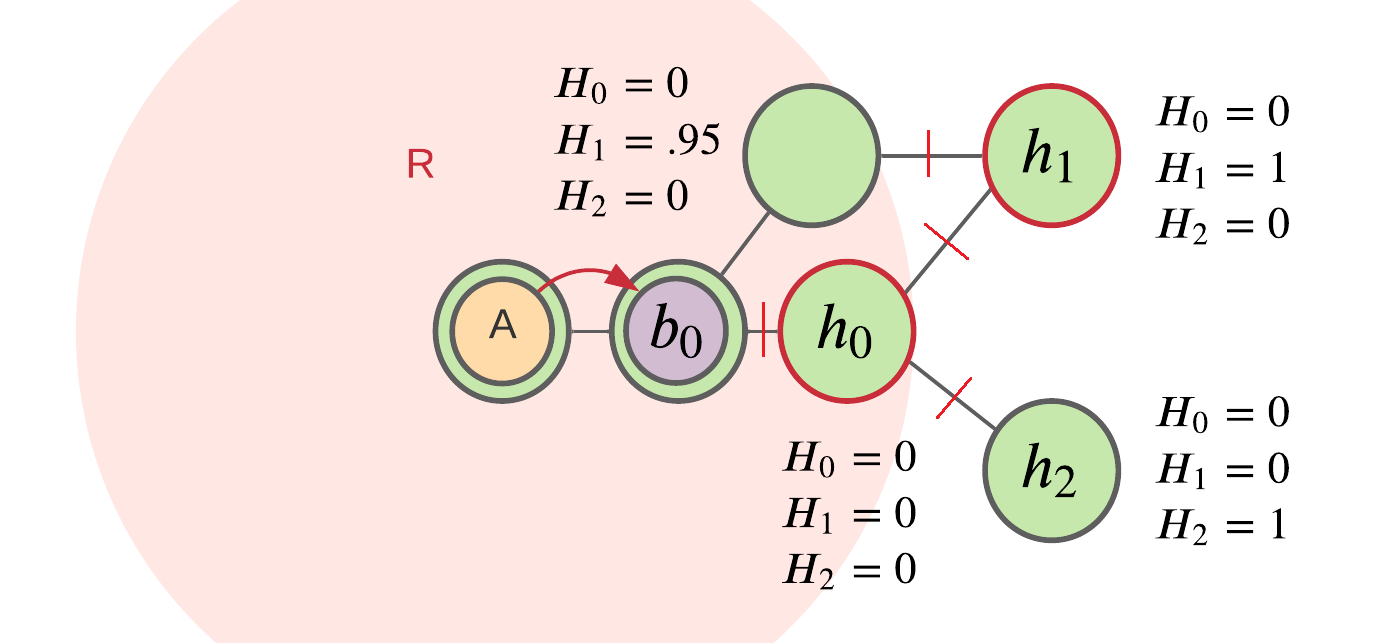}
    \caption{Example of hole candidates. Suppose agent A hopes to push box $b_0$ in a hole. $h_0$ is a candidate because it is within detection radius $R$ of agent A, regardless of $H_0$ concentration. $h_1$ is a candidate because one of $b_0$'s reachable neighbors contains a non-zero concentration of pheromone $H_1$. $h_2$ is not a candidate because it is out of detection radius $R$ and no reachable neighbors of $b_0$ have nonzero concentration of pheromone $H_2$}
    \label{HoleCandidateExample}
\end{figure}

We use another pheromone called Hole Pheromones (H-Pheromones), denoted by the upper-case $H_j$ describing the H-pheromone associated with the $j^{th}$ hole $h_j$. When an agent travels to a node, it is able to detect holes within a radius $R$. For each of the holes $h_j$ detected, the agent places associated H-pheromone $H_j = 1$ on the current node.

The goal for every hole $h_j$ is to construct a trail of pheromones $H_j$ of increasing concentration to $h_j$, or to a node that is within detection radius of $h_j$. Here, we outline an iterative process in which an agent updates H-Pheromones on its current node by observing neighboring nodes. 

\begin{figure}[h]
    \centering
    \includegraphics[width=1\linewidth]{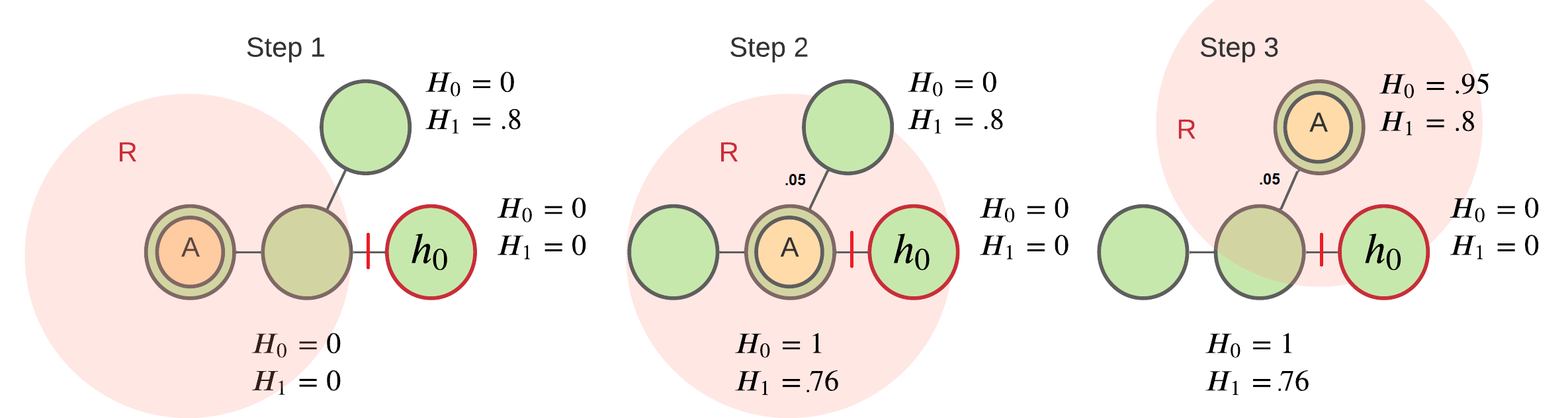}
    \caption{Example H-Pheromone update. In Step 1, we see the current distribution of H-pheromones associated to holes $h_0$ and $h_1$ (not depicted). When agent A moves to the middle node in Step 2, it detects $h_0$ within its radius $R$ and changes $H_0$ to 1. Furthermore, it detects the pheromone $H_1 = .8$ of a neighboring, reachable node. It sets its node's current $H_1$ by decaying $.8$ by $exp(-.05) \approx .95$ to yield $H_1 = .8 * .95 = .76 $. In Step 3, the agent travels to the upper node. It does not update $H_1$ as its current concentration is greater than that of all reachable neighbors. However, $H_0$ gets updated similarly by decaying $1$ by $.95$ and setting the node's concentration appropriately.}
    \label{HPheromoneUpdate}
\end{figure}

Suppose an agent is at node $n_i$. Then, let $N(n_i)$ be all reachable, neighboring node and $H_j^{n_i}$ denote the concentration of H-pheromone associated to $h_j$ at node $n_i$. Let $S_j = \{H_j^{n_j}, n_j \in N(n_i)\}$ represent the set of neighboring H-pheromone concentrations associated with hole $h_j$. Keeping in mind the concentration $H_j^{n_i}$ is inversely proportional to the distance from $n_i$ to hole $h_j$, we update $H_j^{n_i}$ in a similar manner as we update the D-pheromone $D_i$. We let $H_j^{n_i} = \min_{n}(\{H_j^{n_j} * exp(-dist(n_j, n_{i})), n_j \in N(n_i)\})$. Suppose there exists an arbitrarily indexed path of nodes $\{n_k\}_{k=1}^{K}$. Assuming the first node in this path contains the hole, notice the resultant concentration $H_j^N$ is the following: $\prod_{i}^{N-1} exp(-dist(n_i, n_{i+1})) = exp(\sum_{i}^{N-1}dist(n_i, n_{i+1}))$. As a result, $H_j$ is inversely proportional to the path distance to a given location. An example of an H-Pheromone update sequence is shown in Figure \ref{HPheromoneUpdate}. Furthermore, the Algorithm \ref{UpdateHPheromoneAlg} outlines the process of updating H-pheromones.

Through iterative updates, all nodes will house the appropriate concentration of $H_j$ to lead agents to the respective hole $h_j$. To reach a particular hole $h_j$, an agent repeatedly, greedily chooses nodes housing higher values of $H_j$. This method allows agents to indirectly notify each other of key environment junctions. An example of an agent determining hole candidates for a particular box is shown in Figure \ref{HoleCandidateExample}, coupled with the formal algorithm outlined in Algorithm \ref{GetHoleCandidateAlg}.

As a reminder, the purpose of the H-pheromone is to notify agents of holes outside of their detection radius $R$ that they can push boxes to. As a result, we establish two conditions for a hole $h_j$ to be considered a candidate to be filled by box $b_k$: if $b_k$'s neighboring node $n_i$ has nonzero value $H_j^{n_i}$ or hole $h_j$ is within detection radius $R$ from the agent. This rule allows the agent to have access to global information using local pheromones. 

\begin{algorithm}[H]
    \SetAlgoLined
     Get set of nodes with boxes $N_b$ within radius $R$\\
     Let node $n_{target}$ = $argmin_{n_j}{path(n_{i}, n_j), n_j \in N_b}$\;
     Let $b_k$, the box at node $n_{target}$, be the box candidate\\
     \caption{Get Box Candidate}
     \label{GetBoxCandidateAlg}
\end{algorithm}

\begin{algorithm}[H]
\SetAlgoLined
  Denote $n_i$ as the current node, $\mathcal{H}$ as the set of all holes\\
  \If{episode is finished}{
    $H_j^{n_i} = 0$,  $\forall j$ \\
    \textbf{return} \\
  }
  Get set of holes $\mathcal{H}_R$ within radius $R$\\
  \For{$h_j \in \mathcal{H}_R$}{
    $H_j^{n_{i}} = 1$\\
  } 
  Get set of reachable, neighboring nodes $N(n_i)$\\
  \For {$h_j \in \mathcal{H}$}{
    Retrieve concentrations $S_j = \{H_j^{n_j}, n_j \in N(n_i)\}$ \\
    $H_j^{n_{i}} = \min(\{H_j^{n_j} * exp(-dist(n_j, n_i)), n_j \in N(n_i)\})$
  } 
 \caption{Update Hole Pheromones}
 \label{UpdateHPheromoneAlg}
\end{algorithm}

\begin{algorithm}[H]
    \SetAlgoLined
     Let $n_{box}$ be the node with a box \\
     
     Get $S_r = \{h_j; dist(n_{box}, h_j) \leq R, h_j \in \mathcal{H}\}$\\
     Get reachable neighbors $N(n_{box})$ of $n_{box}$ \\
     Get holes with pheromones $S_p = \{h_j; H_j^{n_j} > 0, h_j \in \mathcal{H}, n_j \in N(n_{box})\}$\\
     $\mathcal{H}_{candidate} = S_r \cup S_p$\\
     \caption{Get Hole Candidates}
     \label{GetHoleCandidateAlg}
\end{algorithm}

\subsubsection{Exploration Pheromone}
In this section, we introduce the final pheromone type: the Exploration Pheromone (E-Pheromone). We denote $E_i$ as the concentration of E-Pheromone at node $n_i$. This pheromone encourages exploration of nodes that have not been visited recently. This pheromone prevents agent overcrowding and quicker development of useful pheromone signals by encouraging agents to travel to nodes not visited recently.

\begin{figure}[h]
    \centering
    \includegraphics[width=1\linewidth]{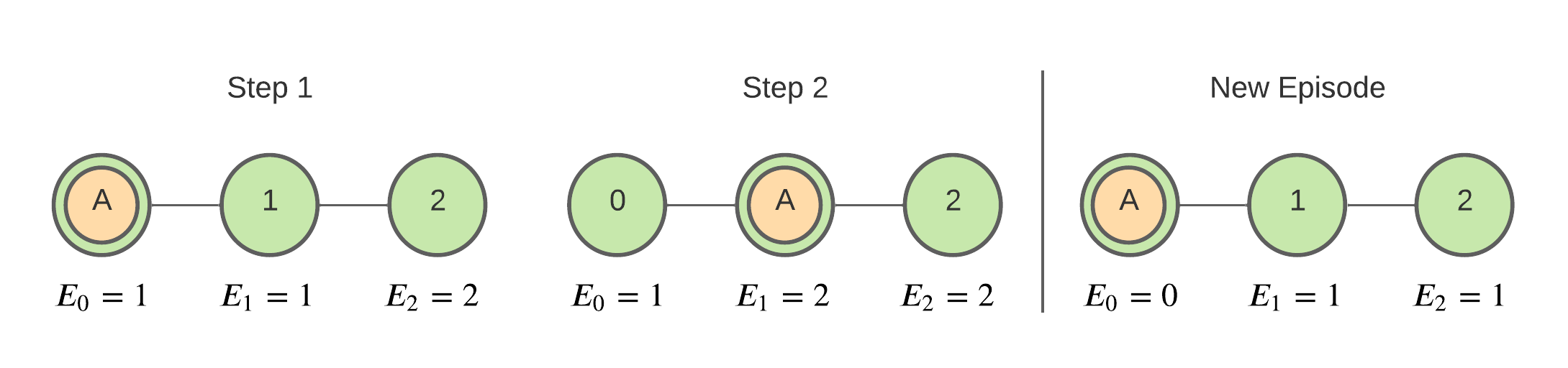}
    \caption{Example E-Pheromone update. In Steps 1 and 2, agent A traverses to a new node, incrementing the current pheromone $E_1$ from $1$ to $2$. In Step 3, we see a new episode starting, where all E-Pheromones are normalized by subtracting all pheromones by $\min\{E_i, n_i \in N\} = 1$.}
\end{figure}

Unlike previous pheromones, an agent is encouraged to travel towards nodes with lower, as opposed to higher, concentrations of E-pheromone. When an agent visits a $n_i$, it will increase $E_i$ by 1. When agents explore the environment, it would be more ideal to travel to nodes that have not been visited recently, therefore equating to nodes with lower concentrations of $E_i$.

\begin{algorithm}[H]
\SetAlgoLined
  Let $n_{i}$ be the current node of the agent\\
  $E_i = E_i + 1$\\
  \If{the episode has ended}{
    Let $N$ denote all nodes in the environment \\
    $E_i = E_i - \min\{E_i, n_i \in N\}$\;
   }
 \caption{Update E Pheromones}
 \label{UpdateEPheromonesAlg}
\end{algorithm}

At the end of each episode, exploration pheromones of all nodes are "re-normalized" so that the lowest E-pheromone concentration is 0 for numerical stability. Algorithm \ref{UpdateEPheromonesAlg} shows the process of updating E Pheromones at a given node.

\subsubsection{Collision Detection}

\begin{figure}[h]
    \centering
    \includegraphics[width=1\linewidth]{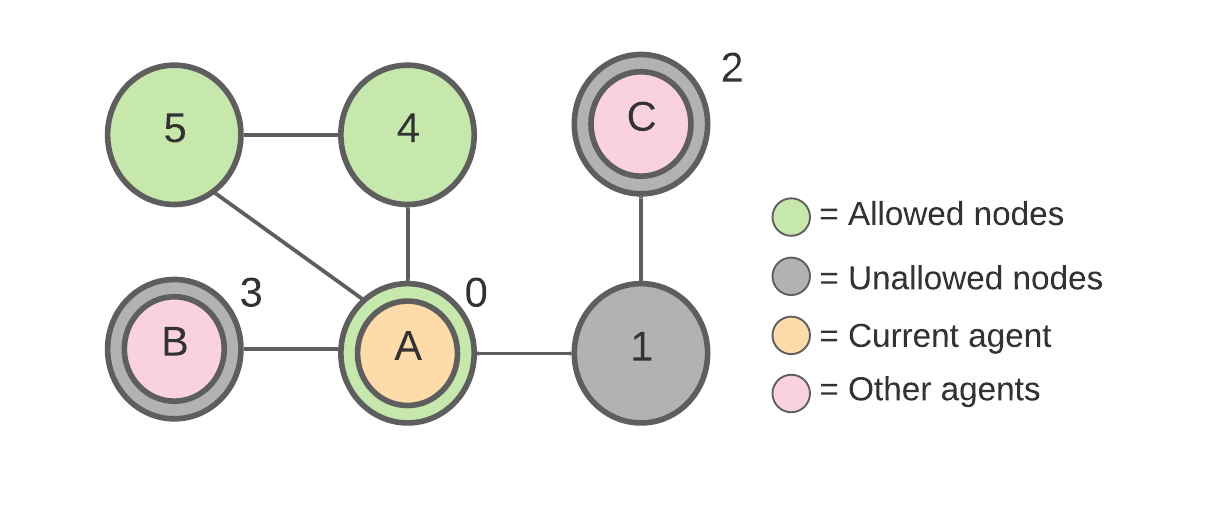}
    \caption{Collision rule example. The green nodes denote nodes in which agent A may travel to next while gray nodes denote unallowed nodes. Nodes $n_3$ and $n_2$ are unallowed because agent B and agent C, respectively, reside there. Agent A is unable to travel to $n_1$ because $n_2$ is a neighboring node with a different agent, agent C. Nodes $n_5$ and $n_4$ are allowed because neither have agents residing there and neither have neighboring nodes with agents present.}
    \label{CollisionRuleExample}
\end{figure}

Lastly, we handle collision between agents. Collisions occur when agents inhabit or travel to the same node at the same time step. Agents should hope to avoid these collisions to prevent unexpected control situations. To prevent these collisions, an agent is disallowed from traveling to nodes that contain an agent or are adjacent to nodes containing a different agent. In other words, agents decide where to travel to using only a subset of adjacent nodes. As a result, agents are effectively guaranteed to never collide. Agents without immediate non-collision nodes to travel towards wait until nodes are available. Any agents moving a box whose path has an incoming collision waits until the next node along the path becomes available. An example of determining unallowed nodes is shown in Figure \ref{CollisionRuleExample}. The process is formalized in Algorithm \ref{CollisionRuleAlg}.

\begin{algorithm}[H]
\SetAlgoLined
  Let $n_i$ denote the current node and $N(n_i)$ be the set of reachable neighbors\\
 \For{$n_j \in N(n_i)$}{
     Let $N(n_j) \setminus n_i$ denote the neighbors of $n_j$ except current node $n_i$\\
    \eIf {$n_j$ has agent $\parallel$ any node $n_k \in N(n_j)$ has an agent}{
        $n_j$ is not allowed \\
    }{
        $n_j$ is allowed\\
    }
 }
 \caption{Collision Detection}
 \label{CollisionRuleAlg}
\end{algorithm}

An edge case remains where two agents each push one box, both along the same path. As a result, their paths are bound to intersect. If both agents attempt to push to a particular node at the same time, both agents will indefinitely wait for the other to move and clear up the target node of collisions. However, this case only occurs when agent policies and locations are resolved synchronously. In practice, particularly when incorporating V-REP, we handle each agent step asynchronously, where an agent only nominally changes its location once it's fully reached a different node. Since one of the two agents is nearly guaranteed to reach its target node and determine its next action before the other, this edge case is highly unlikely.

\subsection{Agent Local Policy}
An agent at a particular node has three pheromones it may choose to follow: D, B and E (Note: H-Pheromones help determine hole candidates but are not used explicitly in the policy).

With a fixed probability $\epsilon$, the agent may choose to explore, following the E-pheromone. If the agent chooses the E-pheromone, it will choose the next node to travel to using the E-pheromones among all neighboring, reachable nodes, or the set $\{E_j; n_j \in N(n_i)\}$ with lower concentrations corresponding to higher probabilities. Immediately neighboring nodes with boxes are included as candidates, and, if chosen, the agent uniformly selects holes to push the box to from the candidates outlined in Algorithm \ref{GetHoleCandidateAlg}.

With probability $1 - \epsilon$, the agent does not explore. Hence, the agent decides between the D pheromones among neighboring nodes or the B-pheromone of the immediately closest box within detection range $R$. The agent uses Softmax over these concentrations to decide which pheromone to follow, as shown in Algorithm \ref{OverallAlgorithm}.

If it follows the D-pheromone, the agent subsequently uses Softmax over all neighboring nodes proportional to the amount of D-pheromone. Then, it will travel to the chosen node.

If the agent follows the B-Pheromone, the decision is a bit more involved. If the target box $b_k$ is able to be pushed immediately (meaning it is an immediate neighbor of the box), the agent will apply Softmax over the values $B_k(h_j)$ to determine which hole to push the box to among all hole candidates $h_j$. Then, we follow the B-pheromone update and path planning outlined in Section 3.3.2. If the box is not an immediate neighbor, the agent will travel one, single node along the shortest path towards the box. Keep in mind the shortest path is accessible to the agent because target boxes are always within detection radius $R$.

\subsubsection{Incorporating Controls}

When traveling between nodes, there are two cases to consider when incorporating controls. First, there is the case where there is no box to push, so an agent is solely transporting itself to a goal location. In this case, the controls are manually specified because of its simplicity. Referencing Table \ref{table:AllControlsTable}, the agent first uses the "Angle Towards Goal" action, described by Equation \ref{ReorientDiagram}, until its orientation with respect to the goal node is within some threshold value. Then, the agent uses the "Push In" action, described by Equation \ref{TravelDiagram}, to travel towards the node until within some fixed distance.

The second case involves pushing a box to some location. To incorporate the policy from Section 2, we require some method of telling whether the policy is able to push a box to a requested location before attempting to do so. As a result, we train a binary classifier to determine this given its current state, as defined in Equation \ref{StateSpace}. If the state is classified as "Yes," then the agent will continue to push the box towards the location. Otherwise, the agent chooses a different pheromone, leaving the box behind. Notice, the B-Pheromones $B_k^{h_j}$ associated to the box $b_k$'s target $h_j$ is decayed by $\frac{1}{B_{decay}}$ the moment an agent begins to push it. In other words, an agent may leave a box after decaying $B_k^{h_j}$, signalling that it is unable to complete its path at the moment. 

An agent leaves the box with the intention that a different agent, or itself, can return from an angle or location in which it is able to push the box to the next desired node. In this way, we eliminate the need for more complex planning or reorientation. Instead, we rely on a simple rule and the availability of multiple agents. 

To train the classifier, we must provide an appropriate data set. This dataset should reflect a variety of states in which an agent is able to or unable to push a box to a particular location. These data points are gathered using a post-training test environment with the trained policy. The post-training environment features the three training environments discussed in Section 2.3. Given a trajectory $\tau = \{s_t\}_{t=1}^{T}$ with a limited number of steps $T$, if the agent is successful, all data points in $\tau$ are added to the data set $\mathcal{S}$ with label $1$. Otherwise, the data points in $\tau$ are inserted with label $0$. We gather data until $|\mathcal{S}| \geq 10,000$.

Then, we train a random forest binary classifier using $\mathcal{S}$ with 100 decision tree instances, each with a max depth of 10. In this way, the classifier is able to determine, based on state information, whether or not the agent policy can push the box to a particular location.

\begin{figure}[h]
    \centering
    \includegraphics[width=1\linewidth, height=10cm]{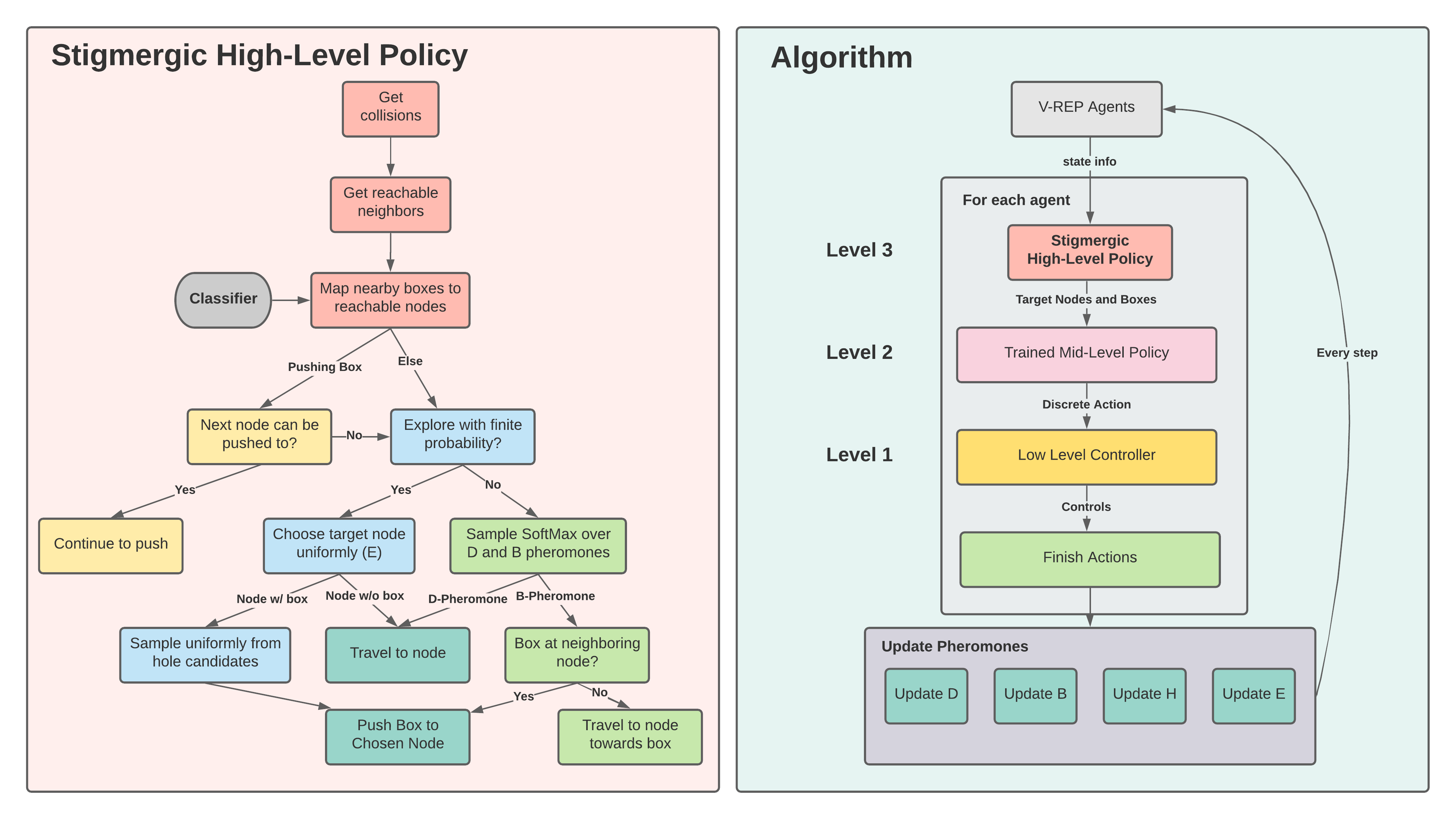}
    \caption{Overall stigmergic algorithm presented in a decision tree diagram on the left. Processes that occur every step are shown on the right. Information regarding episodic restarts and tracking maximum number of steps is not included.}
    \label{OverallAlgorithmDiagram}
\end{figure}

The overall algorithm with controls incorporated is shown in Algorithm \ref{OverallAlgorithm} and Figure \ref{OverallAlgorithmDiagram}. 

\begin{algorithm}[!htbp]
\footnotesize
\SetAlgoLined
\textbf{Note:} ** $:=$ Lines are included when controls are accounted for but excluded when controls are ignored\\
\textbf{Initialize} Max number of steps $S$, binary classifier $M$, Boltzmann constant $\beta$, exploration decay $d_{\epsilon}$, initial exploration $\epsilon_{0}$, minimum exploration $\epsilon_{min}$, box decay $B_{decay}$, detection radius $R$, max distance $d_{max}$\\
\For{every episode}{
     \While{steps < S}{
      \For{every agent}{
         Get collisions $C$ using Algorithm 8\\
         Let $N = N \setminus C$ denote reachable neighbors without collisions \\
         ** Get nodes with boxes $N_b \subseteq N$\\
         ** Map each box node $n \in N_b$ to set of nodes $N_k$ that the agent is able to push box $b_k$ to according to classifier $M$\\
        \If{agent is pushing a box}{
            ** If next node not in $N_k$, stop pushing this box. Move somewhere else \\
            Continue pushing along path if next node not in $C$\\
            Otherwise, wait for next step.\\
            Go to next agent \\
        }
        
        ** Let $N = N \setminus \{n: |N_k| = \varnothing, n \in N_b\}$ \\
        \eIf{explore with probability $\epsilon$}{
            Sample next target node $n_t \sim Softmax(-\beta * E_i; n \in N)$\\
            \eIf{$n_t$ has some box $b_k$}{
                Get hole candidates $\mathcal{H} =\{h_j\}$ of $b_k$ using Algorithm 6 and R\\
            ** Filter $\mathcal{H}$ to only include $h_j$ s.t. first node in paths to $h_j$ are $\in N_k$ \\
                Sample new hole target $h_j \sim U(\mathcal{H})$\\
                Follow path to push box $b_k$ to hole $h_j$\\
            }{
                Travel to $n_t$\\
            }
            $\epsilon = max(\epsilon_{min}, \epsilon * d_{\epsilon}$)\\
        }{
            Get box candidate $b_k$ using Algorithm 5 and R\\
            Get hole candidates $\mathcal{H} =\{h_j\}$ of $b_k$ using Algorithm 6 and R\\
            ** Filter $\mathcal{H}$ to only include $h_j$ s.t. first node in paths to $h_j$ are $\in N_k$ \\
            Get box value $V_k \sim SoftMax(\{ V_k(h_j); h_j \in \mathcal{H}\})$\\
            Get set of neighboring D-pheromones $D = \{D_i; n_i \in N\}$\\
            Get next target node $n_t \sim Softmax(V_k \cup D)$\\
            \eIf{Chose box value $V_k$}{
                Get path $\{n\}_{i=1}^{T}$ to box \\
                \eIf{box $b_k$ is at node $n_1$}{
                    Follow path to push box $b_k$ to hole associated with value $V_k$\\
                }{
                    Travel to node $n_1$\\
                }
    
            }{
                Travel to $n_t$\\
            }
        }
      }
      Allow all agents to move to respective assigned locations \\
      Update D-Pheromones by applying Algorithm 2 using $d_{max}$ with ( to all nodes $n_i$\\
      Update B-Pheromones according to Algorithm 3 using $B_{decay}$\\
      Update H-Pheromones by applying Algorithm 4 to all nodes $n_i$\\
      Update E-Pheromones by applying Algorithm 7 to all nodes $n_i$\\
      \If{all agents are at goal or reached max steps}{
        Go to next episode \\
       }
     }
 }
 \caption{Overall Stigmergic Algorithm}
 \label{OverallAlgorithm}
\end{algorithm}

\subsection{Experiments and Results}
\subsubsection{Without Controls}

To test pheromone behavior and the stigmergic policy, we remove dependence on the trained control policy from Section 2 by restricting this section of tests to a virtual node world. This simplification is shown in Figure \ref{EasyEnvNodes}. All visualizations are generated using NetworkX, a graph visualization library. 

When not accounting for controls, agents execute policies and update locations synchronously. As a result, all agents, after determining target nodes using Algorithm \ref{OverallAlgorithm}, simultaneously teleport to immediate nodes chosen. An agent moving a box is represented as an agent and box sharing the same node, moving across nodes together. Note there is no dependence on controls, meaning it is assumed that an agent is able to push the box to any neighboring, reachable node. As a result, Algorithm \ref{OverallAlgorithm} is executed without the lines marked by $**$.

\begin{figure}[h]
    \centering
    \begin{tabular}{cc}
        \includegraphics[width=.4\textwidth]{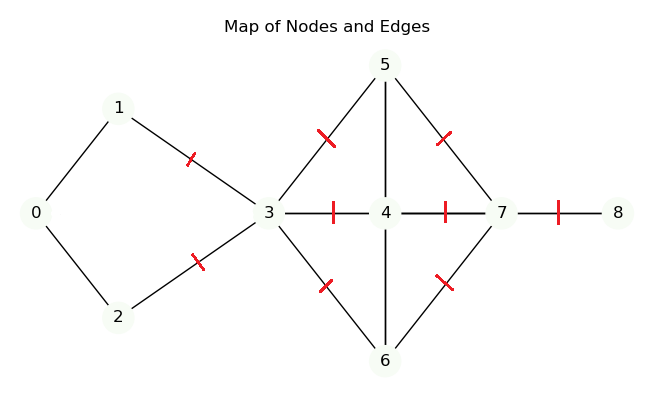} & \includegraphics[width=.4\textwidth]{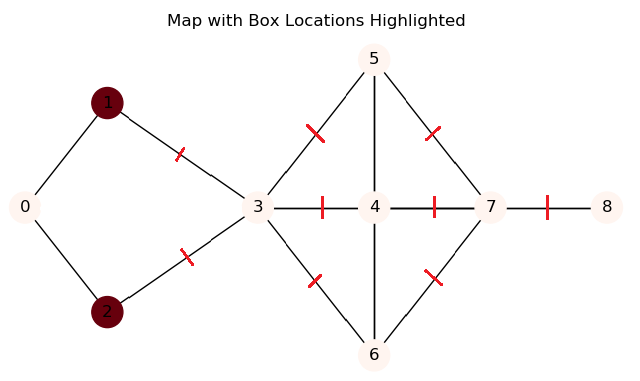} \\
        (a) & (b) \\
        \includegraphics[width=.4\textwidth]{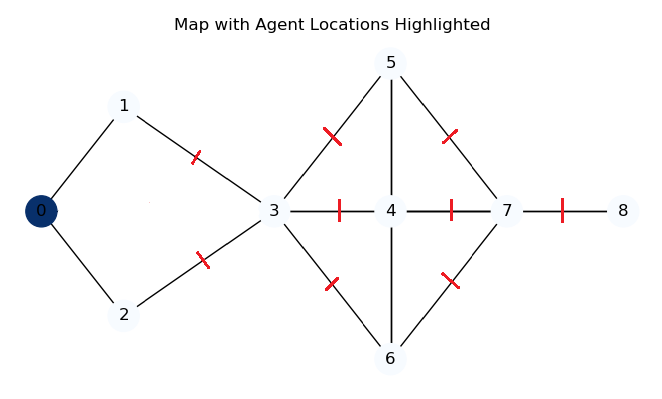}& \includegraphics[width=.4\textwidth]{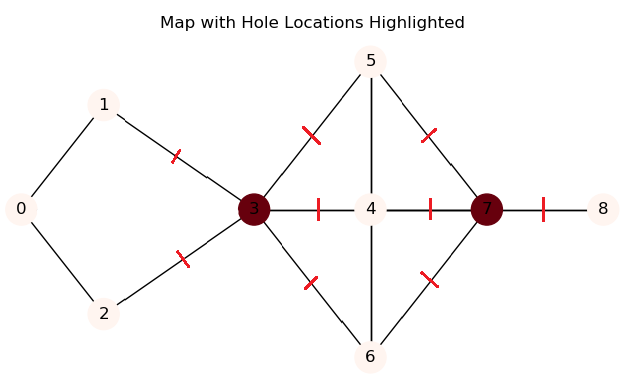} \\
        (c) & (d) \\
    \end{tabular}
    \caption{Sanity check environment with one agent, two boxes, two holes and goal node $n_8$. }
    \label{SanityCheckEnv}
\end{figure}

First, a very simple, sanity-check environment is constructed consisting of one agent, nine nodes, two boxes and two holes. The environment is shown below, with the agent, boxes and nodes highlighted in their respective images in Figure \ref{SanityCheckEnv}. In this environment, the goal is for the agent to traverse across the map from the start node $n_0$ to the goal node, which, in this case, is $n_8$. The box initially at node $n_1$ has a height of 1, and the box at node $n_2$ has a height of 3. Furthermore, the hole at $n_3$ has a depth of 1 and the hole at $n_7$ has a depth of 3. Therefore, it makes immediate sense that the agent should learn to push a box from $n_1$ to $n_3$ and from $n_2$ to $n_7$. Any other combination would prevent passage to the goal $n_8$ due to unfilled holes or insurmountable hills (ex: if both boxes were placed at $n_3$). Since stacking boxes on top of each other is considered an option, the agent must learn which of the four box configurations to choose, all the while attempting to navigate towards node $n_8$.

\begin{figure}[H]
    \centering
    \begin{tabular}{ccc}
        \includegraphics[width=.3\textwidth]{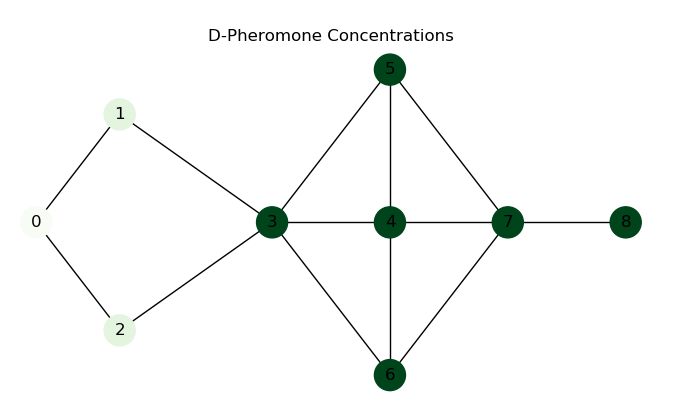} &  
        \includegraphics[width=.3\textwidth]{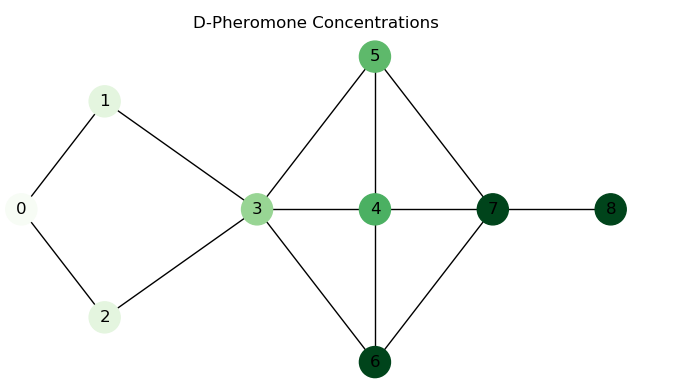} & 
        \includegraphics[width=.3\textwidth]{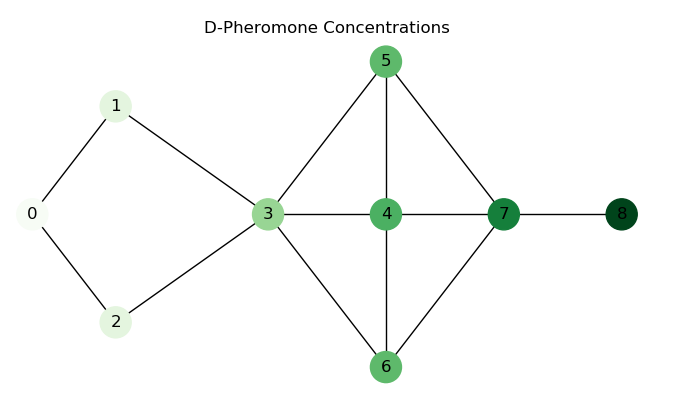} \\
        (a) & (b) & (c) \\
        \includegraphics[width=.3\textwidth]{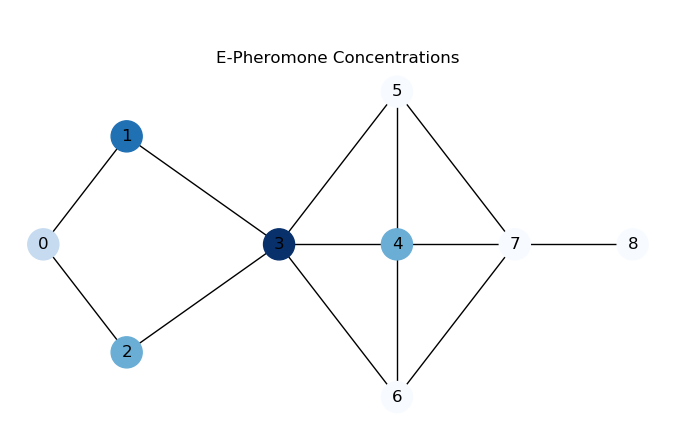} &  
        \includegraphics[width=.3\textwidth]{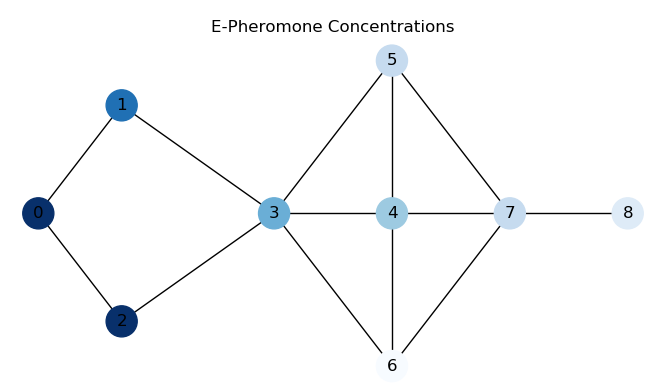} & 
        \includegraphics[width=.3\textwidth]{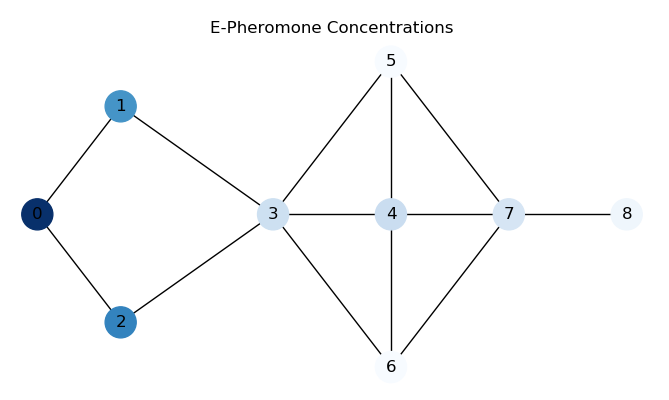} \\
        (d) & (e) & (f) \\
        \includegraphics[width=.3\textwidth]{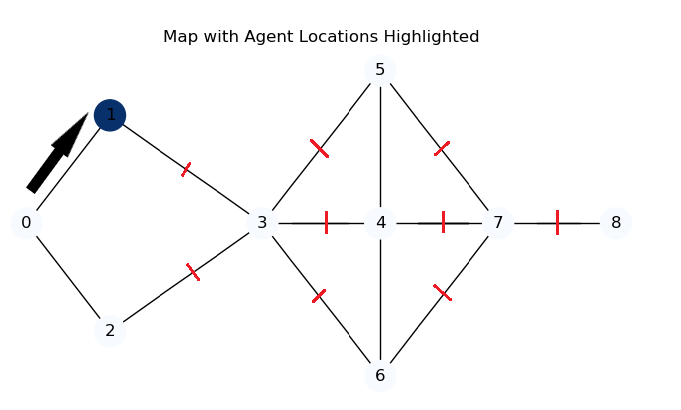} &  
        \includegraphics[width=.3\textwidth]{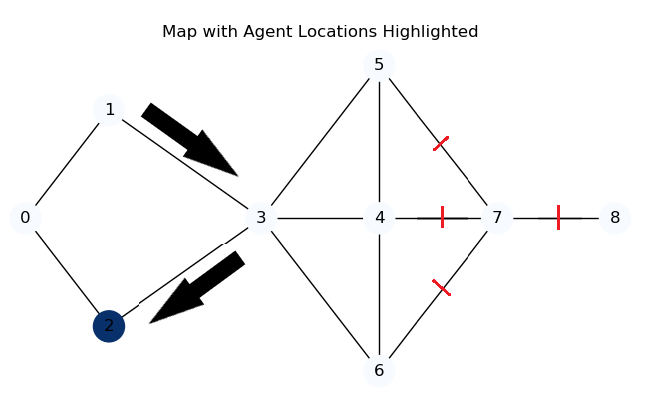} & 
        \includegraphics[width=.3\textwidth]{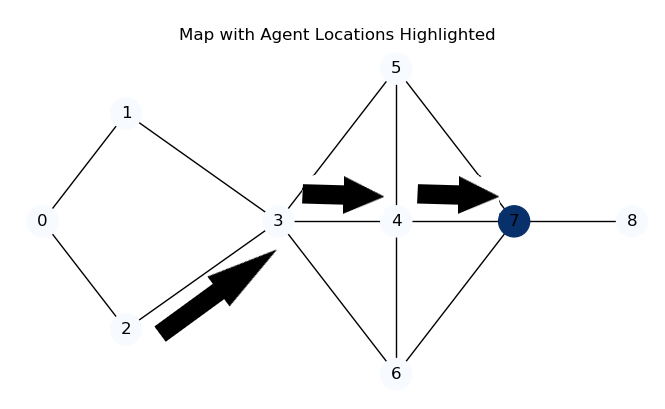} \\
        (g) & (h) & (i) \\
    \end{tabular}
    \begin{tabular}{cc}
        \includegraphics[width=.3\textwidth]{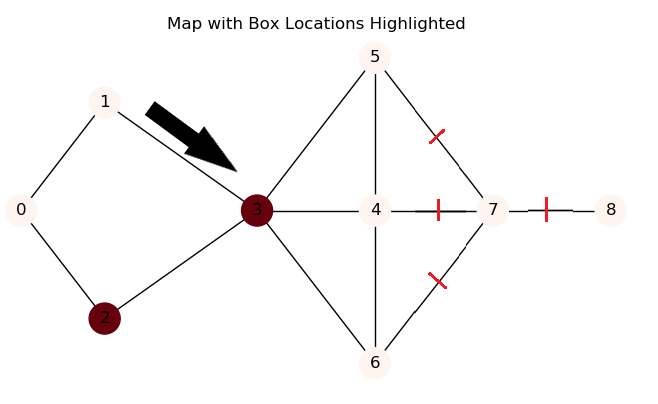} &  
        \includegraphics[width=.3\textwidth]{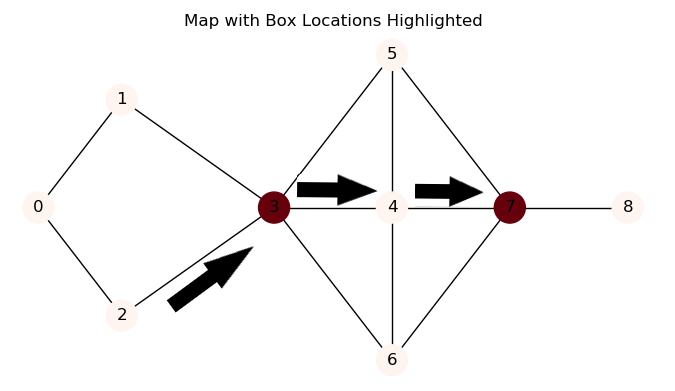} \\
        (j) & (k) \\ 
        \includegraphics[width=.3\textwidth]{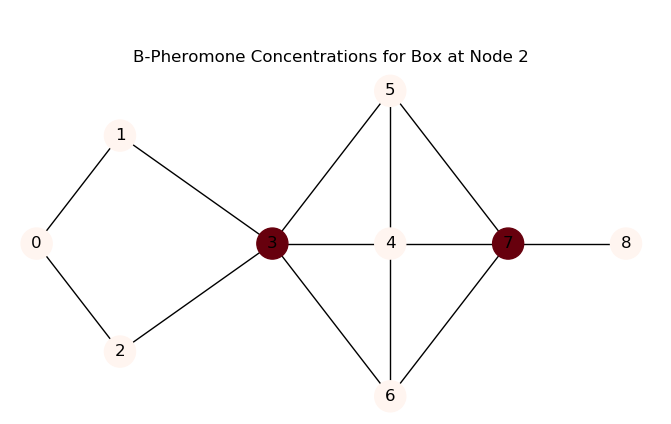} & 
        \includegraphics[width=.3\textwidth]{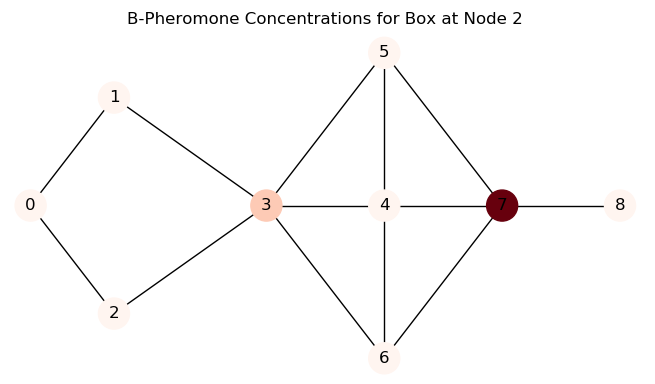} \\
        (l) & (m) \\
    \end{tabular}
    \caption{Progression and development of pheromones across different episodes in the sanity check environment. In (a) and (b), we see the agent update D-pheromones according to Euclidian distance in the first episode. In (c), all nodes become Official by episode 4 with D-Pheromones calculated using path distance. Figures (d), (e), and (f) show the development of E-Pheromones, where higher concentrations are found on the left side near the agent's initial position. This occurs because, in cases where the agent is unable to traverse past $n_3$, the agent is trapped at nodes $n_0$, $n_1$, and $n_2$, forcing it to place more E-Pheromones. In Figures (g), (h), and (i) we describe the motion of the agent and how it moves the appropriate boxes in (j) and (k). It appropriately learns to move the box from $n_1$ to $n_3$ and another box from $n_2$ to $n_7$. Furthermore, the B-Pheromones associated with the box at $n_2$ is shown in (l) and (m), where it slowly learns to favor the hole at $n_7$ over $n_3$.}
    \label{SanityDevelopment}
\end{figure}

\begin{figure}[h]
    \begin{tabular}{ccc}
        \includegraphics[width=.33\textwidth]{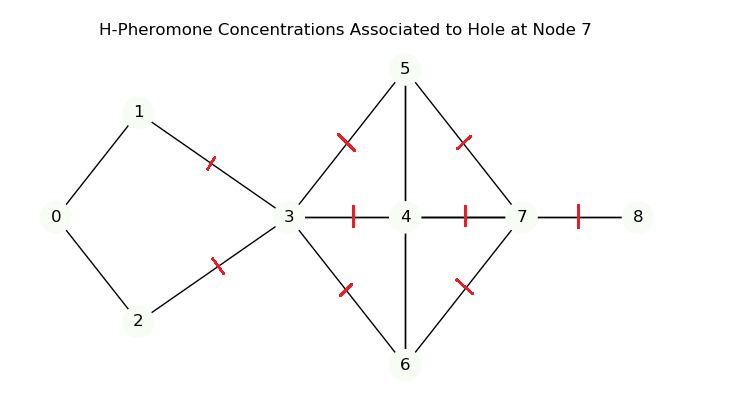} & \includegraphics[width=.33\textwidth]{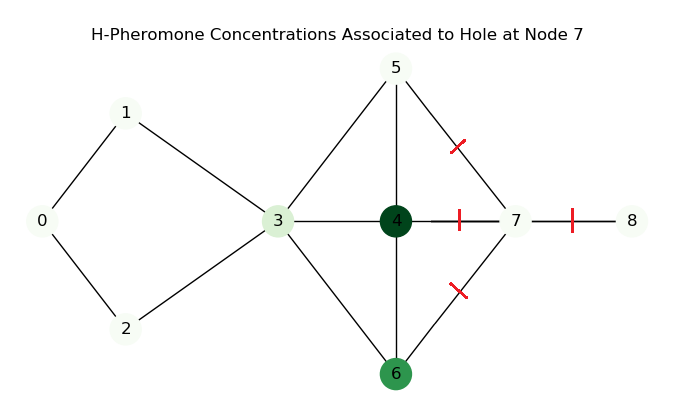} & \includegraphics[width=.33\textwidth]{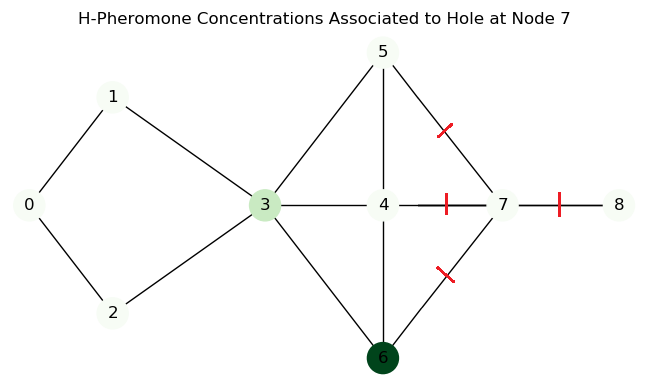} \\
        (a) & (b) & (c) \\
    \end{tabular}
    \caption{When detection radius is limited, the agent can only detect the hole at $n_7$ from $n_4$, $n_5$, or $n_6$. As a result, the box at $n_2$ requires a path of H-Pheromones towards $n_7$ to know where to push it. Figure (b) shows the development of H-Pheromones providing the shortest path from $n_2$ to $n_7$ . However, Figure (c) shows an inefficient path that may be generated by the H-pheromones.}
    \label{HPheromoneEdgeCase}
\end{figure}

We show a sample progression of the D-Pheromones, B-Pheromones, and E-Pheromones in Figure \ref{SanityDevelopment}. After approximately three to four episodes, each of which had 20 max steps, the D-pheromones appropriately converge to values calculated using path distances to $n_8$. B-Pheromones associated to the box at $n_2$ appropriately reflect that it should be pushed to $n_7$ as opposed to $n_3$. Lastly, when detection radius is short enough such that an agent cannot detect the hole at $n_7$ from far away, the H-Pheromones appropriately give the agent the information needed to navigate from $n_3$ to $n_7$, as shown in Figure \ref{HPheromoneEdgeCase}. 

It is important to comment on the optimality of the H-Pheromones update rule. While a given trail of H-Pheromones is guaranteed to appropriately lead an agent to a requested hole, the trail is not guaranteed to be the shortest path. This guarantee is only achieved when the nodes in question have been visited in the appropriate order or have been visited a sufficient number of times. An example of this suboptimality is given in Figure \ref{HPheromoneEdgeCase}. Although this condition is sometimes difficult to fulfill with limited number of agents and exploration, the development of these H-Pheromones are more likely as the number of agents increased, which is shown in Section 3.5.2.

\begin{figure}[H]
\centering
\begin{tabular}{cc}
 \includegraphics[width=.45\textwidth]{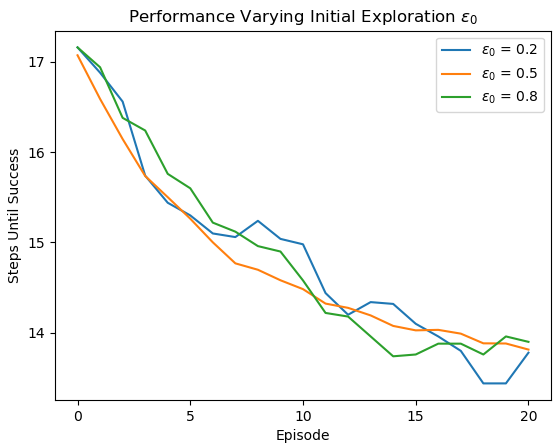} &
  \includegraphics[width=.45\textwidth]{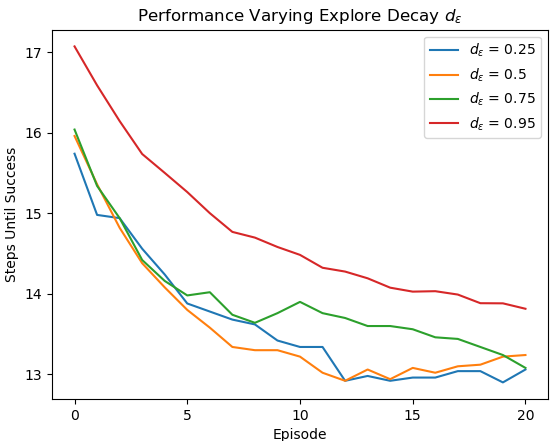} \\
   \includegraphics[width=.45\textwidth]{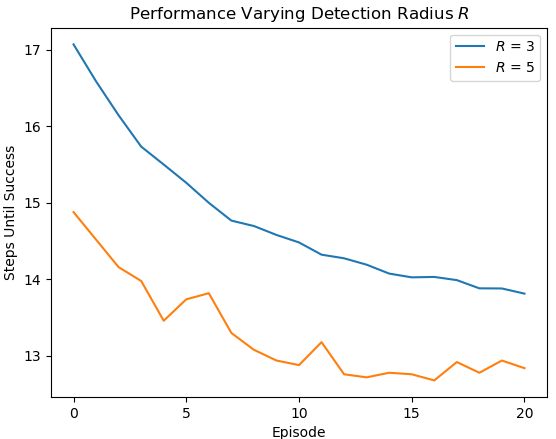} &
    \includegraphics[width=.45\textwidth]{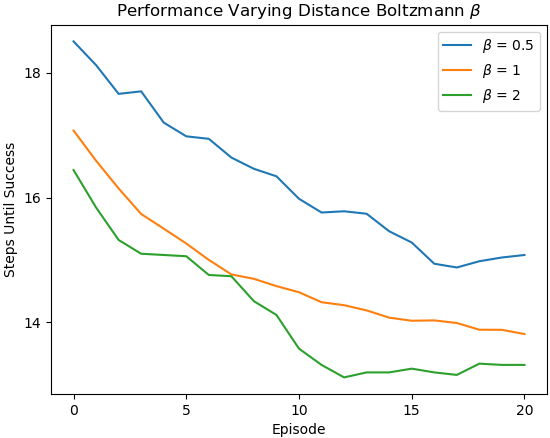} \\
    \multicolumn{2}{c}{\includegraphics[width=.45\textwidth]{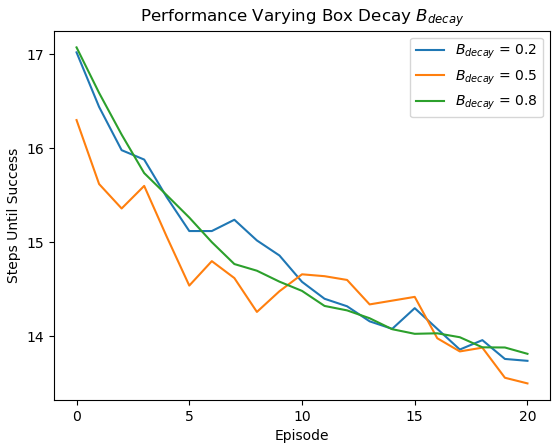}}\\
\end{tabular}
\caption{Ablation testing using NetworkX in sanity check environment}
\label{AblationSanity}
\end{figure}

Next, we test sensitivity to hyper-parameters through ablation tests. Sequestering five particular hyper-parameters: initial exploration $\epsilon_0$, exploration decay $d_{\epsilon}$, detection radius $R$, distance Boltzmann $\beta$, and box decay $B_{decay}$. Limiting each episode to 20 steps, we analyzed the number of steps it took for the agent to successfully reach $n_8$. The results are shown in Figure \ref{AblationSanity}, where we graph performance across 30 episodes of training, with a moving average of 10.

Due to the simplicity of the environment, performance differences are negligible when varying $B_{decay}$ and $\epsilon_0$. However, performance is often bottle-necked by low values of $R$ or $\beta$. The former is expected as detection radius is a representation of how much information any particular agent has access to. With detection radius $R=3$, more steps are required to reach $n_8$ because an agent must also create an H-pheromone trail from $n_2$ to $n_7$ while with a detection radius $R=5$, the agent can directly push the box from $n_2$ to $n_7$. The distance Boltzmann, $\beta$, determines the degree of certainty agents utilize when choosing which nodes to travel to. Less stochastic decisions lead to better performance when the pheromones have appropriately converged, as shown.

\begin{figure}[!ht]
    \centering
    \begin{tabular}{ccc}
        \includegraphics[width=.3\textwidth]{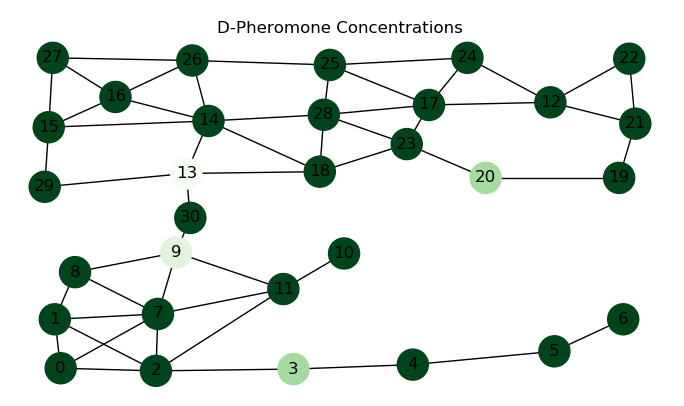} & \includegraphics[width=.3\textwidth]{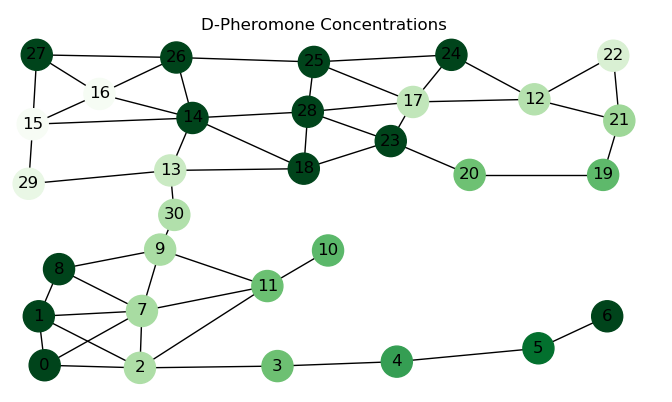} & 
        \includegraphics[width=.3\textwidth]{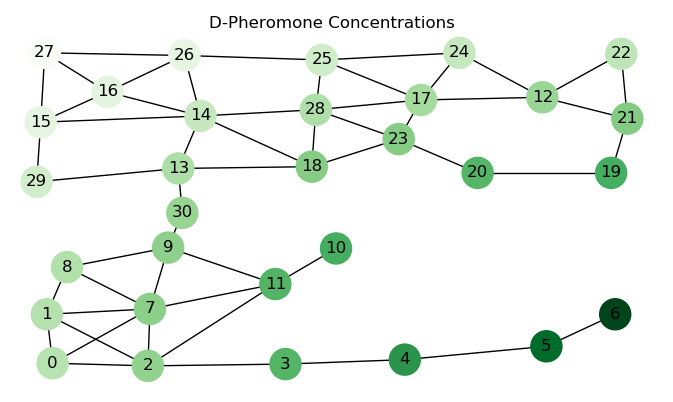} \\
        (a) & (b) & (c) \\
        \includegraphics[width=.3\textwidth]{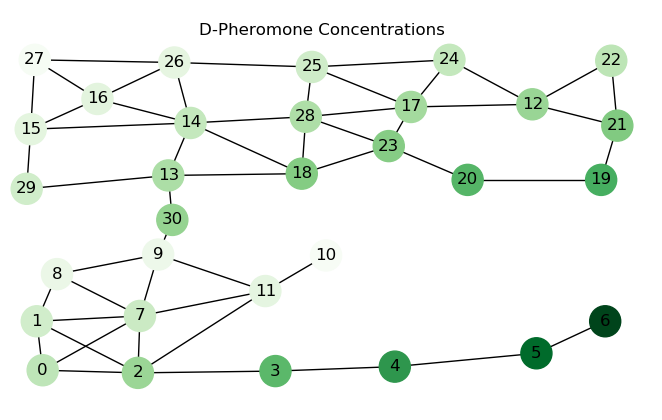} & \includegraphics[width=.3\textwidth]{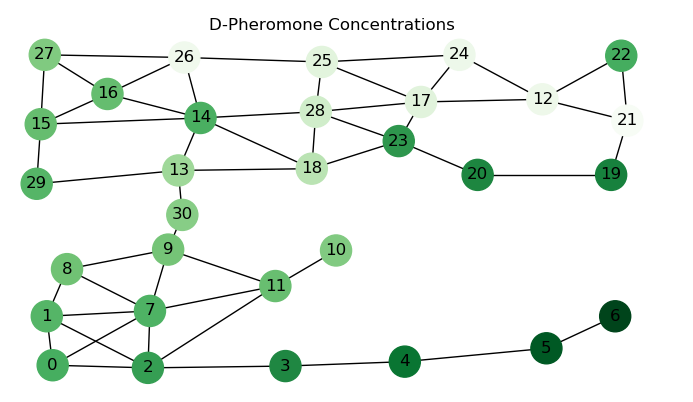} & 
        \includegraphics[width=.3\textwidth]{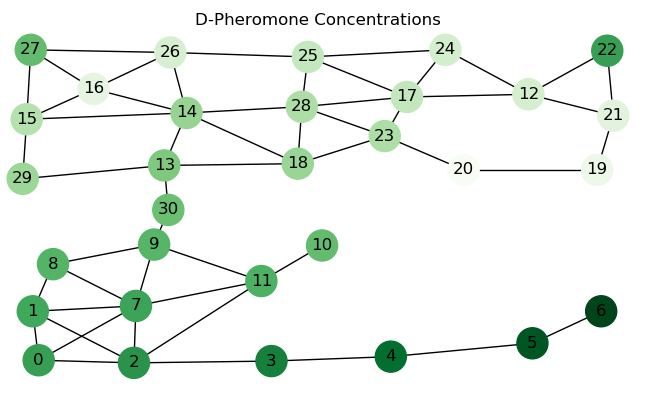} \\ 
        (d) & (e) & (f) \\
    \end{tabular}
    \caption{D-Pheromone progress across episodes 1 to 3. Initial D-Pheromone calculations using Euclidian distance are shown in parts (a)-(c). Then, official path distances are used to calculate D-Pheromones in parts (d)-(f). The near-converged D-Pheromone concentrations are shown in Figure 31c.}
    \label{EasyDPheromone}
\end{figure}

\begin{figure}[!ht]
    \centering
    \begin{tabular}{cc}
        \includegraphics[width=.4\textwidth]{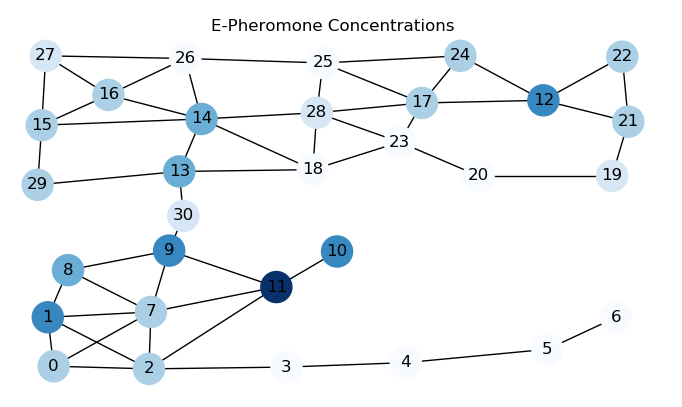} & \includegraphics[width=.4\textwidth]{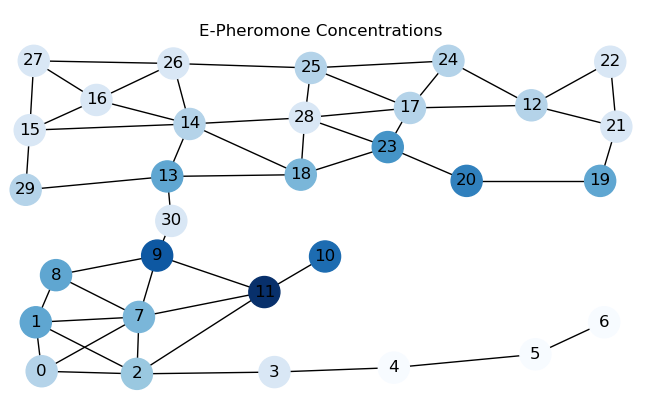}  \\
        \includegraphics[width=.4\textwidth]{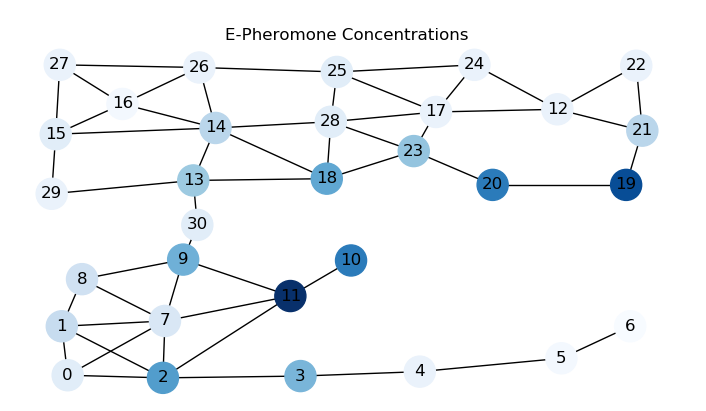} & \includegraphics[width=.4\textwidth]{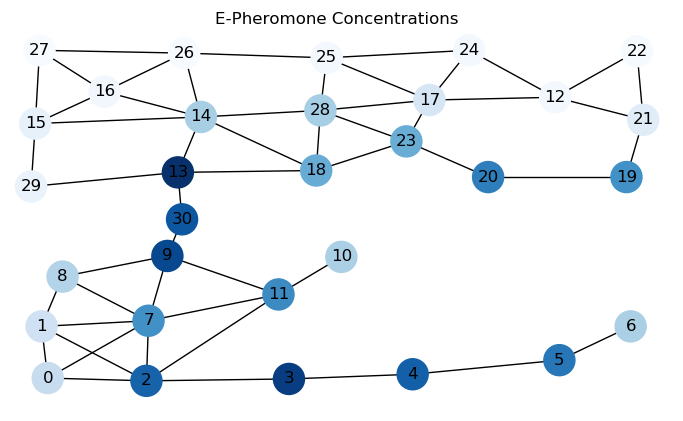}
    \end{tabular}
    \caption{E-Pheromone Progress}
    \label{EasyEPheromone}
\end{figure}

Next, a larger environment is constructed to demonstrate the effectiveness of collision detection and scalability to multiple agents. We name this environment the Easy environment, as it is a reconstructed node version similar to the V-REP environment shown in Figure \ref{EasyEnvNodes}. Similar to the sanity check environment, we analyze the progression of D-Pheromones and E-Pheromones. Due to this environment's  simplicity, B-Pheromones and H-Pheromones were not analyzed, as agents typically had one option to push boxes to and boxes were at nodes immediately neighboring holes. Instead, we analyze the interplay between the D-Pheromones, E-Pheromones and collision detection rules. The D-Pheromones, shown in Figure \ref{EasyDPheromone}, appropriately reflect path distances to the goal node $n_6$. Note the lack of perfect convergence in the final image. This behavior remains possible as agents may already know the shortest path to $n_6$ and do not travel to peripheral locations as they are irrelevant to reaching the goal location. E-Pheromones, shown in Figure \ref{EasyEPheromone}, initially show agent hovering around nodes that had small Euclidian distance to $n_6$. For example, $n_{20}$, $n_{19}$, and $n_{11}$ were high in concentration initially. However, as agents learned to successfully navigate to $n_6$, such pathways began to have higher concentrations of E-pheromone.

\begin{figure}[!ht]
\centering
\begin{tabular}{cc}
 \includegraphics[width=.45\textwidth]{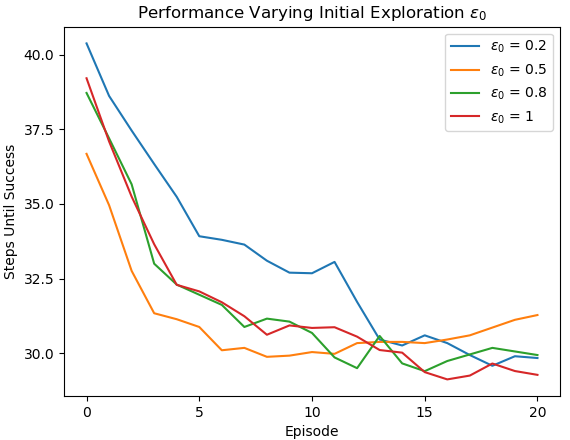} &
  \includegraphics[width=.45\textwidth]{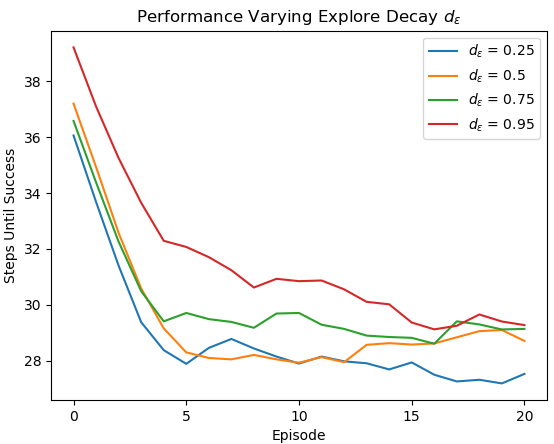} \\
   \includegraphics[width=.45\textwidth]{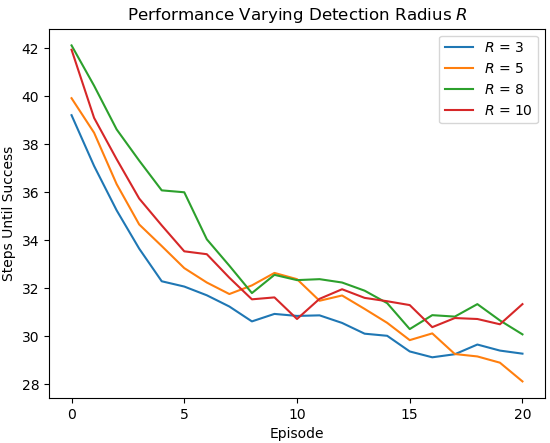} &
  
    \includegraphics[width=.45\textwidth]{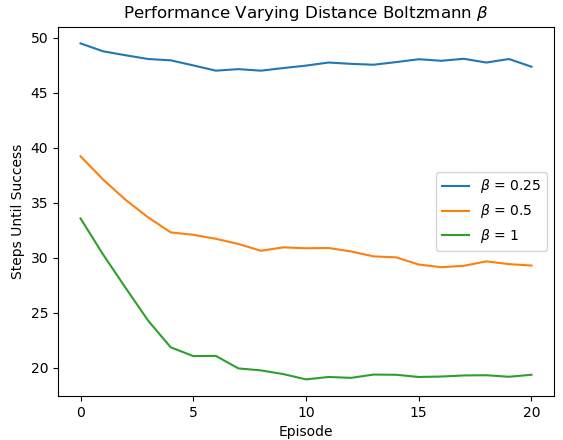} \\
    
  \multicolumn{2}{c}{\includegraphics[width=.45\textwidth]{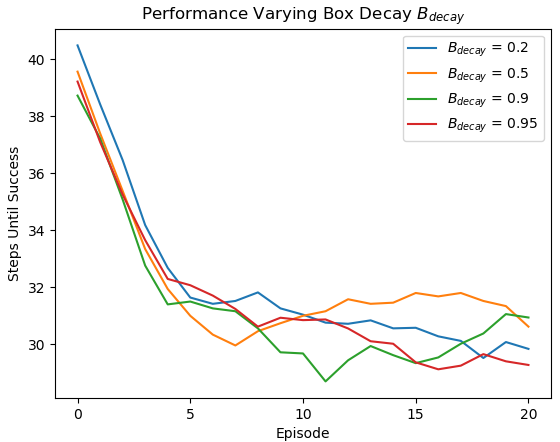}}\\
\end{tabular}
\caption{Ablation testing in easy environment}
\label{AblationEasy}
\end{figure}

Next, we analyze the ablation performance across the same five hyper-parameters in the Easy environment. The results are shown in Figure \ref{AblationEasy}

In this environment, performance is largely similar between all hyper-parameter ablations except for the distance Boltzmann parameter. Similar to before, this parameter is a measure of certainty, which deservedly so should yield better performance the higher it is. All other parameters have relatively similar performance, a promising indication of stability. The detection radius does not affect performance as much as the previous environment as boxes are already initially close to useful, relevant holes.

After analyzing the effects of the pheromone-induced policy in more straightforward instances, we test the entire control-incorporated algorithm in even more difficult environments. The results for those are shown in the subsequent section.

\subsubsection{With Controls}

We use Algorithm \ref{OverallAlgorithm} to incorporate the control policy trained in Section 2. In order to test the effectiveness of the algorithm, we use V-REP environments as opposed to the node environments from the previous section. While performance graphs were appropriate before, the trained policy yields high variance in performance, making such representations yield minimal information about the algorithm but rather the weaknesses of the trained policy. As a result, we display several charts showing statistics regarding the number of steps required to solve the environment, capped at the max number of steps, and the proportion of episodes that all agents were able to reach the goal location. 

Furthermore, we analyze the development of various pheromones using node representations similar to the previous section. Untraversable edges denoted by red slashes are not shown in these graphs as pheromone concentrations are typically independent of what part of the episode the agents are currently in. Instead, the changes in traversability are shown in figures depicting the explicit V-REP simulation, where automobiles are shown interacting with the environment.

We test the control-incorporated algorithm in three disparate environments. The Easy environment is shown in Figure \ref{EasyEnvNodes}. Because of the edge cases regarding Collision Detection mentioned in Section 3.3.5, the subsequent environments can only be tested in these control-incorporated algorithm experiments, not using the previously synchronous node world. In the control-incorporated version of the algorithm, agent policies and location updates are asynchronous. In other words, each agent chooses its next action once it completes its previous, independent from the other agents. This is due to the fact that multiple boxes on a single platform leads to incongruities with the collision detection rule, where, if two agents are pushing boxes to the same node, the collision detection rule will disallow movement by either agent, causing an infinite loop where they wait for each other. However, in asynchronous decision-making, it is nearly guaranteed that one or the other will arrive at a neighboring node first, meaning the agent who appears later is forced to wait for the agent who appeared earlier.

\begin{figure}[!ht]
    \centering
    \begin{tabular}{ccc}
        \includegraphics[width=.3\textwidth, height=3.5cm]{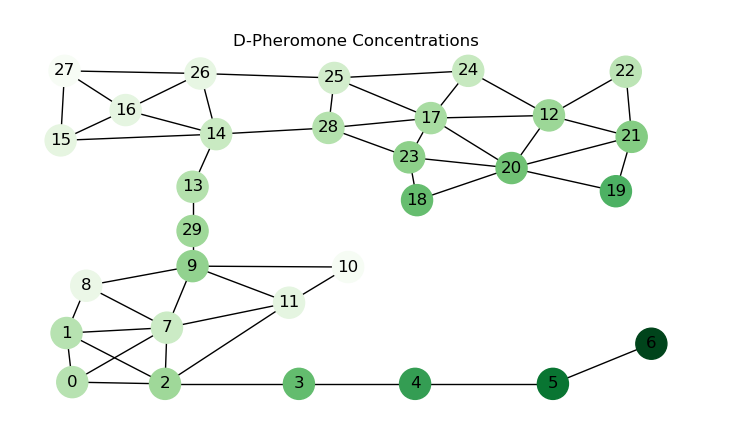} &
        \includegraphics[width=.3\textwidth, height=3.5cm]{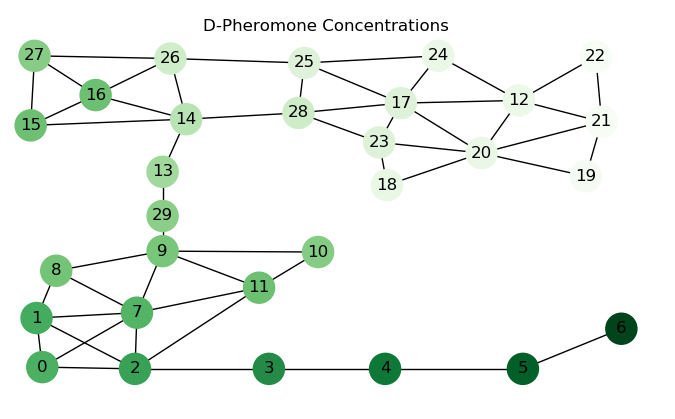} &
        \includegraphics[width=.3\textwidth, height=3.5cm]{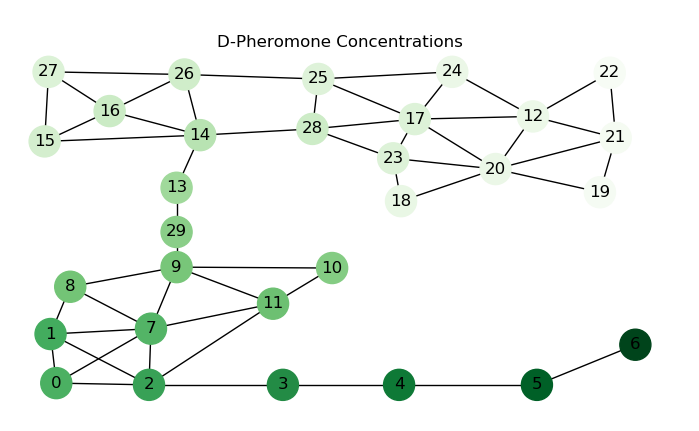} \\
        (a) & (b) & (c) \\
        \includegraphics[width=.3\textwidth, height=3.5cm]{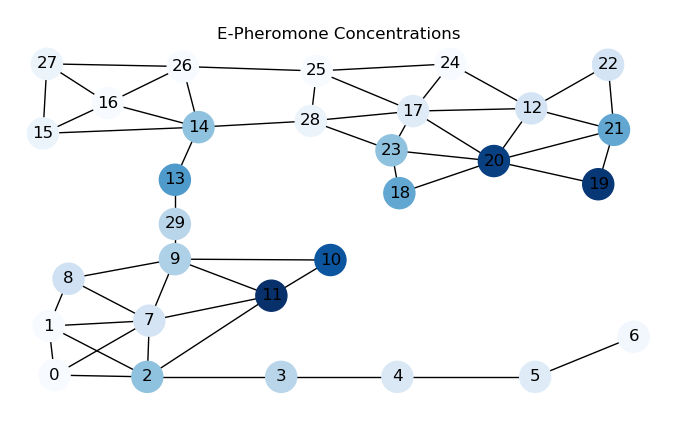} &
        \includegraphics[width=.3\textwidth, height=3.5cm]{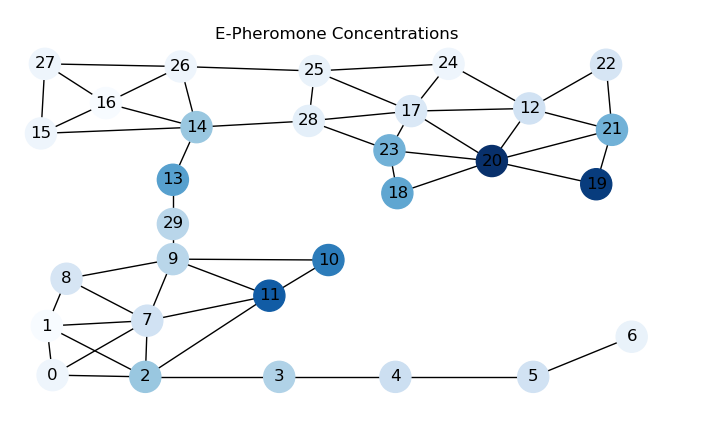} &
        \includegraphics[width=.3\textwidth, height=3.5cm]{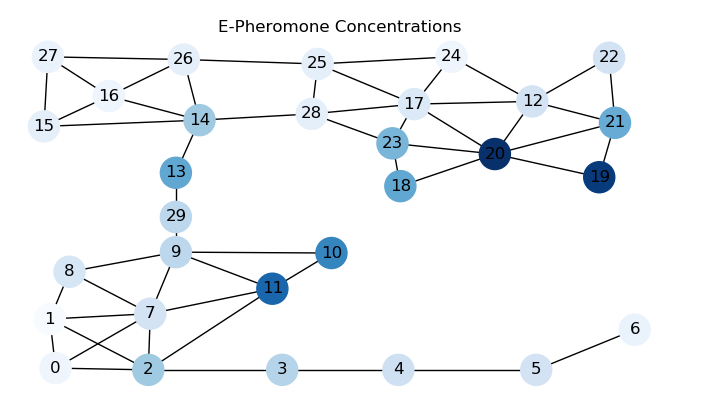} \\
        (d) & (e) & (f) \\
    \end{tabular}
    \caption{D-Pheromones and E-Pheromones during the second, third, and fourth episodes. D-pheromone concentrations calculated using path distance are shown in c. The top right nodes have high E-Pheromone concentrations because agents used Euclidian distances to calculate their respective D-pheromones, encouraging agents to hover around that area. However, as episodes progress, E-Pheromones along the path to the goal increase in concentration.}
    \label{EasyVREPPheromones}
\end{figure}

\begin{figure}[!ht]
    \centering
    \setlength{\tabcolsep}{1pt}
    \begin{tabular}{cc}
        \includegraphics[width=.45\textwidth, height=3.5cm]{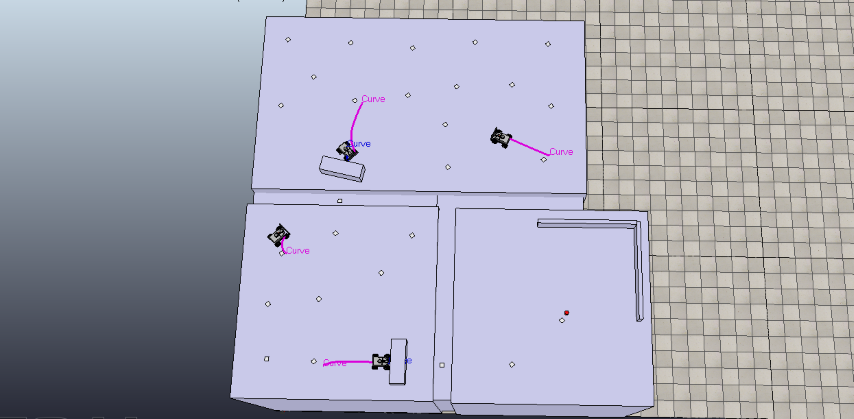} &
        \includegraphics[width=.45\textwidth, height=3.5cm]{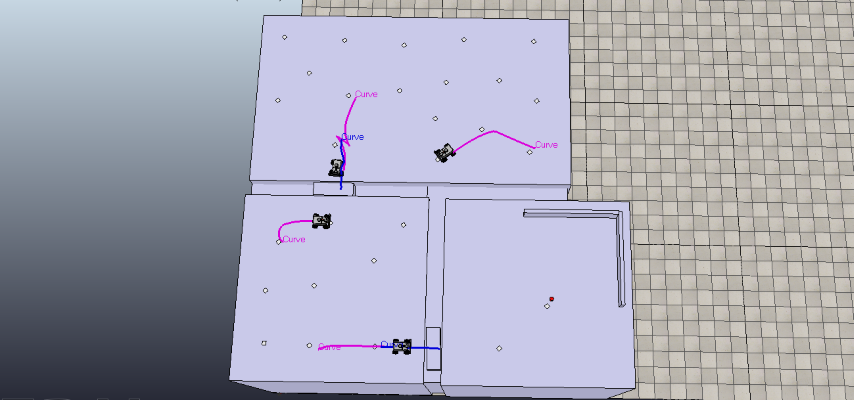} \\
        \includegraphics[width=.45\textwidth, height=3.5cm]{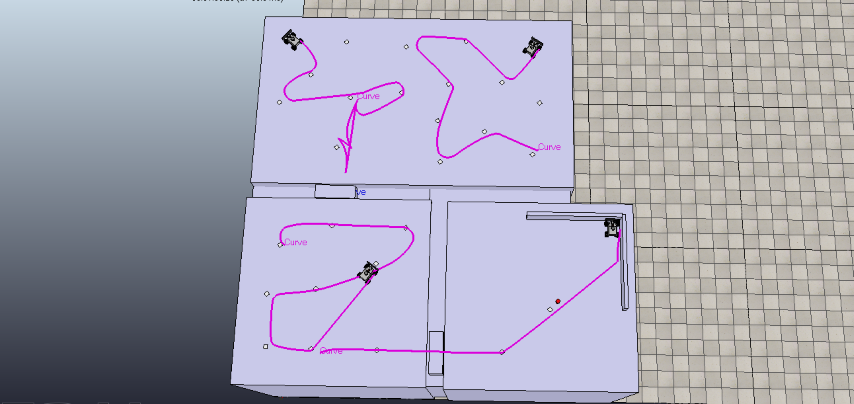} &
        \includegraphics[width=.45\textwidth, height=3.5cm]{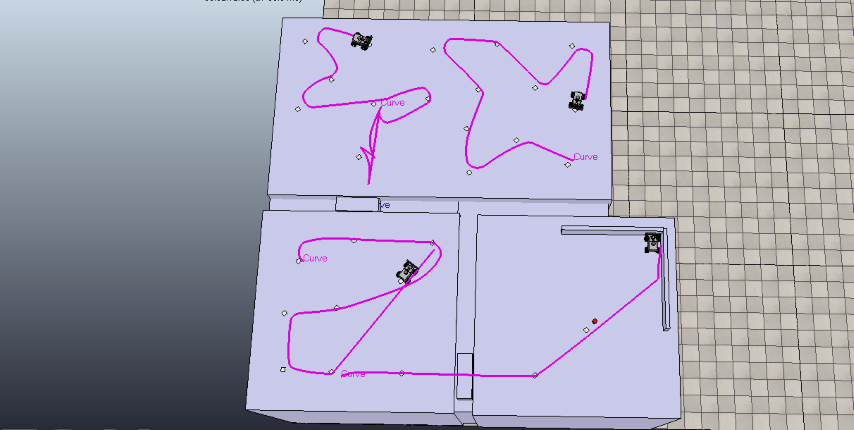} \\
    \end{tabular}
    \caption{Sample trajectory during first episode. Notice that while both agents successfully push the boxes into their respective holes, only one of the four agents succeed in traversing to the goal node in the bottom right. This is credited to nodes lacking the Official status and D-pheromones giving improper signals.}
    \label{EasyTrajectoryFirst}
\end{figure}

In conjunction with showing pheromone concentration progressions, shown in Figure \ref{EasyVREPPheromones}, we also display the agents' trajectories in the V-REP environment. Similar to previous diagrams, pink lines represent agent trajectories while blue lines represent box trajectories. In Figure \ref{EasyTrajectoryFirst}, we show agent and box trajectories in the first episode of training. After both boxes are appropriately used and pushed using the policy trained in Section 2, agents spend more their time updating node D-Pheromones using Euclidian distance. This is seen in Figure \ref{EasyTrajectoryFirst}d as agents visit various nodes.

\begin{figure}[!ht]
    \centering
    \setlength{\tabcolsep}{1pt}
    \begin{tabular}{ccc}
        \includegraphics[width=.3\textwidth, height=3.5cm]{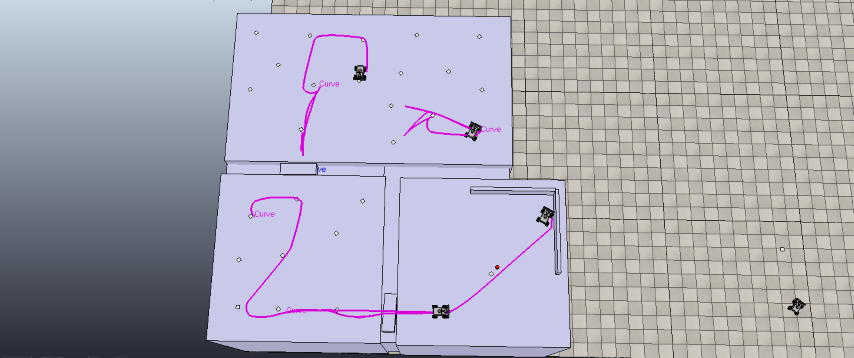} &
        \includegraphics[width=.3\textwidth, height=3.5cm]{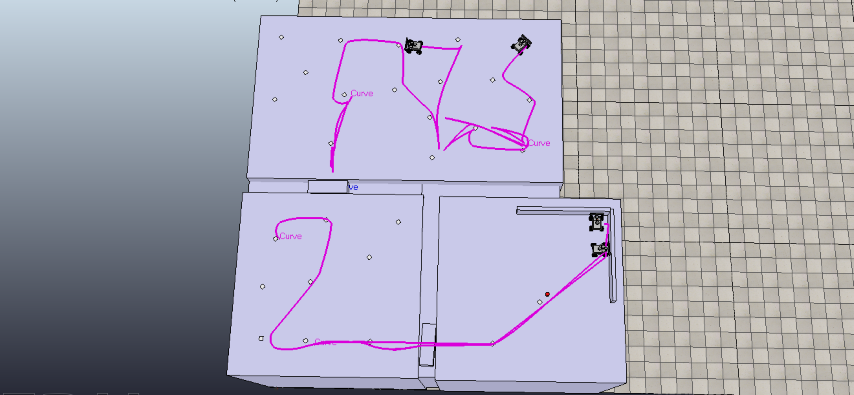} &
        \includegraphics[width=.3\textwidth, height=3.5cm]{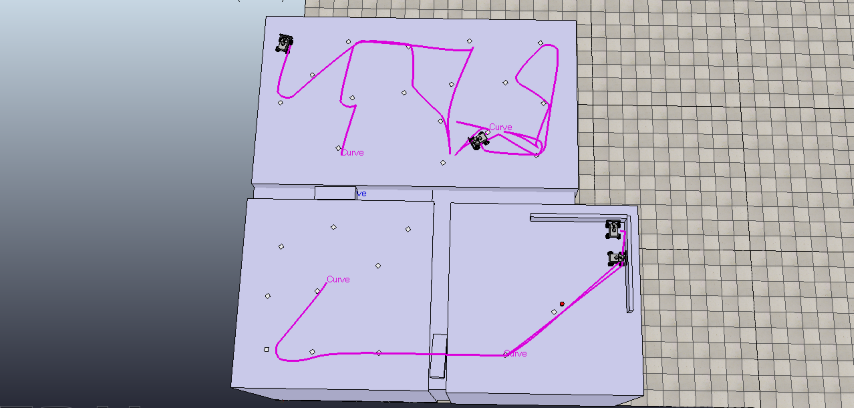} \\
    \end{tabular}
    \caption{Sample trajectory during third episode. This episode signifies the pervasion of the Official status among all nodes in the upper section. This is shown as the two agents in the top platform begin to assign path distances to nodes. Furthermore, a second agent successfully navigates to the goal location.}
    \label{EasyTrajectoryThird}
\end{figure}

Then, in Figure \ref{EasyTrajectoryThird}, we show trajectories associated with D-Pheromone concentrations in the process of becoming Official. As path distances are calculated, agents are more motivated to update node D-Pheromones that were previously calculated using Euclidian distances, as they are perceived as being closer to the goal than they actually are. 

\begin{figure}[!ht]
    \centering
    \setlength{\tabcolsep}{1pt} 
    \begin{tabular}{ccc}
        \includegraphics[width=.32\textwidth, height=3.5cm]{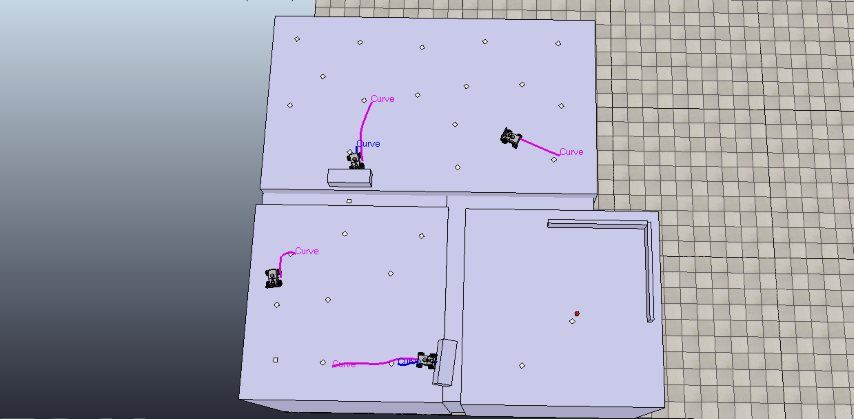} &
        \includegraphics[width=.32\textwidth, height=3.5cm]{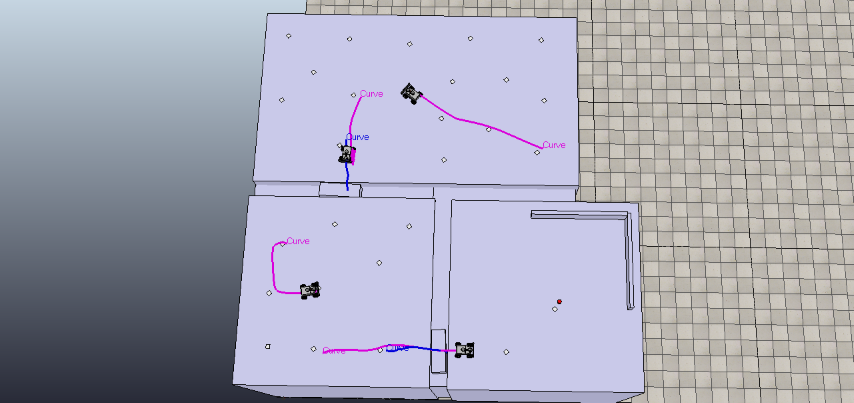} &
        \includegraphics[width=.32\textwidth, height=3.5cm]{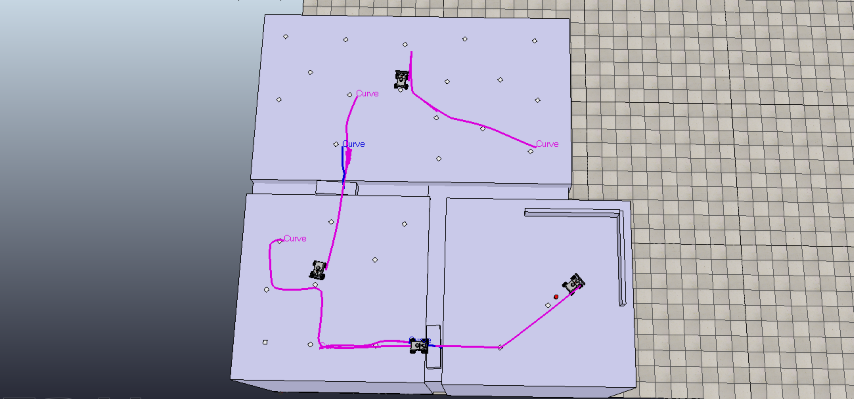} \\
        \includegraphics[width=.32\textwidth, height=3.5cm]{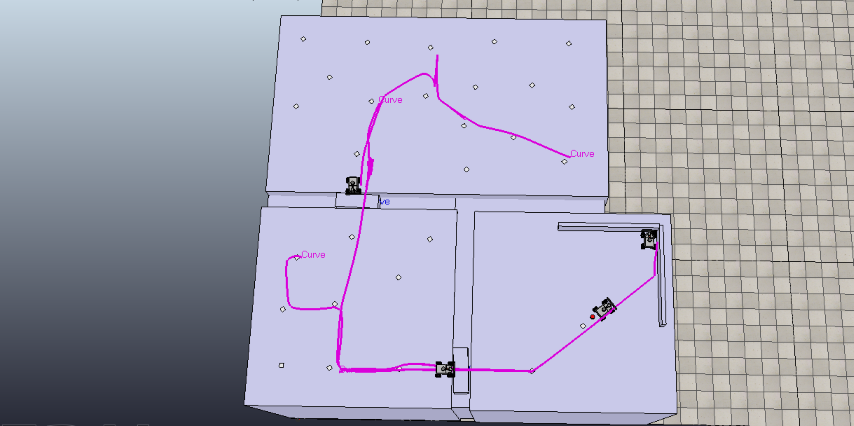} &
        \includegraphics[width=.32\textwidth, height=3.5cm]{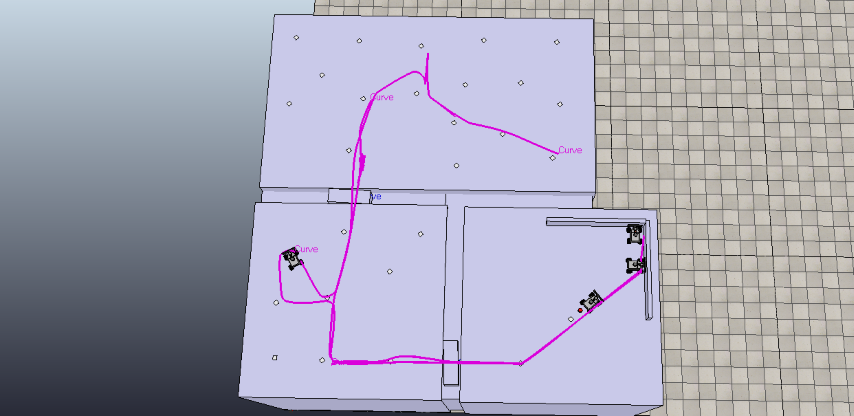} &
        \includegraphics[width=.32\textwidth, height=3.5cm]{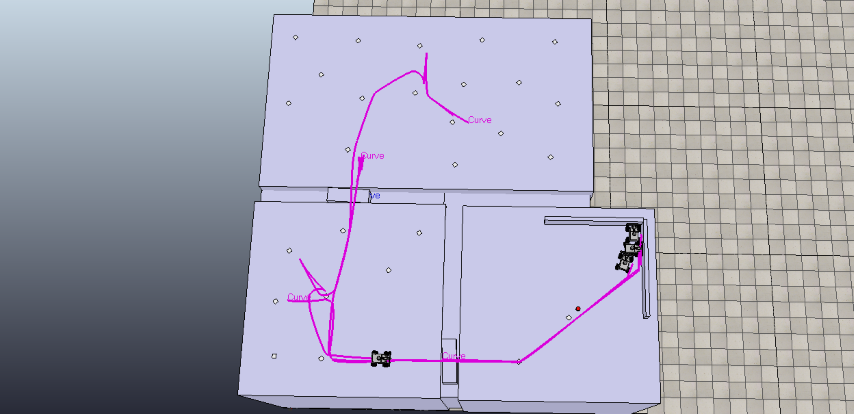} \\
    \end{tabular}
    \caption{Sample trajectory during the fourth episode. This trajectory represents a policy using converged D-pheromones, calculated using path distance. After boxes are successfully pushed to their respective holes, all agents successfully and efficiently navigate to the goal location.}
    \label{EasyTrajectoryFourth}
\end{figure}

Lastly, in Figure \ref{EasyTrajectoryFourth}, we show a sample trajectory once D-Pheromone convergence is complete. The trained control policy exhibits notably reliable results, pushing boxes into the holes consistently. These results are reported in Figure \ref{PerformanceEasyMedium}. Sources of failure include boxes being pushed into holes translationally offset from desired positions. As a result, the box does not appropriately cover the white square denoting the hole's node, making robots more likely to fall into the hole and unable to reach the goal location.

\begin{figure}[!ht]
    \centering
    \setlength{\tabcolsep}{1pt} 
    \begin{tabular}{ccc}
        \includegraphics[width=.32\textwidth, height=3.5cm]{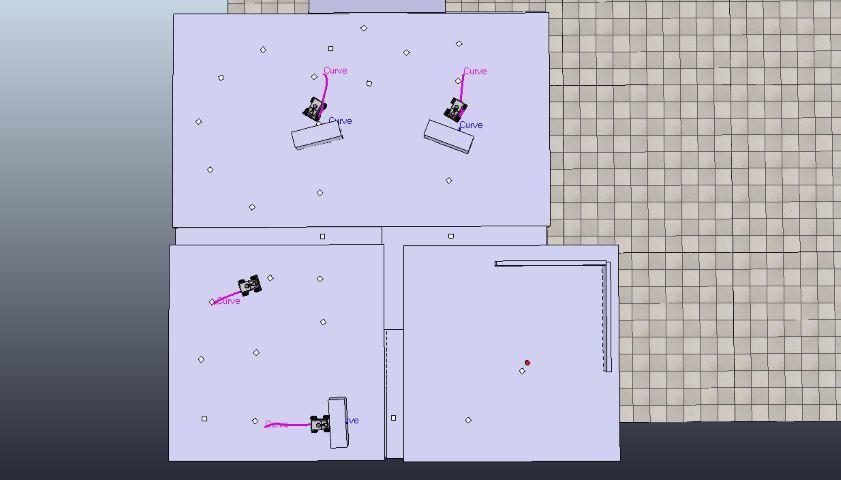} &
        \includegraphics[width=.32\textwidth, height=3.5cm]{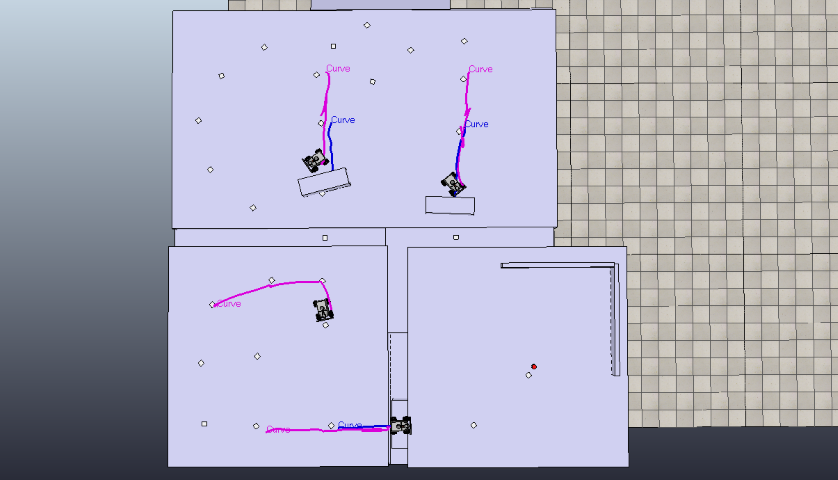} &
        \includegraphics[width=.32\textwidth, height=3.5cm]{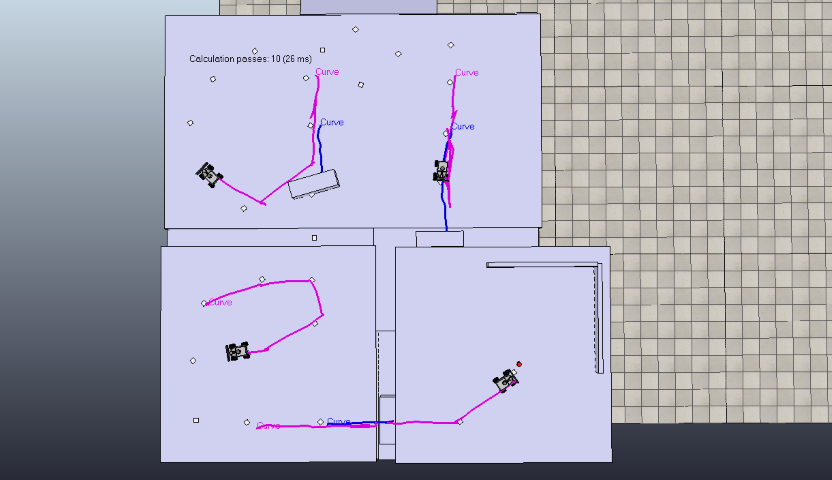} \\
        (a) & (b) & (c) \\
        \includegraphics[width=.32\textwidth, height=3.5cm]{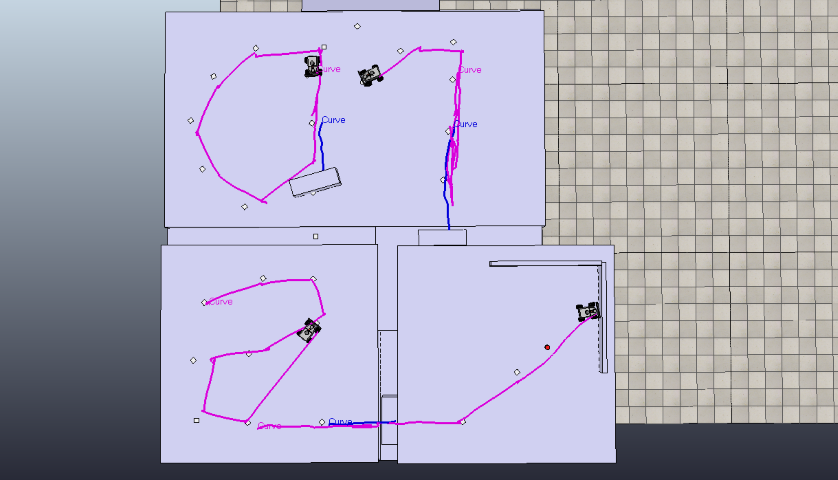} &
        \includegraphics[width=.32\textwidth, height=3.5cm]{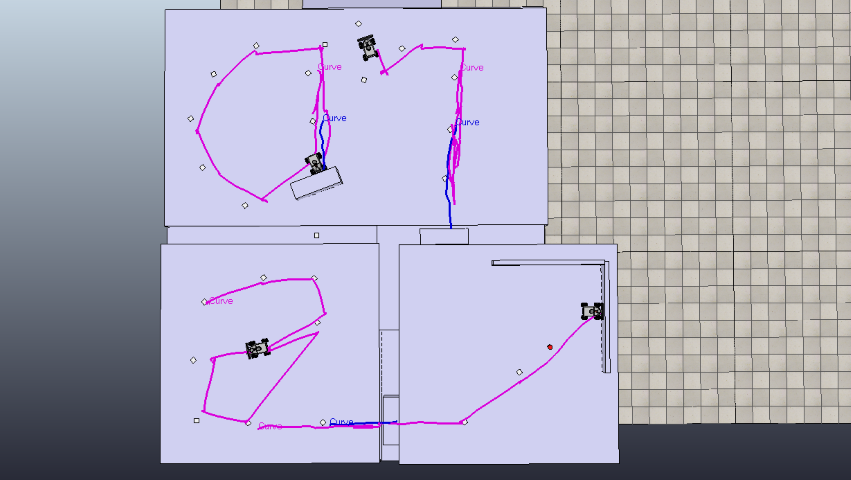} &
        \includegraphics[width=.32\textwidth, height=3.5cm]{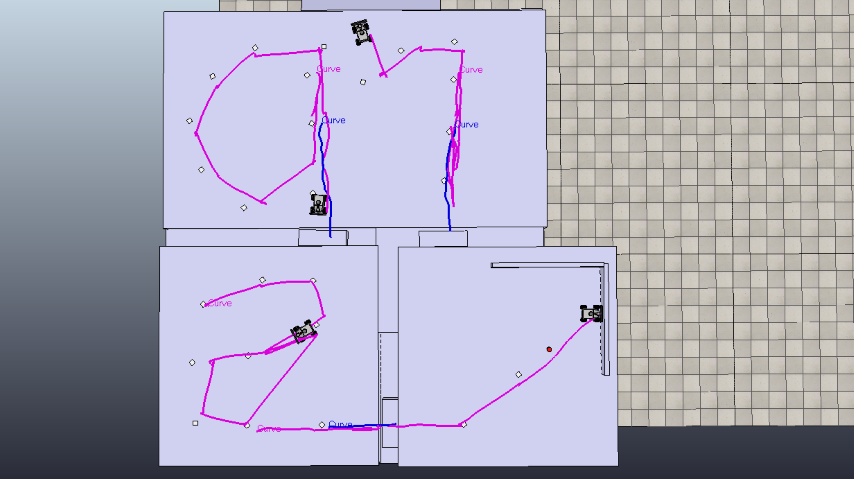} \\
        (d) & (e) & (f) \\
    \end{tabular}
    \caption{Sample trajectory during 1st episode in Medium environment. After an agent determines it is unable to push a box into a hole given its current position, it explores other nodes and stochastically arrives at a position where it is able to continue pushing the box into the hole.}
    \label{MediumTrajectoryFirst}
\end{figure}

The Medium environment features a middle ground of difficulty between the two as shown in Figure \ref{MediumTrajectoryFirst}. It is important to note that the top two boxes must be pushed across two nodes instead of one like in the previous environment. As a result, this endeavor is not only a test of the stigmergic algorithm but also a test of the control policy's composability. In other words, the rate of success in this environment is dependent on whether the same policy can be executed invoked twice for a given path. 

\begin{figure}[!ht]
    \centering
    \setlength{\tabcolsep}{1pt} 
    \begin{tabular}{cc}
        \includegraphics[width=.32\textwidth, height=3.5cm]{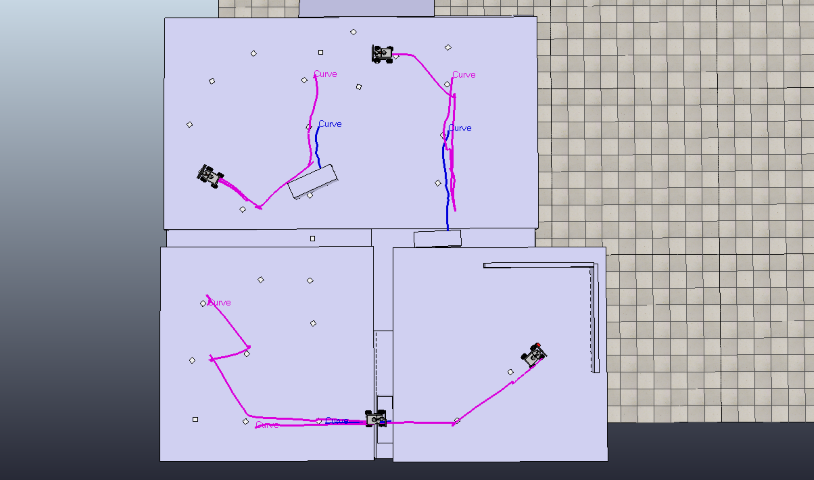} &
        \includegraphics[width=.32\textwidth, height=3.5cm]{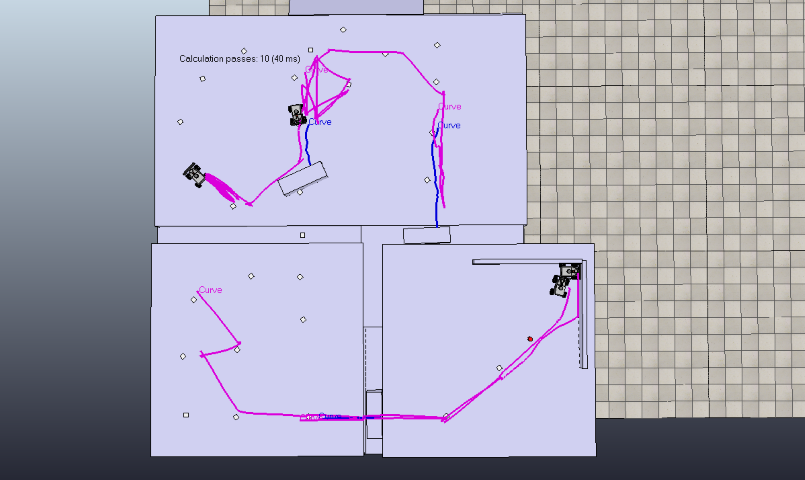} \\
        (a) & (b)\\
        \includegraphics[width=.32\textwidth, height=3.5cm]{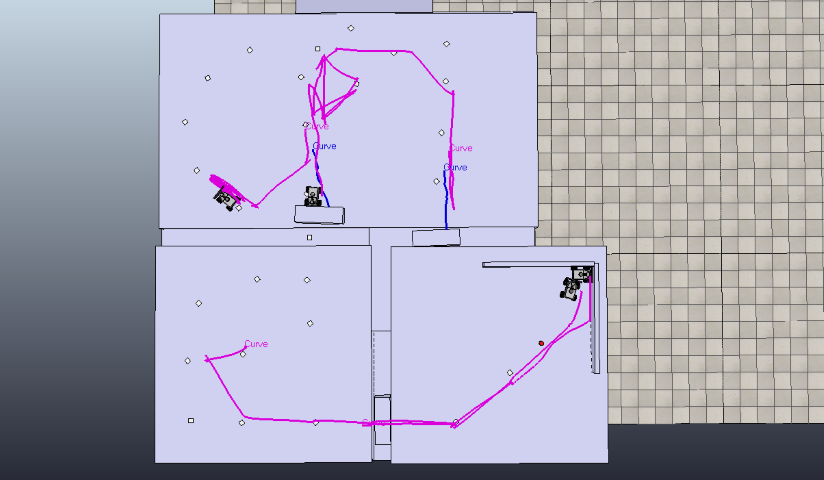} &
        \includegraphics[width=.32\textwidth, height=3.5cm]{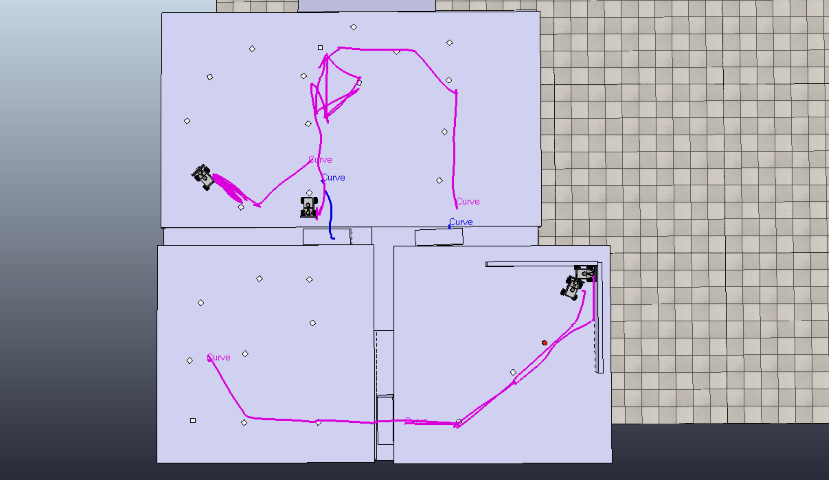} \\
        (d) & (e) \\
    \end{tabular}
    \caption{Sample trajectory during 3rd episode in Medium environment. After an agent determines it is unable to push a box into a hole given its current position, it moves aside to other nodes while a second agent continues pushing the attempted box into the hole. This demonstrates that an agent can easily continue another agent's work on a box being pushed.}
    \label{MediumTrajectoryThird}
\end{figure}

In Figure \ref{MediumTrajectoryFirst}, we show the trajectories for the first episode of training, where agents visit all nodes and update the D-pheromones using Euclidian distance. Furthermore, Figure \ref{MediumTrajectoryFirst} gives an example of the stigmergic policy and classifier at work. Analyzing the agent pushing the box in the upper left, it determines in Figure \ref{MediumTrajectoryFirst}c that it is unable to push the box into the hole given its current position. As a result, it decides to explore other nodes, update their D-pheromones, and stochastically arrive back at a location in Figure \ref{MediumTrajectoryFirst}d where it is able to push the box into the hole. This demonstrates a simple stigmergic rule that allows agents to reposition themselves if certain controls are unfeasible. Furthermore we show in Figure \ref{MediumTrajectoryThird} that this re-positioning and further attempts at pushing in a failed box can just as easily be achieved by another agent. This trajectory was sampled close to D-pheromone convergence, making the agent who originally pushed the box inclined to stay within the hole's vicinity while the other agent finished pushing it in.

\begin{figure}[!ht]
    \centering
    \includegraphics[width=.5\textwidth]{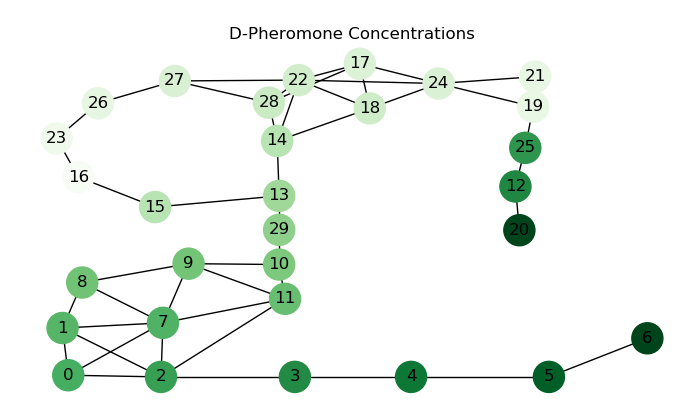}
    \caption{Converged D-Pheromone concentrations in Medium environment. Nodes $n_{25}$ and $n_{12}$ are often left without converged values as the agent must push the box into $n_{20}$ in order to reach these nodes. However, since use of this box is appropriately discouraged, as it serves no purpose for the goal objective, $n_{25}$ and $n_{12}$ are not updated.}
    \label{MediumPheromones}
\end{figure}

The associated D-Pheromones and E-Pheromones in this environment are similar to those of the Easy environment. As a result, only the D-Pheromones after convergence are shown in Figure \ref{MediumPheromones}.

\begin{figure}[!ht]
    \centering
    \setlength{\tabcolsep}{1pt} 
    \begin{tabular}{cc}
        \includegraphics[width=.45\textwidth, height=5cm]{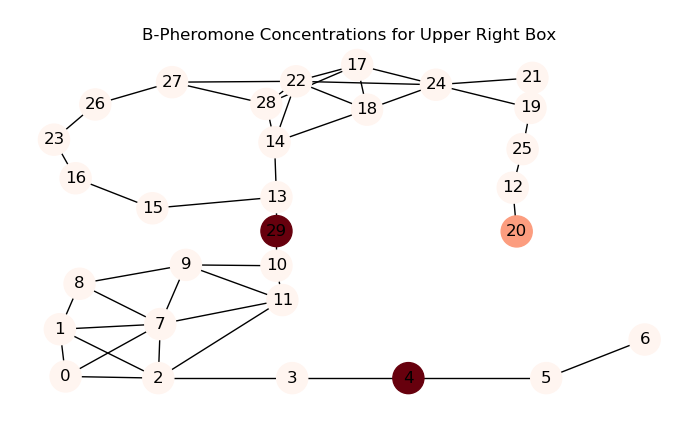} &
        \includegraphics[width=.45\textwidth, height=5cm]{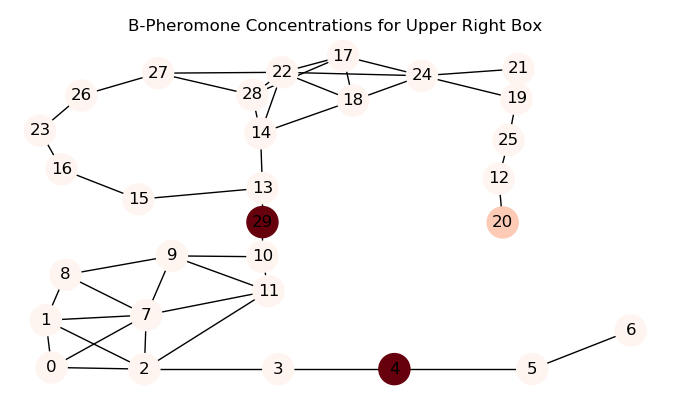} \\
        (a) & (b)\\
    \end{tabular}
    \caption{After attempting to push in the box into hole at $n_{20}$, the agent learns that doing so provides no utility to the given objective. As a result, the B-Pheromone associated with it is decayed. All other holes remain at their original levels because the classifier recognizes it is unable to push it to these locations.}
    \label{MediumBPheromone}
\end{figure}

\begin{figure}[!ht]
    \centering
    \setlength{\tabcolsep}{1pt} 
    \begin{tabular}{ccc}
        \includegraphics[width=.32\textwidth, height=3.5cm]{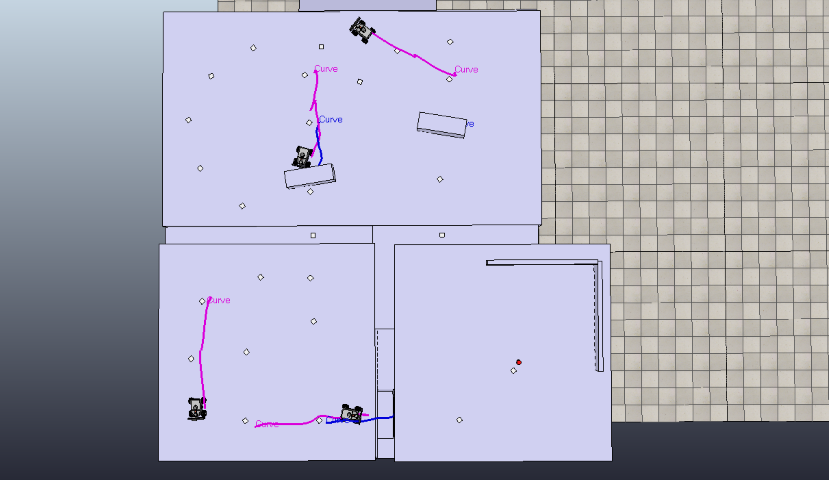} &
        \includegraphics[width=.32\textwidth, height=3.5cm]{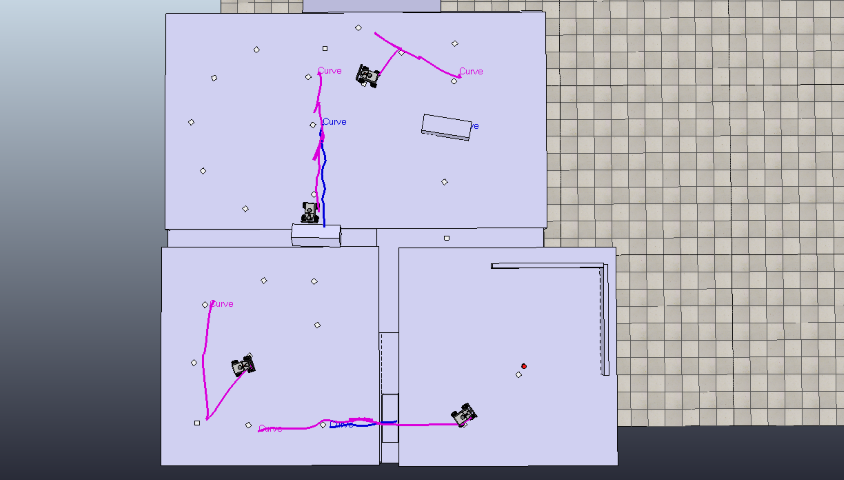} &
        \includegraphics[width=.32\textwidth, height=3.5cm]{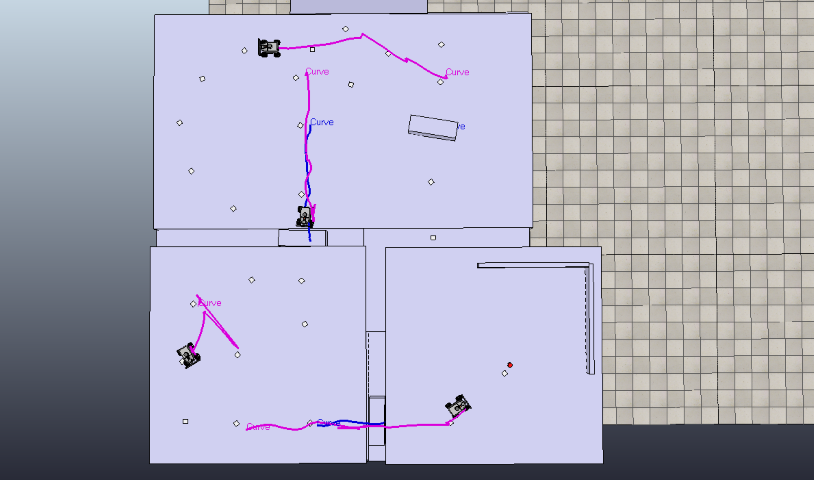} \\
        (a) & (b) & (c) \\
        \includegraphics[width=.32\textwidth, height=3.5cm]{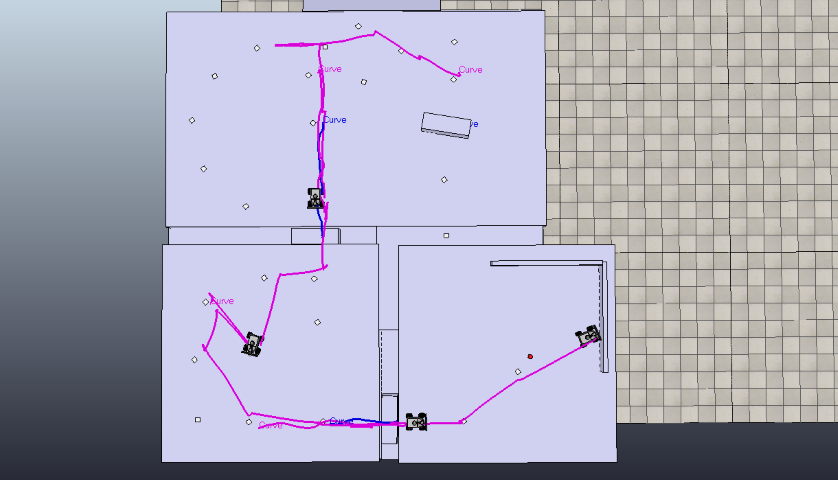} &
        \includegraphics[width=.32\textwidth, height=3.5cm]{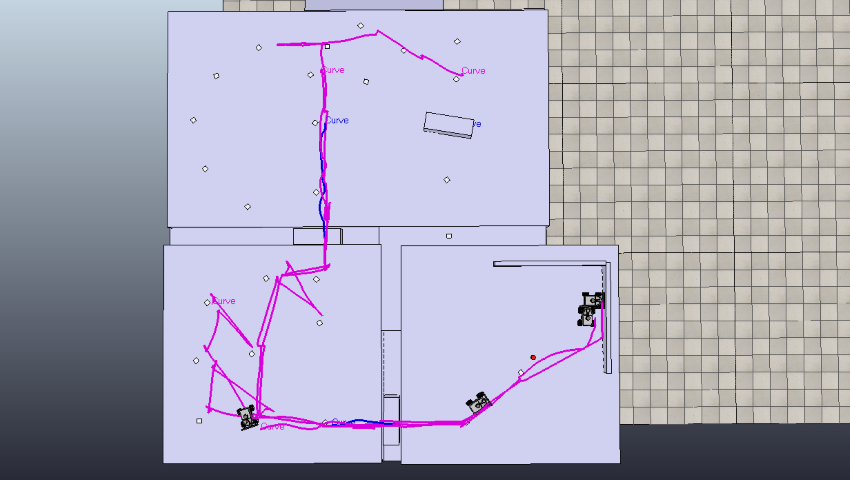} &
        \includegraphics[width=.32\textwidth, height=3.5cm]{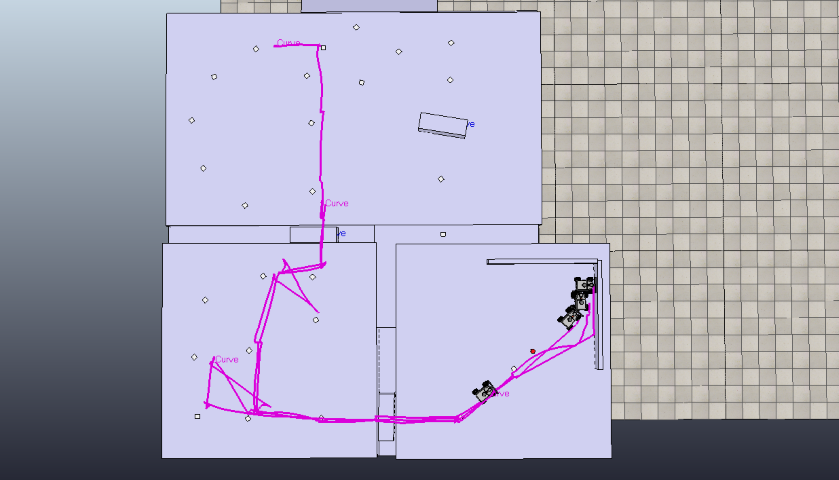} \\
        (d) & (e) & (f) \\
    \end{tabular}
    \caption{Sample trajectory after convergence in Medium environment. The upper right agent learns to ignore the upper left box, as it provides no utility in traveling to the goal location. The bottom two agents easily navigate to the goal location after the bottom box is pushed into the bottom hole. After the upper left agent successfully pushes the box into the left hole, all agents navigate to the goal location and complete the task.}
    \label{MediumTrajectoryEnd}
\end{figure}

However, it is interesting to analyze the B-Pheromones associated with the box in the upper right. This box is initially at $n_{25}$. As the classifier usually determines that an agent is unable to push the box to $n_{29}$ nor $n_4$, the only hole option is at $n_{20}$. However, $n_{20}$ provides no utility when traveling to $n_6$, the goal node. As a result, it makes sense for B-Pheromones associated to $n_{20}$ for this box would decay over time, and, eventually, agents would learn to ignore the box altogether. The B-Pheromones of this box are shown in Figure \ref{MediumBPheromone}. Furthermore, the resulting trajectory towards the end of training is shown in Figure \ref{MediumTrajectoryEnd}, where the upper-right box is discouraged from being used, and all agents navigate to the goal location appropriately. 

Although the algorithm yields a substantial amount of variability compared to results encountered in the previous section, integrating the trained policy with the stigmergic algorithm features limited success. These results are featured in Figure \ref{PerformanceEasyMedium}.

\begin{table}[!ht]
    \centering
    \begin{tabular}{ |p{2cm}|| p{4cm}| p{4cm} |}
     \hline
     \multicolumn{3}{|c|}{Performance in Varying Environments} \\
     \hline
     Difficulty & Number Steps & Proportion Success \\
     \hline
     Easy & 22.3625 $\pm$ 16.0708 & .9125 $\pm$ .1634 \\
     \hline
     Medium & 27.425 $\pm$ 18.521 & .8125 $\pm$ .2355 \\
     \hline 
    \end{tabular}
    \caption{Performance}
    \label{PerformanceEasyMedium}
\end{table}

\begin{figure}[!ht]
    \centering
    \setlength{\tabcolsep}{1pt} 
    \begin{tabular}{cc}
        \includegraphics[width=.45\textwidth, height=5cm]{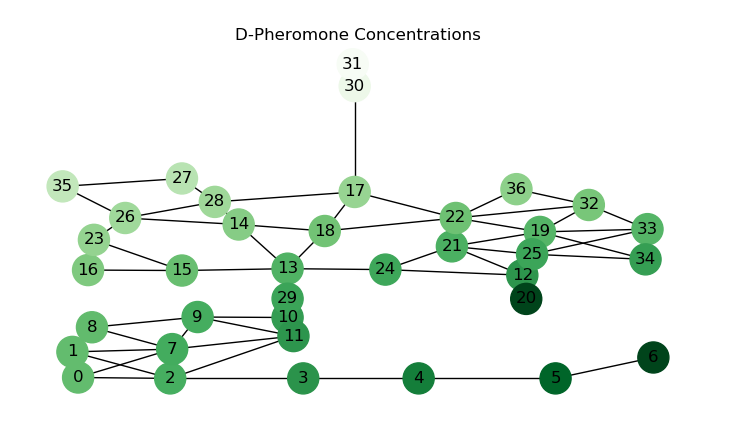} &
        \includegraphics[width=.45\textwidth, height=5cm]{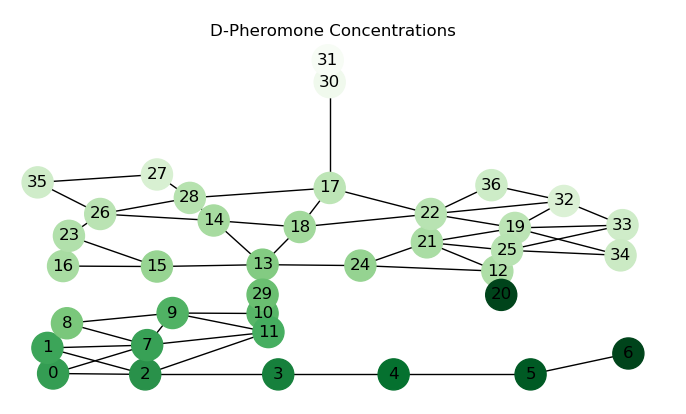} \\
        (a) & (b)\\
    \end{tabular}
    \caption{D-Pheromone concentrations in the Hard environment. Figure (a) displays D-Pheromone concentrations using Euclidian distance whereas Figure (b) displays D-Pheromone concentrations using path distances.}
    \label{HardDPheromone}
\end{figure}

\begin{figure}[!ht]
    \centering
    \setlength{\tabcolsep}{1pt} 
    \begin{tabular}{ccc}
        \includegraphics[width=.32\textwidth, height=3.5cm]{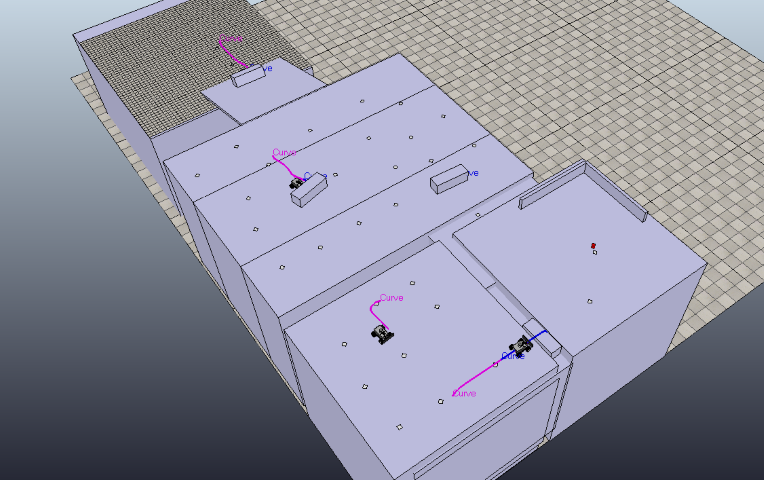} &
        \includegraphics[width=.32\textwidth, height=3.5cm]{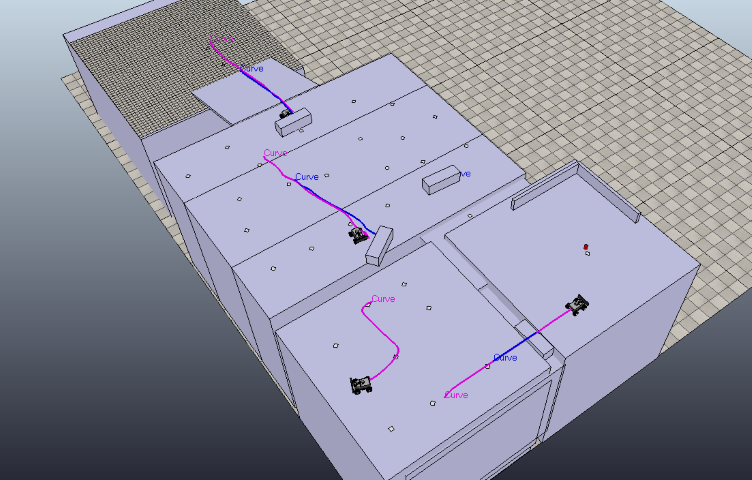} &
        \includegraphics[width=.32\textwidth, height=3.5cm]{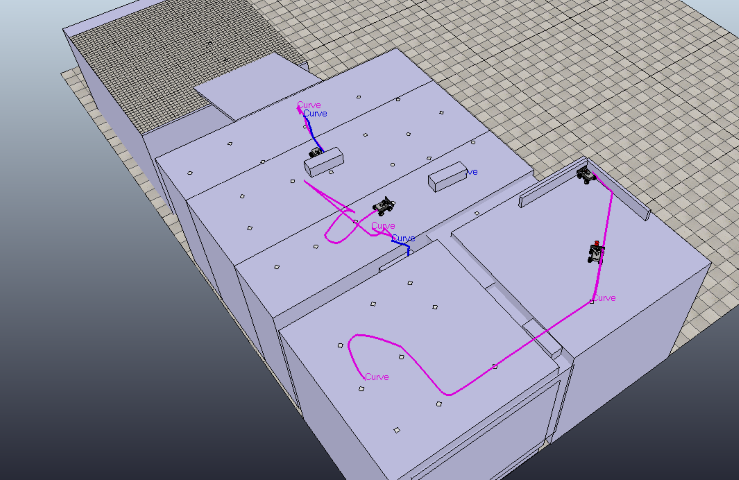} \\
        (a) & (b) & (c) \\
        \includegraphics[width=.32\textwidth, height=3.5cm]{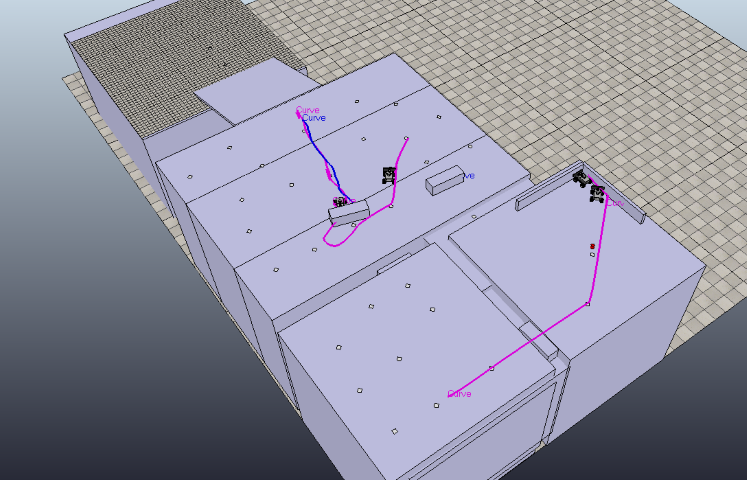} &
        \includegraphics[width=.32\textwidth, height=3.5cm]{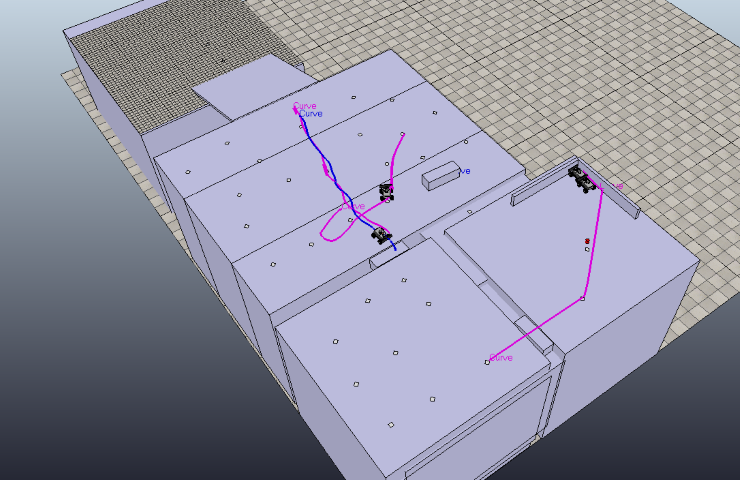} &
        \includegraphics[width=.32\textwidth, height=3.5cm]{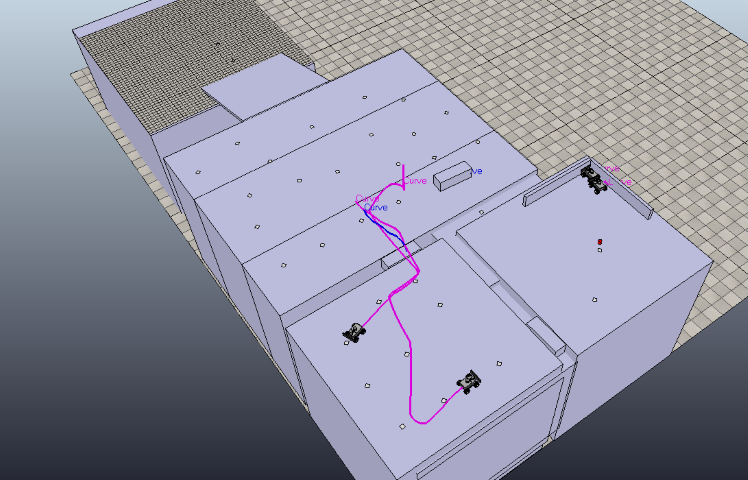} \\
        (d) & (e) & (f) \\
    \end{tabular}
    \caption{Converged agent trajectories in the Hard environment are shown for four agents using a detection radius of $R=20$, meaning agents can detect the presence of faraway holes. The agents successfully stack two boxes into the hole at node $n_{29}$ after pushing the box across long paths. Furthermore, the agent appropriately moves away from $n_{13}$ to not block the second box from being pushed in. Afterwards, all agents successfully traverse to the goal location.}
    \label{HardTrajectoriesEnd}
\end{figure}

In the Hard environment, we add an additional box. This box needs to travel up a ramp, through a series of nodes and into the hole at $n_{29}$. In particular, the task requires two boxes to be stacked at $n_{29}$ in order to traverse it. The two boxes required are at $n_{30}$ and $n_{18}$. We test this environment with 4 agents and large detection radius $R=20$. As a result, the agent initially at $n_31$ is able to detect the hole at $n_{29}$ and subsequently navigate towards it. D-Pheromone concentrations associated with this environment are shown in Figure \ref{HardDPheromone}, where we display D-pheromones calculated using Euclidian distance and Official path distance, respectively. A sample trajectory after convergence is shown in Figure \ref{HardTrajectoriesEnd}. Similar to the Medium environment, agents learn to ignore the box at $n_{12}$ as it is irrelevant to the task at hand. 

\begin{figure}[!ht]
    \centering
    \setlength{\tabcolsep}{1pt} 
    \begin{tabular}{cc}
        \includegraphics[width=.45\textwidth, height=5cm]{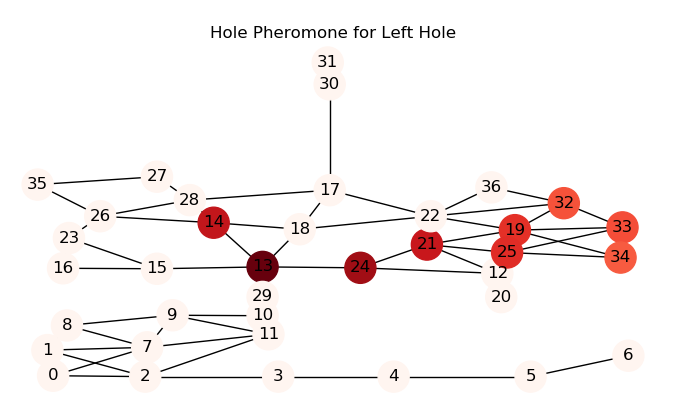} &
        \includegraphics[width=.45\textwidth, height=5cm]{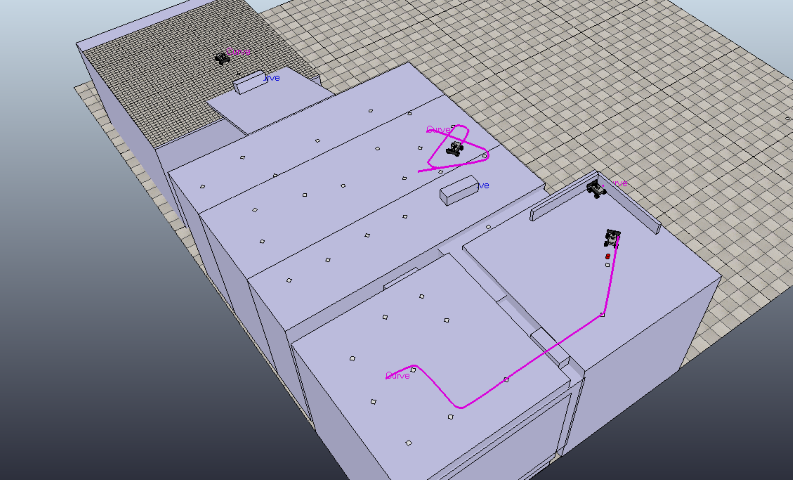} \\
        (a) & (b)\\
    \end{tabular}
    \caption{In the Hard environment with four agents and a reduced detection radius $R=5$, the agent in the upper lefthand corner is unable to detect the hole at $n_{29}$. As a result, it requires the presence of associated H-Pheromones at node $n_{17}$ and a trail to lead it towards $n_{29}$. However, the one agent on the upper platform is unable to provide that information, because it rarely traverses near $n_{17}$. Instead, in the first few episodes, it exploits Euclidian distance-calculated D-Pheromones, making it hover in the upper right side of the map. This behavior prevents the goal from being achieved by all agents.}
    \label{Hard4HPheromone}
\end{figure}

While performance is promising with large detection radius, decreasing detection radius to $R=5$ dramatically decreases performance and typically precludes the agents from traveling to the goal node $n_6$. When $R$ is small, the agent at $n_{31}$ relies on H-Pheromones being placed at $n_{17}$ in order to notify the agent of a path to some hole. However, given the current hyper-parameters, it is unlikely that these H-Pheromones are developed, as the middle agent is unable to traverse the nodes appropriately and provide the necessary pheromone updates. This is shown in Figure \ref{Hard4HPheromone}, where the agent hovers in the upper right section of the map as opposed to traveling towards $n_{17}$. As a result, the H-Pheromones associated to $n_{29}$ do not develop properly. 

\begin{figure}[!ht]
    \centering
    \setlength{\tabcolsep}{1pt} 
    \begin{tabular}{ccc}
        \includegraphics[width=.32\textwidth, height=4cm]{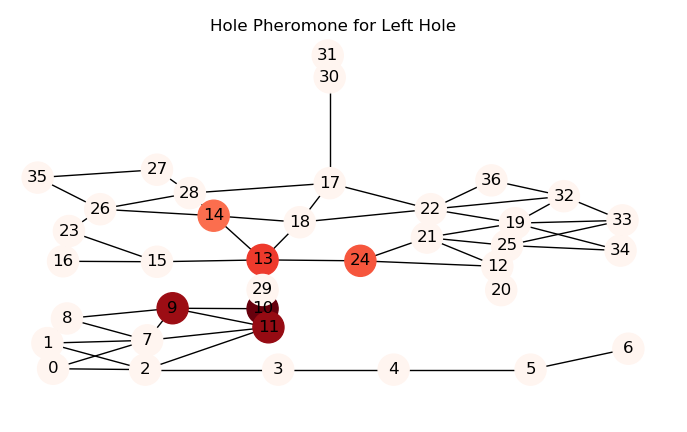} &
        \includegraphics[width=.32\textwidth, height=4cm]{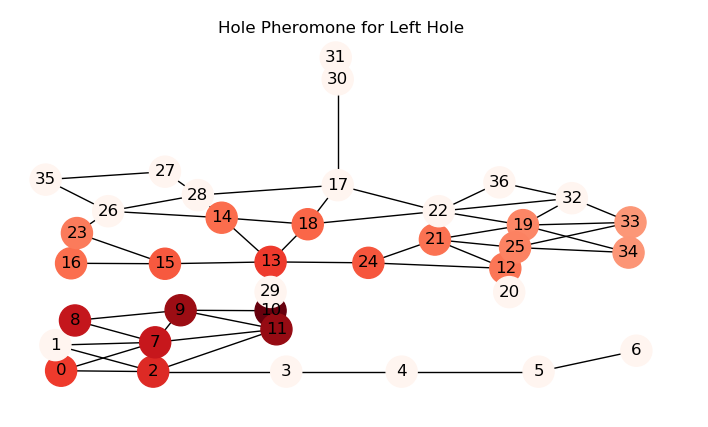} &
        \includegraphics[width=.32\textwidth, height=4cm]{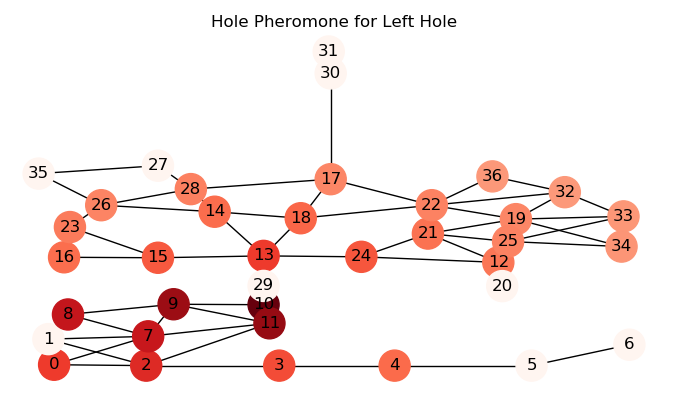} \\
        (a) & (b) & (c) \\
        \includegraphics[width=.32\textwidth, height=4cm]{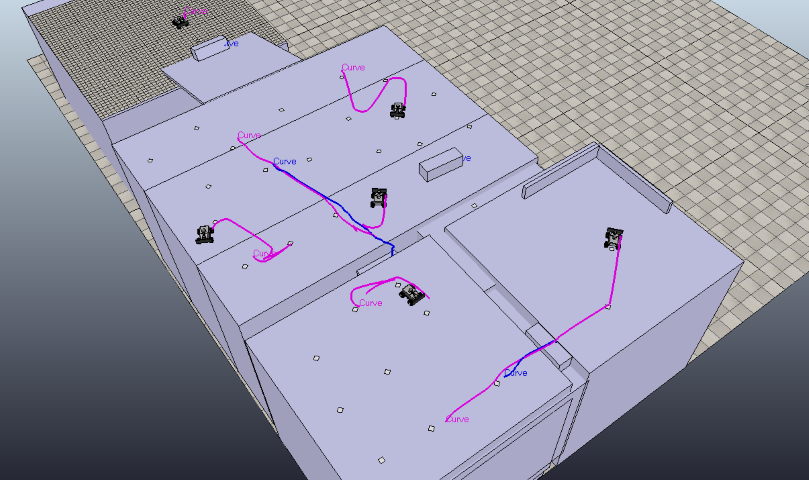} &
        \includegraphics[width=.32\textwidth, height=4cm]{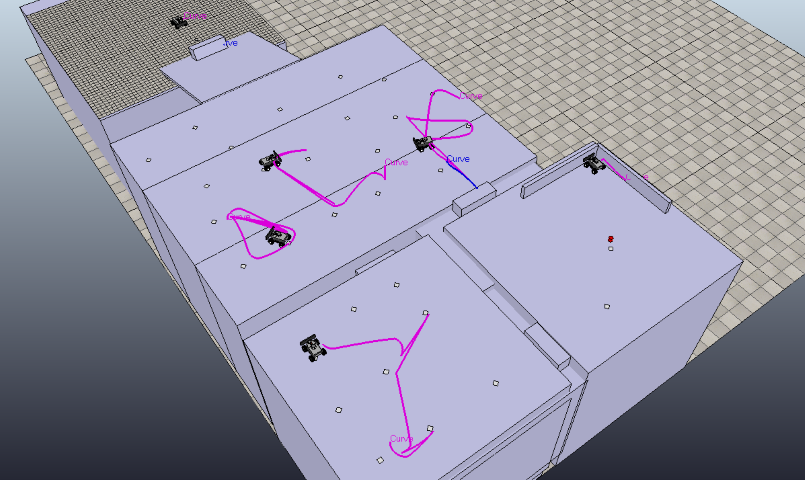} &
        \includegraphics[width=.32\textwidth, height=4cm]{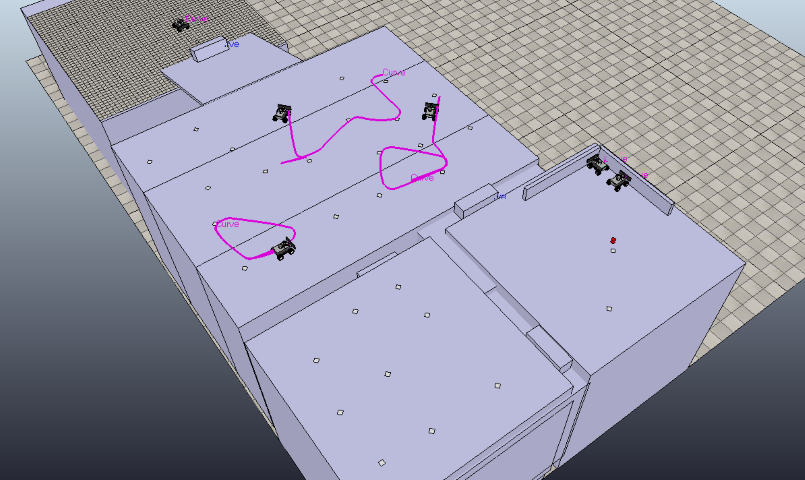} \\
        (d) & (e) & (f) \\
    \end{tabular}
    \caption{Hole Pheromones associated to the hole at $n_{29}$ are successfully propagated both before and after convergence with the addition of two more agents. The H-Pheromones and trajectories shown are from agent behavior after D-Pheromone convergence.}
    \label{Hard6HPheromone}
\end{figure}

\begin{figure}[!ht]
    \centering
    \setlength{\tabcolsep}{1pt} 
    \begin{tabular}{ccc}
        \includegraphics[width=.4\textwidth, height=5cm]{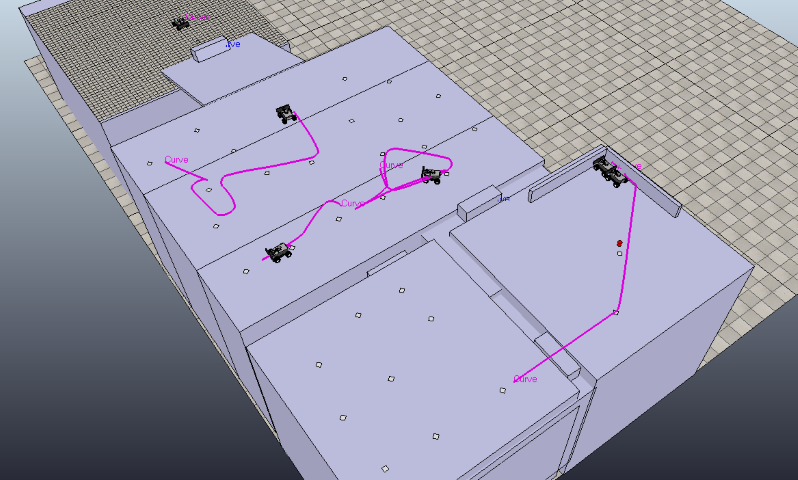} &
        \includegraphics[width=.4\textwidth, height=5cm]{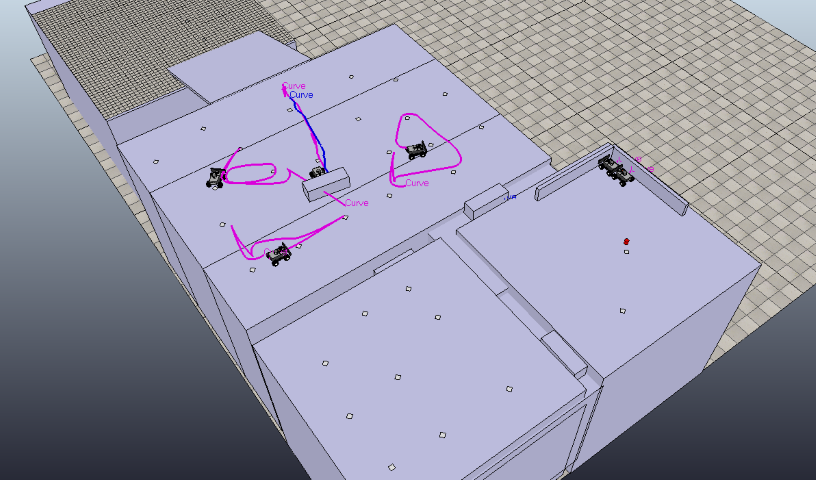} \\
        (a) & (b) \\
        \includegraphics[width=.4\textwidth, height=5cm]{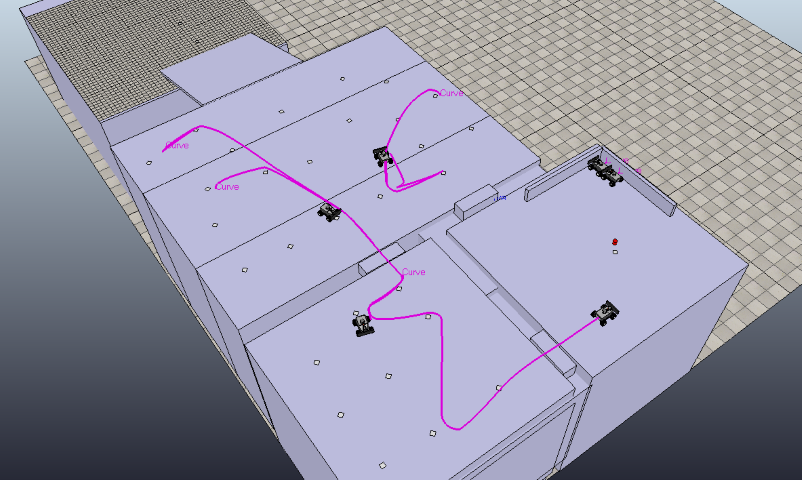} &
        \includegraphics[width=.4\textwidth, height=5cm]{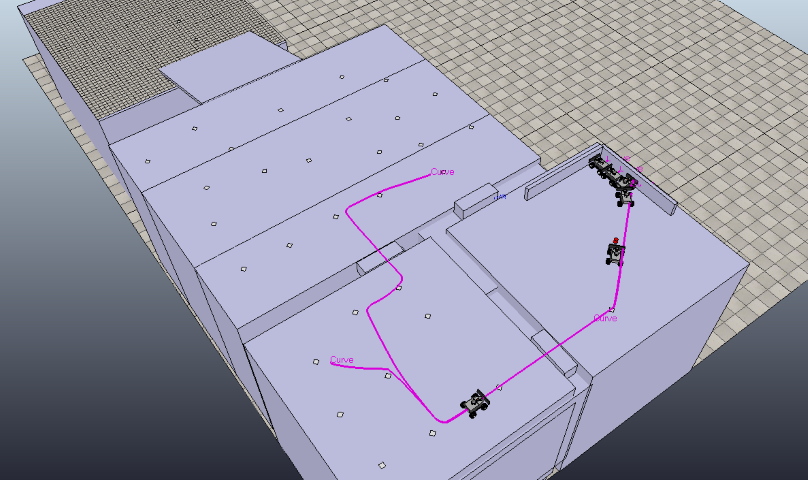} \\
        (c) & (d)\\
    \end{tabular}
    \caption{After the first box is pushed into $n_{29}$ and H-Pheromones are propagated to $n_{17}$ (a), a sample trajectory with six agents in Hard Environment is shown. All six agents successfully navigate to the goal location.}
    \label{HardTrajectoriesEnd2}
\end{figure}

To remedy this issue, we could tune exploration probability $\epsilon$ or distance Boltzmann $\beta$ to encourage more stochastic decisions. That way, the agent is more likely to naturally travel to $n_{17}$ and develop the appropriate H-Pheromones. However, higher $\beta$ and $\epsilon$ corresponds to less determinism, meaning limited performance after convergence as agents are more susceptible to making decisions that don't contribute directly to the task. As a result, we demonstrate the effectiveness of another solution: adding additional agents while keeping other hyper-parameters fixed. In doing so, determinism is preserved and the H-pheromones are more likely to be developed. This is a direct consequence of increase node coverage and spread from the presence of more agents. The results are shown in Figure \ref{Hard6HPheromone}. Equipped with these two additional agents and the appropriate H-Pheromone updates, Figure \ref{HardTrajectoriesEnd2} shows a sample trajectory where all agents reach $n_6$.

The performance results are summarized in Table \ref{HardResults}. The results reflect the advantages of equipping agents with more information, or high values of $R$. However, when such information is not accessible, an increased number of agents compensates for such losses, with limited performance. Furthermore, although the results reflect promising results and reliability, the largest bottleneck is the composability of the trained control policy. As mentioned before, it remains a possibility that boxes are unsuccessfully pushed into holes despite the classifier determining that an agent is able to. In addition, agents may never attain positions or orientations that make pushing a box into a hole achievable. The proportion of agents that are successful in each episode reflects this uncertainty. 

\begin{table}[!ht]
    \centering
    \begin{tabular}{ |p{3cm}| p{3cm}|| p{4cm}| p{4cm} |}
     \hline
     \multicolumn{4}{|c|}{Performance in Hard Environment} \\
     \hline
     Detection Radius & Agents & Steps & Proportion Success \\
     \hline
     20 & 4 & 28.25 $\pm$ 17.975 & .7875 $\pm$ .2655 \\
     \hline
     5 & 4 & 80 $\pm$ 0 & .5 $\pm$ 0 \\
     \hline 
     5 & 6 & 55.825 $\pm$ 26.888 & .73 $\pm$ .29 \\
     \hline
    \end{tabular}
    \caption{Performance in terms of maximum number of steps until all agents reach the goal node and the proportion of agents that reach the goal node}
    \label{HardResults}
\end{table}

\newpage 

\section{Conclusion and Future Work}
While the single-agent RL and stigmergic algorithm perform well independently in their respective environments, performance is limited when integrating the latter into the former. This can be credited to two things: the lack of composability in the lower level policy and miscellaneous environment considerations, such as automobile wheels causing boxes to shift from their originally placed positions after pushing them into holes. This lack of composability fails to be accounted for in the results of Section 2, as shown in Table \ref{PolicyStyleTest}, as we had primarily gauged performance on single episode performance. However, when the policy is used as a mid-level primitive in Section 3, performance is limited as shown in Tables \ref{PerformanceEasyMedium} and \ref{HardResults}, where success rates would be perfect if decoupled from the trained control primitive. 

The algorithm shows promising success even when agents have access to extremely limited information. In particular, performance shown in Figures \ref{MediumTrajectoryEnd} and \ref{HardTrajectoriesEnd2} demonstrate the ability of agents to use pheromones to accomplish an overall, coordinated goal with access to only local information. The distinct properties of the various introduced pheromones provided promising success, particularly when decoupled from the mid-level trained control policy. D-Pheromones were shown to, with sufficient node exploration, converge appropriately to values that were inversely proportional to its path distance to the goal, as shown in Figures  \ref{SanityDevelopment}, \ref{EasyDPheromone}, \ref{MediumPheromones}, and \ref{HardDPheromone}. B-Pheromones appropriately determined not only where to place boxes but also which boxes could be ignored, shown in Figures \ref{SanityDevelopment}, \ref{MediumBPheromone}, \ref{MediumTrajectoryEnd} and \ref{HardTrajectoriesEnd}. E-Pheromones reflected areas that required more exploration, appropriately guiding agents to intelligently explore their environment as depicted in Figures \ref{SanityDevelopment} and \ref{EasyEPheromone}. Lastly, H-Pheromones appropriately notified agents of nearby holes along with paths towards them. While these paths are not guaranteed to be optimal, performance was not hindered by this edge case in the environments tested, as shown in Figures \ref{Hard4HPheromone} and \ref{Hard6HPheromone}. 

The idea of having a hierarchical approach to stigmergic learning, while appealing, as shown yields limited success and fails to address some edge cases, particularly ones regarding H-Pheromones in Figure \ref{HPheromoneEdgeCase} or collision detection that may result in indefinite standstills as explained in Section 3.3.5. Furthermore, we did not explore a more autonomous or smooth method of merging the two approaches apart from the relatively explicit conditional in Algorithm \ref{OverallAlgorithm}, where a random forest classifier is used in an if-statement to help determine which nodes an agent can travel to. This degree of human input, coupled with the explicitness of stigmergic algorithms as a whole, is a characteristic that can be alleviated in future works.

To curb the lack of composability of the control policy and limitations of combining disparate methods, future works may tackle joint learning and local control in one, cohesive algorithm. We saw previously that learning policies at an individual level, as shown in Section 2, and subsequently using an explicit stigmergic algorithm yielded promising yet limited results, as mentioned previously in Table \ref{PerformanceEasyMedium} and \ref{HardResults}. As a result, considering algorithms that learn stigmergic multi-agent policies directly may yield more favorable results, whether it be through MARL or other iterative approaches. It was shown in Section 2 that a hierarchical RL approach can yield sample-efficient control for single-episode trajectories, as shown in Figure \ref{SingleDDQNAblation} and Table \ref{PolicyStyleTest}. Applying similarly spirited approaches to directly to multi-agent domains may provide more reliable, stable performance. Instead of training agents individually and using a different algorithm to connect the pieces, training all agents in a single environment will alleviate ramifications from the composability issue. Furthermore, more generalizable algorithms can be devised to be applicable in a variety of environments, as opposed to specific ones such as those presented here. 

\medskip

\printbibliography[title={References}]

@article{FeudalMARL,
  title={Feudal multi-agent hierarchies for cooperative reinforcement learning},
  author={Ahilan, Sanjeevan and Dayan, Peter},
  journal={arXiv preprint arXiv:1901.08492},
  year={2019}
}

@article{DeepNashQ,
  title={Deep {Q}-learning for Nash equilibria: Nash-{DQN}},
  author={Casgrain, Philippe and Ning, Brian and Jaimungal, Sebastian},
  journal={arXiv preprint arXiv:1904.10554},
  year={2019}
}

@article{AntTSP,
  title={Ant algorithms and stigmergy},
  author={Dorigo, Marco and Bonabeau, Eric and Theraulaz, Guy},
  journal={Future Generation Computer Systems},
  volume={16},
  number={8},
  pages={851--871},
  year={2000},
  publisher={Elsevier}
}

@inproceedings{CuldeSac,
  title={Models of adaptive navigation, inspired by ant transport strategy in the presence of obstacles},
  author={Esterly, Elizabeth E and McCreery, Helen and Nagpal, Radhika},
  booktitle={2017 IEEE symposium series on computational intelligence (SSCI)},
  pages={1--8},
  year={2017},
  organization={IEEE}
}

@inproceedings{Counterfactual,
  title={Counterfactual multi-agent policy gradients},
  author={Foerster, Jakob and Farquhar, Gregory and Afouras, Triantafyllos and Nardelli, Nantas and Whiteson, Shimon},
  booktitle={Proceedings of the AAAI Conference on Artificial Intelligence},
  volume={32},
  number={1},
  year={2018}
}

@article{SNN,
  title={Stochastic neural networks for hierarchical reinforcement learning},
  author={Florensa, Carlos and Duan, Yan and Abbeel, Pieter},
  journal={arXiv preprint arXiv:1704.03012},
  year={2017}
}

@inproceedings{TD3,
  title={Addressing function approximation error in actor-critic methods},
  author={Fujimoto, Scott and Hoof, Herke and Meger, David},
  booktitle={International Conference on Machine Learning},
  pages={1587--1596},
  year={2018},
  organization={PMLR}
}

@article{HierarchicalMARL,
  title={Hierarchical multi-agent reinforcement learning},
  author={Ghavamzadeh, Mohammad and Mahadevan, Sridhar and Makar, Rajbala},
  journal={Autonomous Agents and Multi-Agent Systems},
  volume={13},
  number={2},
  pages={197--229},
  year={2006},
  publisher={Springer}
}

@inproceedings{SAC,
  title={Soft actor-critic: Off-policy maximum entropy deep reinforcement learning with a stochastic actor},
  author={Haarnoja, Tuomas and Zhou, Aurick and Abbeel, Pieter and Levine, Sergey},
  booktitle={International Conference on Machine Learning},
  pages={1861--1870},
  year={2018},
  organization={PMLR}
}

@inproceedings{Opponent,
  title={Opponent modeling in deep reinforcement learning},
  author={He, He and Boyd-Graber, Jordan and Kwok, Kevin and Daum{\'e} III, Hal},
  booktitle={International conference on machine learning},
  pages={1804--1813},
  year={2016},
  organization={PMLR}
}

@article{Survey1,
  title={A survey and critique of multiagent deep reinforcement learning},
  author={Hernandez-Leal, Pablo and Kartal, Bilal and Taylor, Matthew E},
  journal={Autonomous Agents and Multi-Agent Systems},
  volume={33},
  number={6},
  pages={750--797},
  year={2019},
  publisher={Springer}
}

@article{Frisbee,
  title={Stigmergy, self-organization, and sorting in collective robotics},
  author={Holland, Owen and Melhuish, Chris},
  journal={Artificial life},
  volume={5},
  number={2},
  pages={173--202},
  year={1999},
  publisher={MIT Press}
}

@article{NashQ,
  title={Nash Q-learning for general-sum stochastic games},
  author={Hu, Junling and Wellman, Michael P},
  journal={Journal of machine learning research},
  volume={4},
  number={Nov},
  pages={1039--1069},
  year={2003}
}

@article{RFID,
  title={Stigmergic algorithms for multiple minimalistic robots on an RFID floor},
  author={Khaliq, Ali Abdul and Di Rocco, Maurizio and Saffiotti, Alessandro},
  journal={Swarm Intelligence},
  volume={8},
  number={3},
  pages={199--225},
  year={2014},
  publisher={Springer}
}

@inproceedings{Bohg,
  title={Learning hierarchical control for robust in-hand manipulation},
  author={Li, Tingguang and Srinivasan, Krishnan and Meng, Max Qing-Hu and Yuan, Wenzhen and Bohg, Jeannette},
  booktitle={2020 IEEE International Conference on Robotics and Automation (ICRA)},
  pages={8855--8862},
  year={2020},
  organization={IEEE}
}

@inproceedings{M3DDPG,
  title={Robust multi-agent reinforcement learning via minimax deep deterministic policy gradient},
  author={Li, Shihui and Wu, Yi and Cui, Xinyue and Dong, Honghua and Fang, Fei and Russell, Stuart},
  booktitle={Proceedings of the AAAI Conference on Artificial Intelligence},
  volume={33},
  number={01},
  pages={4213--4220},
  year={2019}
}

@article{DDPG,
  title={Continuous control with deep reinforcement learning},
  author={Lillicrap, Timothy P and Hunt, Jonathan J and Pritzel, Alexander and Heess, Nicolas and Erez, Tom and Tassa, Yuval and Silver, David and Wierstra, Daan},
  journal={arXiv preprint arXiv:1509.02971},
  year={2015}
}

@inproceedings{FriendOrFoe,
  title={Friend-or-foe {Q}-learning in general-sum games},
  author={Littman, Michael L},
  booktitle={ICML},
  volume={1},
  pages={322--328},
  year={2001}
}

@incollection{Minimax,
  title={Markov games as a framework for multi-agent reinforcement learning},
  author={Littman, Michael L},
  booktitle={Machine learning proceedings 1994},
  pages={157--163},
  year={1994},
  publisher={Elsevier}
}

@article{MADDPG,
  title={Multi-agent actor-critic for mixed cooperative-competitive environments},
  author={Lowe, Ryan and Wu, Yi and Tamar, Aviv and Harb, Jean and Abbeel, Pieter and Mordatch, Igor},
  journal={arXiv preprint arXiv:1706.02275},
  year={2017}
}

@article{ScarceBox,
  title={Autonomous construction using scarce resources in unknown environments},
  author={Magnenat, St{\'e}phane and Philippsen, Roland and Mondada, Francesco},
  journal={Autonomous robots},
  volume={33},
  number={4},
  pages={467--485},
  year={2012},
  publisher={Springer}
}

@inproceedings{Planner9,
  title={Planner9, a {HTN} planner distributed on groups of miniature mobile robots},
  author={Magnenat, St{\'e}phane and Voelkle, Martin and Mondada, Francesco},
  booktitle={International Conference on Intelligent Robotics and Applications},
  pages={1013--1022},
  year={2009},
  organization={Springer}
}

@inproceedings{EcitonBridge,
  title={Eciton robotica: Design and algorithms for an adaptive self-assembling soft robot collective},
  author={Malley, Melinda and Haghighat, Bahar and Houe, Lucie and Nagpal, Radhika},
  booktitle={2020 IEEE International Conference on Robotics and Automation (ICRA)},
  pages={4565--4571},
  year={2020},
  organization={IEEE}
}

@inproceedings{A3C,
  title={Asynchronous methods for deep reinforcement learning},
  author={Mnih, Volodymyr and Badia, Adria Puigdomenech and Mirza, Mehdi and Graves, Alex and Lillicrap, Timothy and Harley, Tim and Silver, David and Kavukcuoglu, Koray},
  booktitle={International Conference on Machine Learning},
  pages={1928--1937},
  year={2016},
  organization={PMLR}
}

@article{DQN,
  title={Playing atari with deep reinforcement learning},
  author={Mnih, Volodymyr and Kavukcuoglu, Koray and Silver, David and Graves, Alex and Antonoglou, Ioannis and Wierstra, Daan and Riedmiller, Martin},
  journal={arXiv preprint arXiv:1312.5602},
  year={2013}
}

@article{HIRO,
  title={Data-efficient hierarchical reinforcement learning},
  author={Nachum, Ofir and Gu, Shixiang and Lee, Honglak and Levine, Sergey},
  journal={arXiv preprint arXiv:1805.08296},
  year={2018}
}

@inproceedings{Image-Conditioned,
  title={Learning image-conditioned dynamics models for control of underactuated legged millirobots},
  author={Nagabandi, Anusha and Yang, Guangzhao and Asmar, Thomas and Pandya, Ravi and Kahn, Gregory and Levine, Sergey and Fearing, Ronald S},
  booktitle={2018 IEEE/RSJ International Conference on Intelligent Robots and Systems (IROS)},
  pages={4606--4613},
  year={2018},
  organization={IEEE}
}

@inproceedings{ModelFreeTuning,
  title={Neural network dynamics for model-based deep reinforcement learning with model-free fine-tuning},
  author={Nagabandi, Anusha and Kahn, Gregory and Fearing, Ronald S and Levine, Sergey},
  booktitle={2018 IEEE International Conference on Robotics and Automation (ICRA)},
  pages={7559--7566},
  year={2018},
  organization={IEEE}
}

@inproceedings{Self-play,
  title={Modeling others using oneself in multi-agent reinforcement learning},
  author={Raileanu, Roberta and Denton, Emily and Szlam, Arthur and Fergus, Rob},
  booktitle={International Conference on Machine Learning},
  pages={4257--4266},
  year={2018},
  organization={PMLR}
}

@inproceedings{TRPO,
  title={Trust region policy optimization},
  author={Schulman, John and Levine, Sergey and Abbeel, Pieter and Jordan, Michael and Moritz, Philipp},
  booktitle={International Conference on Machine Learning},
  pages={1889--1897},
  year={2015},
  organization={PMLR}
}

@article{PPO,
  title={Proximal policy optimization algorithms},
  author={Schulman, John and Wolski, Filip and Dhariwal, Prafulla and Radford, Alec and Klimov, Oleg},
  journal={arXiv preprint arXiv:1707.06347},
  year={2017}
}

@inproceedings{IQL,
  title={Multi-agent reinforcement learning: Independent vs. cooperative agents},
  author={Tan, Ming},
  booktitle={Proceedings of the Tenth International Conference on Machine Learning},
  pages={330--337},
  year={1993}
}

@inproceedings{DDQN,
  title={Deep reinforcement learning with double q-learning},
  author={Van Hasselt, Hado and Guez, Arthur and Silver, David},
  booktitle={Proceedings of the AAAI Conference on Artificial Intelligence},
  volume={30},
  number={1},
  year={2016}
}

@inproceedings{FeudalNets,
  title={Feudal networks for hierarchical reinforcement learning},
  author={Vezhnevets, Alexander Sasha and Osindero, Simon and Schaul, Tom and Heess, Nicolas and Jaderberg, Max and Silver, David and Kavukcuoglu, Koray},
  booktitle={International Conference on Machine Learning},
  pages={3540--3549},
  year={2017},
  organization={PMLR}
}

@article{SIRL,
  title={Stigmergic Independent Reinforcement Learning for Multiagent Collaboration},
  author={Xu, Xing and Li, Rongpeng and Zhao, Zhifeng and Zhang, Honggang},
  journal={IEEE Transactions on Neural Networks and Learning Systems},
  year={2021},
  publisher={IEEE}
}

@techreport{Survey2,
  title={Multiagent reinforcement learning for multi-robot systems: A survey},
  author={Yang, Erfu and Gu, Dongbing},
  year={2004},
  institution={tech. rep}
}

@inproceedings{Survey3,
  title={A Survey on Multiagent Reinforcement Learning Towards Multi-Robot Systems.},
  author={Yang, Erfu and Gu, Dongbing},
  booktitle={CIG},
  year={2005},
  organization={Citeseer}
}

\end{document}